\documentclass[fleqn,usenatbib]{mnras}

\usepackage{newtxtext,newtxmath}

\usepackage[T1]{fontenc}

\DeclareRobustCommand{\VAN}[3]{#2}
\let\VANthebibliography\thebibliography
\def\thebibliography{\DeclareRobustCommand{\VAN}[3]{##3}\VANthebibliography}

\usepackage{graphicx}	
\usepackage{amsmath}	
\usepackage{orcidlink}

\newcommand{\aman}[1]{\textcolor{red}{{}{\textbf{#1}}{}}}

\title[Colours of galaxies with tidal features]{Investigating the imprints of tidal features on simulated galaxy outskirts in LSST-like mock observations}

\author[A. Khalid et al.]{
Aman Khalid\textsuperscript{\orcidlink{0000-0003-1302-4426}},$^{1,2}$\thanks{E-mail: aman.khalid@unsw.edu.au}
Sarah Brough\textsuperscript{\orcidlink{0000-0002-9796-1363}},$^{1,2}$
Garreth Martin\textsuperscript{\orcidlink{0000-0003-2939-8668}},$^{3}$
Lucas C. Kimmig\textsuperscript{\orcidlink{0009-0006-8337-8712}},$^{4}$ 
Rhea-Silvia Remus\textsuperscript{\orcidlink{0009-0008-9260-7278}},$^{4}$
\newauthor
Claudia del P. Lagos\textsuperscript{\orcidlink{0000-0003-3021-8564}},$^{2,5}$
Louisa Canepa\textsuperscript{\orcidlink{0009-0002-5450-6683}},$^{1,2}$
Alice Desmons\textsuperscript{\orcidlink{0000-0002-4225-1890}}$^{1,2}$
\\
$^{1}$School of Physics, University of New South Wales, NSW 2052, Australia\\
$^{2}$ARC Centre of Excellence for All Sky Astrophysics in 3 Dimensions (ASTRO 3D)\\
$^{3}$School of Physics and Astronomy, University of Nottingham, University Park, Nottingham NG7 2RD, UK\\
$^{4}$Universitäts-Sternwarte München, Fakultät für Physik, LMU München, Scheinerstr. 1, D-81679 München, Germany\\
$^{5}$International Centre for Radio Astronomy Research, The University of Western Australia, 35 Stirling Highway, Crawley, WA 6009, Australia\\
}
\date{Accepted XXX. Received YYY; in original form ZZZ}

\pubyear{2025}

\begin{document}
\label{firstpage}
\pagerange{\pageref{firstpage}--\pageref{lastpage}}
\maketitle

\begin{abstract}
Tidal features provide signatures of recent galaxy mergers, offering insights into the role of mergers in galaxy evolution. The Vera C. Rubin Observatory’s upcoming Legacy Survey of Space and Time (LSST) will allow for an unprecedented study of tidal features around millions of galaxies. We use mock images of galaxies at $z\sim0$ ($z\sim0.2$ for \textsc{NewHorizon}) from \textsc{NewHorizon}, \textsc{eagle}, \textsc{IllustrisTNG}, and \textsc{Magneticum Pathfinder} simulations to predict the properties of tidal features in LSST-like images. We find that tidal features are more prevalent around blue galaxies with intrinsic colours $(g-i)\leq0.5$, compared to redder ones, at fixed stellar mass. This trend correlates with elevated specific star formation rates ($\mathrm{sSFR}>10^{-10}\mathrm{\:yr}^{-1}$), suggesting that merger-induced star formation contributes to the bluer colours. Tidal feature hosts in the red sequence appear to exhibit colour profiles offset to bluer colours for galaxies with stellar masses $10^{10}<M_{\star\mathrm{,\:30\:pkpc}}/\mathrm{M}_\odot<10^{11}$, similarly blue cloud tidal feature host galaxies appear to have their colour profiles offset to bluer colours for $10^{9.5}<M_{\star\mathrm{,\:30\:pkpc}}/\mathrm{M}_\odot<10^{10.5}$. However, the differences in colour profiles in either the red sequence or the blue cloud are not statistically robust and larger samples are needed to test if these differences are real. The predictions across the simulations are quantitatively distinct; therefore, LSST observations will allow us to further constrain the differences between different subgrid physics models.
\end{abstract}




\section{Introduction}

In hierarchical structure formation-based models of the Universe, mergers play an important role in transforming galaxies over cosmological timescales \citep[e.g.][]{pressFormationGalaxiesClusters1974, fallFormationRotationDisc1980, rydenGALAXYFORMATIONGRAVITATIONAL1987, vandenboschUniversalMassAccretion2002, agertzFormationDiscGalaxies2011}. They can alter the baryonic component of the galaxy, through contributing mass, transforming galaxy dynamics, triggering star formation, active galactic nuclei (AGN) feedback and the accretion of gas \citep[e.g.][]{ostrikerGalaxyFormationIntergalactic1981,barnesEncountersDiskHalo1988,hernquistOriginKinematicSubsystems1991,dimatteoEnergyInputQuasars2005,springelModellingFeedbackStars2005,duboisHORIZONAGNSimulationMorphological2016,martinLimitedRoleGalaxy2017,martinRoleMergersDriving2018,davisonEAGLEsViewEx2020,martinRoleMergersInteractions2021,remusAccretedNotAccreted2022,cannarozzoContributionSituEx2023,rutherfordSAMIGalaxySurvey2024}.

Cosmological simulations indicate that massive galaxies initially undergo a phase of rapid ``in-situ'' star formation followed by a phase dominated by the ``ex-situ'' accretion of stars at $z\lesssim3$ \citep[e.g.][]{naabMINORMERGERSSIZE2009,oserTWOPHASESGALAXY2010}. For galaxies with stellar mass, $M_{\star}\sim10^{9.5}\:\mathrm{M}_{\scriptstyle \odot}$ at $z=0$, almost all the stars tend to be formed in-situ, whereas in higher mass objects with $M_{\star}\gtrsim10^{12}\:\mathrm{M}_{\scriptstyle \odot}$, the fraction of stars that are accreted is $\gtrsim70$ per cent \citep[e.g.][]{oserTWOPHASESGALAXY2010,lacknerBuildingGalaxiesAccretion2012,robothamGalaxyMassAssembly2014,rodriguez-gomezStellarMassAssembly2016,davisonEAGLEsViewEx2020,remusAccretedNotAccreted2022,eisertERGOMLInferringAssembly2023}. The growth of mass through accretion since $z\lesssim1$ in massive galaxies is dominated by minor (1:10$\leq$ mass ratio $<$1:4) and mini (mass ratio $<$1:10) mergers, rather than major mergers (mass ratio $>$1:4) \citep[e.g.][]{naabMINORMERGERSSIZE2009,martinRoleMergersDriving2018,lagosQuantifyingImpactMergers2018,lagosConnectionMassEnvironment2018,remusAccretedNotAccreted2022} and these mergers, in-particular mini mergers, tend to deposit their stellar mass at larger radii \citep[e.g.][]{amoriscoContributionsAccretedStellar2017,karademirOuterStellarHalos2019}, increasing the size of the galaxy. These results are reflected in observations of massive early-type galaxies, where massive ($M_\star\gtrsim10^{11}\:\mathrm{M}_{\scriptstyle \odot}$) compact (sizes $\sim1$ kpc) elliptical galaxies were more common at $z\sim2$ and have decreased in frequency since then \citep[e.g.][]{trujilloStrongSizeEvolution2007,vanderwel3DHST+CANDELSEvolutionGalaxy2014}.

To test these predictions, we need to study galaxy mergers observationally. There are several ways to do this, including visible signatures of ongoing or past mergers or investigating the characteristics of close pair galaxies that will likely merge soon. Visible signatures, known as tidal features, are diffuse non-uniform regions of stars that extend out from a galaxy. They offer signatures of proceeding and concluded mergers in the forms of `tails', `streams', `asymmetric halos', `double nuclei' and `shells' \citep[e.g.][]{zwickyMultipleGalaxies1956,malinCatalogEllipticalGalaxies1983,miskolcziTidalStreamsGalaxies2011,bilekCensusClassificationLowsurfacebrightness2020,solaCharacterizationLowSurface2022,desmonsGalaxyMassAssembly2023}. Observations and numerical simulations have found these features to have lifetimes of $\sim1$\textendash$4$ Gyrs \citep[e.g.][]{lotzGalaxyMergerMorphologies2008,jiLifetimeMergerFeatures2014,mancillasProbingMergerHistory2019,yoonFrequencyTidalFeatures2020,huangMassiveEarlyTypeGalaxies2022}, making tidal features crucial probes of a galaxy's recent merger history. Tidal feature detection requires very deep images \citep[e.g.][]{miskolcziTidalStreamsGalaxies2011,ducATLAS3DProjectXXIX2015}, \citet{martinPreparingLowSurface2022} found that at surface brightness limits of $r\sim30$ to 31 mag/arcsec$^{2}$ ($3\sigma$, $10^{\prime\prime}\times10^{\prime\prime}$), observations would resolve $\sim80$ per cent of tidal features around a Milky Way mass galaxy out to $z\sim0.2$.

Close pair detection is important in understanding the role mergers play in driving galaxy evolution \citep[e.g.][]{robothamGalaxyMassAssembly2014,banksGalaxyMassAssembly2021,chamberlainPhysicallyMotivatedFramework2024}. However, this method does have limitations. It cannot detect the presence of a secondary galaxy if that galaxy has been ripped apart by tidal forces, absorbed into a host galaxy, or the secondary galaxy is too low-mass to detect via spectroscopy \citep[e.g.][]{lotzMAJORMINORGALAXY2011,desmonsGalaxyMassAssembly2023}. Studying galaxy tidal features provides insight into these aspects of the merging process as well.

The morphology of the tidal features also provides some information regarding the properties of the merger. \textit{N}-body and hydrodynamic simulations have shown that tail-like tidal features are formed from high angular momentum passages from similar mass galaxies (common in major mergers), while streams form from almost circularly infalling lower mass satellite galaxies \citep[minor mergers; e.g.][]{toomreGalacticBridgesTails1972, hendelTidalDebrisMorphology2015, karademirOuterStellarHalos2019}. Furthermore, shells are thought to form predominantly through radial mergers between galaxies \citep[e.g.][]{hendelTidalDebrisMorphology2015,amoriscoFeathersBifurcationsShells2015, popFormationIncidenceShell2018,karademirOuterStellarHalos2019, valenzuelaStreamComeTrue2024}. \citet{popFormationIncidenceShell2018} found that the majority of the $z=0$ shell population around the most massive elliptical galaxies in the hydrodynamic-cosmological simulation \textsc{Illustris} were formed predominantly through major mergers. 

In addition to the morphological information, the colour difference between the tidal feature and the host could probe the mass ratio of the merger \citep[e.g.][]{kado-fongTidalFeatures0052018} as galaxies follow relations with their colour and mass \citep[e.g.][]{taylorGalaxyMassAssembly2015}. Observational studies of the colours of tidal features relative to their host galaxies have provided some insight into the recent merger histories of galaxies. \citet{kado-fongTidalFeatures0052018} used the differences between the colours of shells and their host galaxies in the Hyper Suprime-Cam Subaru Strategic Program \citep[HSC-SSP][]{miyazakiHyperSuprimeCamSystem2018} to infer the mass ratios of the mergers. They found that the majority of shell galaxies ($85$ per cent) have shells bluer than their hosts, consistent with a minor merger scenario. However, $15$ per cent of their shells have similar colours, consistent with a major merger scenario. The shell colour results are consistent with the analysis of the \textsc{Illustris} simulation \citep{popGalaxiesShellsIllustris2017}. 

To perform a detailed test of galaxy mergers using tidal features, it will be necessary to study a larger observational sample of deep images that probe tidal features around many more galaxies than have been available to date. To fully understand these observations, it will also be essential to forward model mock observations from simulated data for comparison. With the upcoming Vera C. Rubin Observatory's Legacy Survey of Space and Time \citep[LSST;][]{ivezicLSSTScienceDrivers2019a,robertsonGalaxyFormationEvolution2019,broughVeraRubinObservatory2020} it will be possible to study tidal features around millions of galaxies \citep{martinPreparingLowSurface2022}, allowing for the most robust statistical survey of tidal features to date. In \citet{khalidCharacterizingTidalFeatures2024}, we undertook an observationally-motivated analysis of tidal features around galaxies in cosmological simulations. To do this we produced a set of mock observations at the predicted 10-year LSST surface brightness limits and visually identified the tidal features present around galaxies from four state-of-the-art cosmological-hydrodynamical simulations \textsc{NewHorizon} \citep{duboisIntroducingNEWHORIZONSimulation2021}, \textsc{eagle} \citep{schayeEAGLEProjectSimulating2015, crainEAGLESimulationsGalaxy2015}, \textsc{IllustrisTNG} \citep{pillepichFirstResultsIllustristng2018,springelFirstResultsIllustrisTNG2018, nelsonFirstResultsIllustrisTNG2018, marinacciFirstResultsIllustrisTNG2018, naimanFirstResultsIllustrisTNG2018} and \textsc{Magneticum Pathfinder} \citep{tekluConnectingAngularMomentum2015,dolagEncyclopediaMagneticumScaling2025}. This provided theoretical predictions regarding the occurrence of tidal features as a function of the host galaxy's stellar mass and host halo mass.

The colours and colour profiles of galaxies provide information on the stellar population and reflect the growth and accretion histories of galaxies \citep[e.g.][]{marianColorGradientsReflect2018}. In this work, we use the mock photometry produced in \citet{khalidCharacterizingTidalFeatures2024} to study the colours and colour profiles of galaxies hosting tidal features. However, there is inherent scatter in the colour profiles of galaxies \citep[e.g.][]{jedrzejewskiCCDSurfacePhotometry1987,merrittDRAGONFLYNEARBYGALAXIES2016,millerColorGradientsHalfmass2023} and we need to control for this to determine whether we can infer the traits of progenitors. We use the population of non-tidal galaxies in each simulation as a control sample. By comparing the colours and colour profiles of tidal and non-tidal feature hosts in a given stellar mass range, we can infer the contribution of tidal features and therefore, the properties of the recent progenitor from which the tidal features are born.

We introduce our simulations and the derived LSST-like mock images and accompanying catalogue of tidal features in Sections \ref{subsec:simulations} and \ref{subsec:tidal_feature_cat}. Our methods for measuring the colour of the galaxy and the colour profile are described in Section \ref{subsec:measure_size_and_colour}. We present and discuss our results in Sections \ref{sec:results} and \ref{sec:discussion} and draw our conclusions in Section \ref{sec:conclusion}. We use the native cosmology from each simulation for calculating the distances between particles and creating our mock images, which are given in Table \ref{tab:simulations}. For distances, we use a `c' prefix to denote comoving coordinates and a `p' prefix to denote proper coordinates, i.e. ckpc is comoving kiloparsecs and pkpc is proper kiloparsecs.

\section{Data and Methods}
\label{sec:data_methods}

For this work, we use the LSST-like mock images and the catalogue of tidal features around galaxies in the mock images presented in \citet{khalidCharacterizingTidalFeatures2024}. The mock images covered galaxies with $9.5\leq\log_{10}(M_{\star\mathrm{,\:30\:pkpc}}/\mathrm{M}_{\scriptstyle \odot})\leq11.8$, where $M_{\star\mathrm{,\:30\:pkpc}}$ is stellar mass within a 30 pkpc radius spherical aperture centred on the centre of potential of the galaxy. In our observationally-motivated analysis to probe the characteristics of recent progenitors in the colours of tidal feature hosts and their outskirts, we explore four different cosmological simulations, for which we have already quantified the frequency of tidal features and the variance of this frequency with respect to the galaxy stellar mass and host halo mass in a subset of their galaxies (the size of the subsets are given by $N_{\mathrm{catalogue}}$ in Table \ref{tab:simulations}). We briefly describe the relevant aspects of each simulation's model in Section \ref{subsec:simulations}, and the mock images and tidal feature catalogue are described in Section \ref{subsec:tidal_feature_cat}. Lastly, we describe our method for defining the colour of galaxies and their outskirts in Section \ref{subsec:measure_size_and_colour}.

\subsection{Simulations}
\label{subsec:simulations}

\begin{table*}
    \centering
    \caption{Summary of the properties of the four cosmological hydrodynamical simulations and the derived measurements. From left to right, the columns are the simulation name, cosmology selected, simulation box volume, snapshot redshift, and number of galaxies in catalogue with $M_{\star\mathrm{,\:30\:pkpc}}\geq10^{9.5}\:\mathrm{M}_{\scriptstyle \odot}$, number of galaxies hosting tidal features with $conf.\geq2$, number of galaxies with reliable colour measurements within 1 $R_e^\mathrm{maj}$, between 1 and 2 $R_e^\mathrm{maj}$, between 2 and 3 $R_e^\mathrm{maj}$, between 3 and 4 $R_e^\mathrm{maj}$ and between 4 and 5 $R_e^\mathrm{maj}$}
    \begin{tabular}{|p{3.0cm}|p{1.5cm}|p{0.8cm}|p{1.5cm}|p{0.8cm}|p{0.8cm}|p{0.8cm}|p{0.8cm}|p{0.8cm}|p{0.8cm}|p{0.8cm}|p{0.8cm}}
        \hline
        Simulation & Cosmology & $V_\mathrm{box}$ \newline[$\text{cMpc}^3$] & z & $N_\mathrm{catalogue}$ & $N_\mathrm{T}$ & $N_{R_e^\mathrm{maj}}$ & $N_{1\to2\/R_e^\mathrm{maj}}$ & $N_{2\to3\/R_e^\mathrm{maj}}$ & $N_{3\to4\/R_e^\mathrm{maj}}$ & $N_{4\to5\/R_e^\mathrm{maj}}$\\
        \hline
        \textsc{NewHorizon} & WMAP-7 & $16^3$ & 0.260, 0.263 &  62 & 24 & 56 & 56 & 55 & 55 & 54\\
        \textsc{EAGLE RefL0100N1504} & Planck13 & $100^3$ & 0.05 & 1983 & 717 & 1959 & 1953 & 1943 & 1936 & 1933\\
        \textsc{TNG L75n1820TNG} & Planck15 & $111^3$ & 0.05 & 1826 & 593 & 1773 & 1747 & 1712 & 1688 & 1683\\
        \textsc{Magneticum Pathfinder Box4-uhr} & WMAP-7 & $68^3$ & 0.07 & 1989 & 531 & 1902 & 1894 & 1887 & 1874 & 1825\\
        \hline
    \end{tabular}
    \label{tab:simulations}
\end{table*}

The simulations in our analyses are \textsc{NewHorizon}, \textsc{EAGLE RefL0100N1504}, \textsc{IllustrisTNG L75N1820}, \textsc{Magneticum Pathfinder} \textit{Box4-uhr} (from here on, NewHorizon, EAGLE, TNG and Magneticum). These are summarised in Table \ref{tab:simulations}. The differences between the simulations allow us to probe the characteristics of tidal features across a range of simulation resolutions, galaxy environments and subgrid physics models. EAGLE and TNG probe similar volumes of $(100$ cMpc$)^3$ and $(111$ cMpc$)^3$ respectively and therefore the same range of environments (isolated galaxies, groups and a few low-mass clusters $M_{\scriptstyle\mathrm{200,\:crit}}\sim10^{14}\:\mathrm{M}_{\odot}$), where $M_{\scriptstyle\mathrm{200,\:crit}}$ is the total mass contained within the radius at which the mean density is 200 times the critical density of the Universe at that redshift. For studies focused on galaxy dynamics (e.g. galaxy mergers), the dark matter particle resolution is most important in understanding resolution-driven dynamical heating \citep[][]{ludlowSpuriousHeatingStellar2021,ludlowSpuriousHeatingStellar2023}. Therefore, we provide below the dark matter particle masses, the initial gas mass resolution and the mean masses for the stellar particles, as they have varying masses with time.

The NewHorizon simulation \citep{duboisIntroducingNEWHORIZONSimulation2021} is a zoom-in simulation based on \textsc{Horizon-AGN} \citep{duboisDancingDarkGalactic2014}, which adopts the WMAP-7 cosmology \citep[][]{komatsuSevenyearWilkinsonMicrowave2011}. NewHorizon employs a spherical volume with varying dark matter resolution. The initial 10 cMpc radius has its dark matter resolved to $m_{\rm DM}=1.2\times10^6\text{ M}_{\odot}$, this high-resolution patch is embedded in spheres of decreasing mass-resolution of $10^7$, $8\times10^7$ and $6\times10^8\text{ M}_{\odot}$ corresponding to radii of $10.6$, $11.7$, and $13.9$ cMpc. The remaining volume is resolved at $5\times10^9\text{ M}_{\odot}$ \citep{martinPreparingLowSurface2022}. The initial gas mass resolution is $m_\mathrm{gas}=2\times10^5\:\mathrm{M}_{\scriptstyle \odot}$ and the average stellar particle mass at $z=0.26$ is $m_{\scriptstyle\star}\simeq9\times10^3\:\mathrm{M}_{\scriptstyle \odot}$.

Each stellar resolution element represents a Chabrier initial mass function \citep[IMF;][]{chabrierGalacticStellarSubstellar2003} simple stellar population ranging from 0.1 - 150 M$_{\scriptstyle \odot}$. NewHorizon sets star formation to occur in regions where the hydrogen gas number density exceeds $10$ cm$^{-3}$, following the Schmidt law: $\dot{\rho_{\star}}=\epsilon_\star\rho_{\rm{g}}/t_{\rm{ff}}$, where the star formation rate mass density, $\rho_{\star}$, is related to the gas mass density, $\rho_{\mathrm{g}}$, local free fall time of the gas, $t_{\rm{ff}}$, and lastly a varying star formation efficiency, $\epsilon_{\star}$ \citep{kimmImpactLymanAlpha2018,trebitschOBELISKSimulationGalaxies2021}. The star formation efficiency is computed following \citet{hennebelleAnalyticalStarFormation2011}, where the turbulence of the gas is accounted for. H and He gases are modelled using an equilibrium chemistry model with a homogeneous UV background. This gas can cool to $\simeq10^4$K through collisional ionisation, excitation, recombination, Bremsstrahlung, and Compton cooling. If the gas is metal-enriched, it can further cool down to 0.1 K using the tabulated cooling rates from \citet{dalgarnoHeatingIonizationHI1972} and \citet{sutherlandCoolingFunctionsLowDensity1993}. Supernovae (SNe) impact the surrounding gas and therefore star formation \citep[e.g.][]{mckeeTheoryStarFormation2007}. NewHorizon models each SNe explosion by releasing $10^{51}$ erg of energy into the surrounding gas, with a minimum of $6\:\mathrm{M}_{\scriptstyle \odot}$ needed for a supernova (SN). They implement a specific SNe rate of 0.03 $M_\odot^{-1}$ using the mechanical SN feedback scheme described in \citet{kimmEscapeFractionIonizing2014} and \citet{kimmSimulatingStarFormation2015}. NewHorizon implements two modes of active galactic nuclei (AGN) feedback, thermal and kinetic (radio and quasar), depending on the accretion rate of the supermassive black hole \citep[following][]{duboisJetregulatedCoolingCatastrophe2010,teyssierMassDistributionGalaxy2011}. AGN feedback can help remove gas from the galaxy and/or heat it, making it unsuitable for star formation. The details of all subgrid physics are provided in \citet{duboisIntroducingNEWHORIZONSimulation2021}.

The \textsc{eagle} (Evolution and Assembly of GaLaxies and their Environments) project is a large set of cosmological hydrodynamical simulations. \textsc{eagle} contains simulations of cubic volume $12^3$, $25^3$, $50^3$ and $100^3$ cMpc$^{3}$.  The data that will be used in this study is from the reference model (\textsc{RefL0100N1504}), which has a volume of $(100\text{ cMpc})^3$. The reference model is the simulation run with standard parameters and physics as described in \citet{schayeEAGLEProjectSimulating2015} and \citet{crainEAGLESimulationsGalaxy2015}. The simulation uses the cosmological parameters advocated by the Planck 2013 results \citep[][]{adePlanck2013Results2014}. The dark matter particle mass is $m_{\rm DM}=9.7 \times 10^{6} \text{ M}_{\scriptstyle \odot}$, the initial gas mass resolution is $m_\mathrm{gas}=1.8\times10^6$ and the average stellar particle mass at $z=0.05$ is $m_{\scriptstyle\star}=1.1\times10^{6} \text{ M}_{\scriptstyle \odot}$.

\textsc{eagle} implements star formation stochastically, using the pressure law scheme of \citet{schayeRelationSchmidtKennicuttSchmidt2008} that enforces the observed Kennicut-Schmidt \citep{schmidtRateStarFormation1959,kennicuttGlobalSchmidtLaw1998} relation into the simulation. A metallicity-dependent hydrogen mass density threshold, formulated by \citet{schayeStarFormationThresholds2004}, determines the regions where star formation occurs, with metal-rich gas clouds allowing for more efficient radiative transfer. Stellar particles follow a Chabrier IMF, with masses spanning 0.1-100 M$_{\scriptstyle \odot}$. The impact of stellar winds, radiation and SNe feedback on gas is accounted for through thermal heating, distributing the energy produced by stellar particles during a timestep to the neighbouring particles using the stochastic thermal feedback scheme of \citet{dallavecchiaSimulatingGalacticOutflows2012}. \textsc{cloudy} \citep{ferlandCLOUDY90Numerical1998} computes the radiative cooling and heating rates of gas resolution elements at a given density, temperature and redshift. Gas is assumed to be optically thin, in an ionisation equilibrium and exposed to the cosmic microwave background and a spatially homogeneous evolving UV/X-ray background \citep{2001cghr.confE..64H,wiersmaEffectPhotoionizationCooling2009}. \textsc{eagle} implements a single-mode AGN feedback through stochastic heating, following \citet{boothCosmologicalSimulationsGrowth2009}. The details of all subgrid physics are described by \citet{schayeEAGLEProjectSimulating2015} and \citet{crainEAGLESimulationsGalaxy2015}.

\textsc{IllustrisTNG} is a suite of simulations, with the following box volumes: $51.7^3$, $110.7^3$ and $302.6^3\text{ cMpc}^3$. We use simulation \textsc{L75N1820}, which uses a $110.7^3\text{ cMpc}^3$ box. This simulation adopts the $\Lambda$CDM model fit by the Planck 2015 results \citep[]{adePlanck2015Results2016}. The dark matter mass resolution is $m_{\rm DM}=7.5\times10^6\text{ M}_{\scriptstyle \odot}$, the gas mass resolution is $m_\mathrm{gas}=1.4\times10^6\:\mathrm{M}_{\scriptstyle\odot}$ and the average stellar mass resolution at $z=0.05$ is $m_{\scriptstyle\star}=1.1\times10^6\:\mathrm{M}_\odot$.

\textsc{IllustrisTNG} models star formation stochastically, treating the star formation and pressurisation of a multi-phase interstellar medium following \citet{springelCosmologicalSmoothedParticle2003}. Cold gas above a density threshold of $0.1$ cm$^{-3}$ forms star particles following the empirically-defined Kennicutt-Schmidt relation \citep{schmidtRateStarFormation1959,kennicuttGlobalSchmidtLaw1998} and the Chabrier IMF \citep{chabrierGalacticStellarSubstellar2003} with star masses ranging from $0.1$ to $100\:\mathrm{M}_{\scriptstyle \odot}$ \citep[e.g.][]{vogelsbergerModelCosmologicalSimulations2013}. In this model, SNe pressurise gas and may enhance star formation; the stellar winds produced by star formation carry kinetic energy that they transfer into the surrounding gas. They are modelled following \citet{vogelsbergerModelCosmologicalSimulations2013} and \citet{pillepichSimulatingGalaxyFormation2018}. The radiative cooling of gas is modelled, accounting for its metal enrichment following \citet{wiersmaEffectPhotoionizationCooling2009}, and a time-evolving homogeneous UV background with self-shielding corrections in dense interstellar medium following \citet{katzGalaxiesGasCold1992,faucher-giguereNewCalculationIonizing2009}. The radiation fields of nearby AGN can impact the efficiency of radiative cooling following the prescriptions of \citet{vogelsbergerModelCosmologicalSimulations2013}. For accretion rates below 5 per cent of the Eddington limit, the radio-mode feedback is active, injecting bursty thermal energy into a $\sim50$ pc bubble displaced away from the host galaxy \citep{sijackiUnifiedModelAGN2007}. For higher accretion rates, the quasar mode is active, injecting thermal energy continuously into the adjacent gas \citep{springelModellingFeedbackStars2005,dimatteoEnergyInputQuasars2005}. It has been found that at $z=0$ the quasar mode of AGN feedback switches on at $M_\mathrm{BH}\gtrsim10^{8.2}\:\mathrm{M}_{\scriptstyle\odot}$ \citep[e.g.][]{terrazasRelationshipBlackHole2020,lagosDiverseStarFormation2025}. The details of all subgrid physics are described by \citet{weinbergerSimulatingGalaxyFormation2017,pillepichSimulatingGalaxyFormation2018} and \citet{nelsonIllustrisTNGSimulationsPublic2019}.

\textsc{Magneticum Pathfinder} simulations are a suite of cosmological hydrodynamical simulations, ranging in box size from $25.6^3$ to 3818$^3$ cMpc$^3$. We use the \textit{Box4-uhr} simulation, which has a  $68^3$ cMpc$^3$ volume box. The simulation adopts a WMAP-7 fit cosmology \citep{komatsuSevenyearWilkinsonMicrowave2011}. The dark matter resolution is $m_\mathrm{DM}=5.1\times10^7\:\mathrm{M}_{\scriptstyle\odot}$, the initial gas mass resolution is $m_\mathrm{gas}=1.0\times10^7\:\mathrm{M}_{\scriptstyle\odot}$ and the mean stellar mass resolution at $z=0.07$ is $m_{\scriptstyle\star}=1.9\times10^6\:\mathrm{M}_{\scriptstyle\odot}$.

Star formation and the stellar wind-driven kinetic feedback are modelled following \citet{springelCosmologicalSmoothedParticle2003}. Each gas particle can form up to 4 stars with stellar populations following a Chabrier IMF \citep{chabrierGalacticStellarSubstellar2003} from 0.1 to 100 M$_{\scriptstyle \odot}$ \citep{tornatoreChemicalEnrichmentGalaxy2007}. Star formation and metal enrichment from supernova feedback and Asymptotic Red Giant Branch stars are modelled following \citet{springelCosmologicalSmoothedParticle2003} as well as the local metallicity-dependent processes \citep{wiersmaEffectPhotoionizationCooling2009,dolagDistributionEvolutionMetals2017}. \textsc{cloudy} \citep{ferlandCLOUDY90Numerical1998} is used to compute the radiative heating and cooling of gas at a given density, temperature and redshift. The impact of the UV/X-ray background on radiative cooling is accounted for following \citet{2001cghr.confE..64H} and \citet{wiersmaEffectPhotoionizationCooling2009}. The physics of supermassive black holes and their AGN feedback are implemented as described by \citet{fabjanSimulatingEffectActive2010} and \citet{hirschmannCosmologicalSimulationsBlack2014}. The implemented black hole feedback scheme accounts for a transition from quasar to radio mode according to \citet{sijackiUnifiedModelAGN2007}. Note that the black holes in this simulation are not pinned to the potential minimum. Thermal conduction is implemented according to \citet{dolagThermalConductionSimulated2004} but following \citet{arthAnisotropicThermalConduction2017}, with 1/20 of the classical Spitzer value \citep{spitzerPhysicsFullyIonized1962}. The details of all subgrid physics are described by \citet{hirschmannCosmologicalSimulationsBlack2014}, \citet{tekluConnectingAngularMomentum2015} and \citet{dolagEncyclopediaMagneticumScaling2025}.

EAGLE and TNG have similar dark matter, gas and stellar mass resolutions and the largest volumes. The Magneticum simulation is of an intermediate volume and a lower mass resolution than EAGLE and TNG, whereas NewHorizon provides a dataset with the highest mass resolution but the smallest box size and therefore, a more limited sample of galaxies and environments.

\subsection{LSST-mock images and catalogue of tidal features}
\label{subsec:tidal_feature_cat}
\begin{figure}
    \centering
    \includegraphics[width=\linewidth]{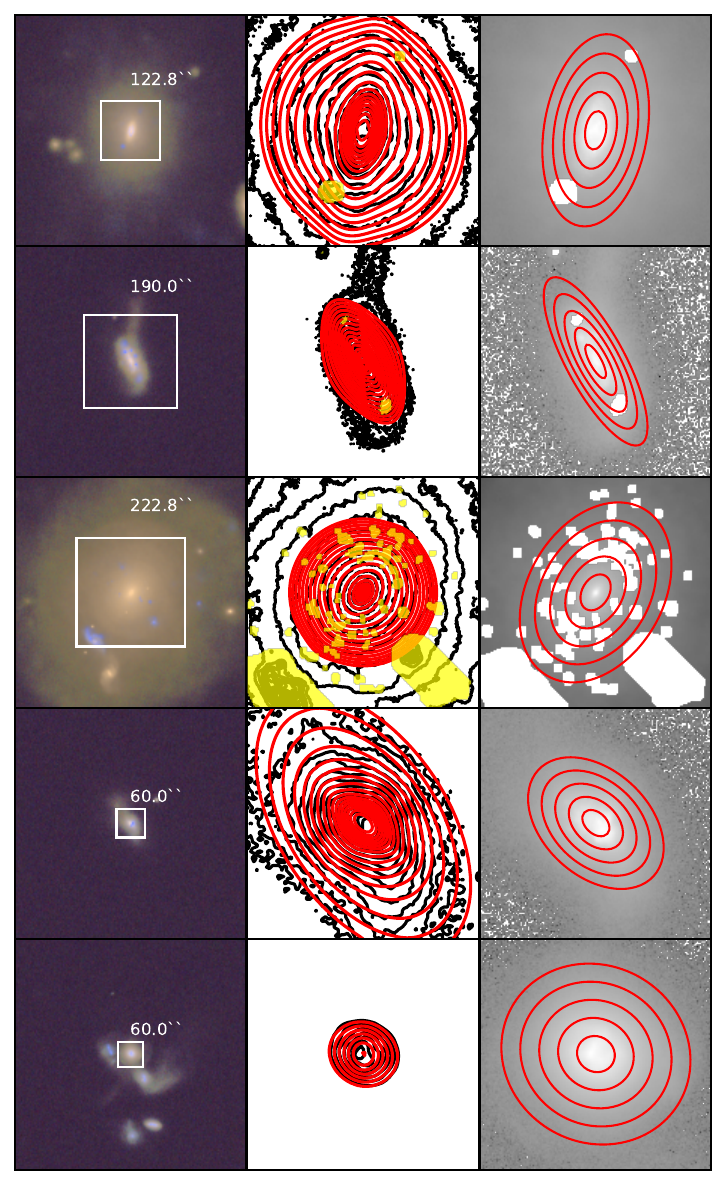}
    \caption{Illustration of the process we followed for measuring galaxy size and colour. The middle column shows the MGE fit (red contours) to the data (black contours), and the yellow shows the masked regions. The right-hand column shows a masked grey-scale \textit{gri} coadded image with the apertures used to measure our colours and colour profiles overplotted. The left-hand column shows the \textit{gri} colour images that were used in the process of classification, the white rectangle shows the box used for the final colour measurements. All the galaxies in our sample are placed at $\sim109\mathrm{\:cMpc}$ away from the observer. From top to bottom, we illustrate fits on a galaxy without any tidal features, a galaxy hosting a stream/tail, a galaxy hosting shells, a galaxy hosting an asymmetric halo and a galaxy hosting a double nucleus.}
    \label{fig:demo_size_measure_mask}
\end{figure}

To measure the colours of galaxies and their outskirts and perform a comparison between tidal feature hosts and non-tidal feature hosts, we use the LSST-like mock images and catalogue of tidal features in the mock images presented in \citet{khalidCharacterizingTidalFeatures2024}. Here, we briefly describe the construction of these images and the catalogue of tidal features.

The mock images were made following \citet{martinPreparingLowSurface2022}. The mock images place the simulated galaxy at a distance corresponding to $z=0.025$ in the cosmology of the simulation (comoving distance of $\sim109$ cMpc). Briefly, the mock images are constructed by extracting all particles associated with a structure finder-identified group (\textsc{AdaptaHOP} for NewHorizon or \textsc{FOF} for the other simulations) within a 1 pMpc box of the galaxy being imaged. The spectral energy distribution (SED) for each stellar particle is computed using simple stellar population models \citep{bruzualStellarPopulationSynthesis2003} interpolated onto the age and metallicity of each stellar particle. Assuming a Chabrier IMF \citep{chabrierGalacticStellarSubstellar2003}, the $g$, $r$ and $i$ filter magnitudes for each particle are computed accounting for the galaxy redshift and the transmission function of the LSST filters \citep{olivierOpticalDesignLSST2008a}. We note that reddening from dust attenuation is not accounted for. Therefore, the colours presented in this work are the intrinsic galaxy colours.

Smoothing of the particle distributions is necessary to remove unrealistic fluctuations in the flux in the final image, this is carried out by subdividing stellar resolution elements into lower mass elements distributed normally around the original's position, with standard deviations equal to the 5th nearest neighbour distance for the particle being smoothed \citep[][see also \citealt{merrittMissingOutskirtsProblem2020}]{martinPreparingLowSurface2022}. The mock images are produced by collapsing the cube along the \textit{z} axis of the simulation box (\textit{x}\textendash\textit{y} projection) with a grid size corresponding to $0.2^{\prime\prime}\times0.2^{\prime\prime}$ \citep[the LSST pixel size;][]{ivezicLSSTScienceDrivers2019a}. The apparent magnitude for each pixel is measured, accounting for cosmological dimming. To model the dispersion of light due to seeing, the images are convolved with the Hyper-Suprime Cam point spread function measured by \citet{montesBuildupIntraclusterLight2021}, rescaled to the LSST pixel scale.

Using the observationally-measured empirical relation for background noise \citep{romanGalacticCirriDeep2020}, Gaussian background noise is inserted into each image, with standard deviation given by:
\begin{equation}
    \sigma_{\rm{noise}}=\frac{10^{-0.4\mu_{\rm band}^{\rm lim}(n\sigma, \Omega\times\Omega)}\text{pix}\times\Omega}{n}.
\end{equation}

Here $\Omega$ is the size of one side of the box over which the surface brightness limit is computed, $=10^{\prime\prime}$, pix is the pixel scale in arcseconds per pixel, $=0.2$ arcsec/pixel, $n$ is the number of Gaussian standard deviations for the surface brightness limits, $=3$ and $\mu_{\rm band}^{\rm lim}$ is the $3\sigma$, $10^{\prime\prime}\times10^{\prime\prime}$ limiting surface brightness modelled for a particular LSST-photometric band \footnote{Peter Yaochim, \href{https://smtn-016.lsst.io/}{LSST surface brightness limit derivations}} ($\mu_{g}^{\mathrm{lim}}=30.3$ mag/arcsec$^{2}$, $\mu_{r}^{\mathrm{lim}}=30.3$ mag/arcsec$^{2}$, $\mu_{i}^{\mathrm{lim}}=29.7$ mag/arcsec$^{2}$).

The Gaussian noise is included to account for spatially invariant sources, including instrumental noise, unresolved background sources, scattered light, etc., without the need to explicitly model them. However, spatially varying sources are unaccounted for in these images, e.g. sky gradients, diffraction spikes, imperfect sky-subtraction or galactic-cirrus. Correcting for these effects will be crucial in achieving the lowest surface brightness limits with the data, and the necessary corrections are being studied in detail in preparation for LSST \citep[e.g.][]{watkinsStrategiesOptimalSky2024,bazkiaeiBrightStarSubtraction2024}.

A catalogue identifying tidal features and their morphologies in LSST-mock images was constructed for a subsample of galaxies from each simulation. The sample was constructed by randomly selecting galaxies from each simulation with a 30 pkpc spherical aperture stellar mass, $M_{\star\mathrm{, 30 pkpc}}>10^{9.5}\:\mathrm{M}_{\scriptstyle \odot}$. The first author carried out a visual identification and classification of tidal features using single \textit{g}, \textit{r}, and \textit{i} band images and a combined colour image of fixed contrast and brightness. Following a classification scheme similar to those used by \citet[]{bilekCensusClassificationLowsurfacebrightness2020} and \citet{desmonsGalaxyMassAssembly2023}, we classify them into the following categories (examples in Fig.~\ref{fig:demo_size_measure_mask}):
\begin{itemize}
    \item \textbf{Streams/Tails}: Prominent, elongated structures orbiting or expelled from the host galaxy. These usually have similar colours to the host galaxy
    \item \textbf{Shells}: Concentric radial arcs or ring-like structures around a galaxy.
    \item \textbf{Plumes or Asymmetric Stellar Halos}: Diffuse features in the outskirts of the host galaxy, lacking well-defined structures like stellar streams or tails.
    \item \textbf{Double nuclei}: Two clearly separated galaxies within the mock image where merging is evident through the presence of tidal features.
\end{itemize}

In our classification, a galaxy is allowed to have multiple tidal features and feature types associated with it. A galaxy is considered to host a tidal feature if the feature is connected to (e.g. tidal tail expelled from a galaxy; a stream being accreted onto a galaxy) or centred on the object (e.g. shells around a galaxy). In cases where the tidal feature is associated with two objects, e.g. a tail/stream between two objects, the tidal feature can be associated with both objects. Furthermore, for each tidal feature classification, the catalogue provides a confidence level ranging from zero to three.
\begin{enumerate}
    \item[(0)] No tidal feature detected.
    \item[(1)] Hint of tidal feature detected, classification difficult.
    \item[(2)] Even chance of correct classification of tidal feature presence and/or morphology.
    \item[(3)] High likelihood of the tidal feature being present and morphology being obvious.
\end{enumerate}

To strike a balance between the completeness of the sample of tidal features and the accuracy of the classification, we use tidal features of confidence level 2 or above. The tidal feature fractions for the sample given this classification level are shown in Table \ref{tab:tf_classifications}.

In \citet{khalidCharacterizingTidalFeatures2024}, we found that the simulations generally agreed regarding the trends in tidal feature occurrence with both stellar and halo mass, with NewHorizon showing slightly higher tidal feature occurrence as a function of stellar and halo mass than the other three simulations. This is likely explained by a combination of the higher mass resolution and the higher normalization in its stellar mass-halo mass relation. We also found good agreement with previous results regarding the occurrence of tidal features and their relationship with galaxy stellar mass in simulations \citep[e.g.][]{martinPreparingLowSurface2022,valenzuelaStreamComeTrue2024} and observations \citep[e.g.][]{atkinsonFAINTTIDALFEATURES2013,kado-fongTidalFeatures0052018,huangMassiveEarlyTypeGalaxies2022,desmonsGalaxyMassAssembly2023}.

The observational results for total tidal feature fraction varied between $f_{\mathrm{Tidal}}\sim0.06$ \citep{kado-fongTidalFeatures0052018} to $f_\mathrm{Tidal}\sim0.4$ \citep{atkinsonFAINTTIDALFEATURES2013}. These variances were driven by observational factors such as the surface brightness limits of the surveys and the distribution of the galaxies in redshift. The range of total tidal feature fractions from the mock images in \citet{khalidCharacterizingTidalFeatures2024} spanned $f_\mathrm{Tidal}\sim0.2$ (for confidence level 3) to $f_\mathrm{Tidal}\sim0.5$ (for confidence level 1), which covered most of the observational results we compared to \citep{atkinsonFAINTTIDALFEATURES2013,huangMassiveEarlyTypeGalaxies2022,desmonsGalaxyMassAssembly2023}. The lowest observational fraction of $f_\mathrm{Tidal}\sim0.06$, from \citet{kado-fongTidalFeatures0052018}, included galaxies reaching $z\sim0.45$ and galaxy stellar masses reaching down to $10^8\:\mathrm{M}_\odot$. Given the $28.4$ mag/arcsec$^2$ \textit{i} band surface brightness limit of their data, it is likely that they were not resolving the faintest tidal features around low-mass galaxies or those at higher redshifts. \citet{kado-fongTidalFeatures0052018} found that using the same magnitude and redshift cuts as \citet{atkinsonFAINTTIDALFEATURES2013}, they achieve a similar fraction.

\subsubsection{Uncertainties on fractions}
We estimate uncertainties on our feature fractions using the $1\sigma\simeq0.683$ binomial confidence levels estimated using a Bayesian beta distribution generator for binomial confidence intervals described in detail in \citet{cameronEstimationConfidenceIntervals2011}. This approach is robust to small sample sizes, which can be the case for subsamples of our catalogue used to measure how tidal features relate to galaxy colours.

\begin{table*}
    \centering
    \caption{Total and specific tidal feature fractions for each simulation from \citet{khalidCharacterizingTidalFeatures2024}. The fractions quoted here are for confidence level 2 or above classifications only.}
    \begin{tabular}{c|c|c|c|c|c}
    \hline
    Simulation &  All tidal features & Stream/Tail & Shell & Asymmetric halo & Double nucleus\\
    \hline
      NewHorizon   & $0.40\pm0.06$ & $0.11^{+0.05}_{-0.03}$ & $0.016^{+0.04}_{-0.005}$ & $0.31^{+0.06}_{-0.05}$ & $0.26^{+0.06}_{-0.05}$\\
      \\
      EAGLE & $0.37\pm0.01$ & $0.065^{+0.008}_{-0.006}$ & $0.006^{+0.003}_{-0.001}$ & $0.32\pm0.01$ & $0.130^{0.01}_{-0.009}$\\
      \\
      TNG & $0.32\pm0.01$ & $0.077^{+0.008}_{-0.007}$ & $0.0023^{+0.002}_{-0.0007}$ & $0.25\pm0.01$ & $0.133^{0.01}_{-0.009}$\\
      \\
      Magneticum & $0.32\pm0.01$ & $0.040^{+0.006}_{-0.005}$ & $0.004^{+0.003}_{-0.001}$ & $0.27\pm0.01$ & $0.16\pm0.01$\\
      \hline
    \end{tabular}
    \label{tab:tf_classifications}
\end{table*}

\subsection{Measuring galaxy size and colours}
\label{subsec:measure_size_and_colour}

To measure galaxy sizes and colours, it is necessary to know which light is associated with the galaxy of interest and mask light that is associated with other sources. We use coadded \textit{gri} image cutouts of our galaxies that are $240\text{\:pkpc}\times240\text{\:pkpc}\simeq2400\mathrm{\:pixels}\times2400\mathrm{\:pixels}=480\:^{\prime\prime}\times480\:^{\prime\prime}$ for our masks to undertake our image analysis.

In Section \ref{subsubsec:masks}, we describe our process for producing masks for our images. In Section \ref{subsubsec:mge_model}, we measure galaxy size and shape. Finally, in Section \ref{subsubsec:measure_colour} we describe how we measure the colours of the galaxy and their outskirts colours using the mask, galaxy shape and size measurements.

\subsubsection{Image masks}
\label{subsubsec:masks}

Due to the size of our sample, our masking routine needed to be fully automated, such that the segmentation and deblending parameters used remain the same across the entire sample. To detect and deblend sources in our images, we use the \textsc{photutils} \citep{larry_bradley_2023_8248020} segmentation and watershed deblending tools on coadded \textit{gri} images to produce our masks. From \textsc{photoutils} we use the \texttt{detect\_sources} and \texttt{deblend\_sources} functions. The \texttt{detect\_sources} function locates all the sets of connected pixels with flux above a specified \texttt{threshold}. Pixels are considered connected if joined along one side or corner of the pixel. It takes as input \texttt{npixels}, the minimum number of connected pixels above \texttt{threshold}, which is the minimum flux that the source must have to be detected. The \texttt{deblend\_sources} function takes as input the segmentation map of detections output by \texttt{detect\_sources} and uses multi-thresholding and watershed deblending to separate overlapping sources. For deblending, \texttt{npixels} is kept the same, we specify the number of thresholding levels between the minimum and maximum values in the image to detect separate peaks with \texttt{nlevels}, and we specify the fraction of the total flux that a local peak must have to be detected as a separate source.

In low-surface brightness images, galaxies often have a diffuse low-surface brightness component. To capture both the smaller sources (e.g. compact dwarf galaxies) and the more extended sources (galaxies and galaxies with diffuse components), we use a hot and a cold mask \citep[e.g.][]{martinez-lombillaGalaxyMassAssembly2023}. The hot mask is used to capture the more compact sources, and the cold mask is used to capture the more extended sources.  

To produce the hot masks, the coadd is convolved with a Gaussian kernel with $\sigma_\mathrm{G}=40$ pixels (we present our tests of different values in Appendix \ref{app:masking}) to produce a smoothed image. This smoothed image is subtracted from the original, producing an image where the compact sources are highlighted. The mask is created by running \texttt{detect\_sources} on this subtracted image. To produce the cold mask, \texttt{detect\_sources} and \texttt{deblend\_sources} are run on the coadded images. We tested a range of detection and deblending parameters (presented in Appendix \ref{app:masking}) and found these worked best for our automated approach:

\begin{itemize}
    \item Hot mask: \texttt{detect\_sources} parameters, \texttt{npixels}$=40$ pixels, \texttt{threshold}=$2\sigma_\mathrm{subtracted}$.
    \item Cold mask: \texttt{detect\_sources} parameters, \texttt{npixels}$=400$ pixels, \texttt{threshold}$=6\sigma_{\mathrm{coadd}}$, and \texttt{deblend\_sources} parameters of \texttt{nlevels}$=30$, \texttt{contrast}$=0.001$.
\end{itemize}

The mock-image data produced give fluxes in zero-point normalized units, such that the AB-magnitude is $m_\mathrm{AB}=-2.5\log_{10}(f)$, where $f$ is the zero-point normalized flux. In these units, $\sigma_\mathrm{coadd}=1.133\times10^{-12}$ is the background noise level of the coadded image \bigg($\sigma_\mathrm{coadd}=\sqrt{\sigma_{g}^2+\sigma_i^2+\sigma_r^2}$\bigg), $\mathbf{im}_\mathrm{coadd}$, and $\sigma_\mathrm{subtracted}=1.1325\times10^{-12}$ zero-point normalised flux. We start by calculating the standard deviation of the smoothed image, $\mathbf{im}_\mathrm{smoothed}=\sum_{i=-\infty}^\infty\sum_{j=-\infty}^{\infty}\frac{1}{2\pi\sigma_\mathrm{G}^2}e^{-\frac{i^2+j^2}{2\sigma_\mathrm{G}^2}}\mathbf{im}_\mathrm{coadd}$, where $\sigma_\mathrm{G}$ is the standard deviation of the Gaussian convolution kernel. By the variance of linear combinations, we get the following:

\begin{align}
    \sigma_\mathrm{smoothed}^2&=\sum_{i=-\infty}^{\infty}\sum_{j=-\infty}^\infty(\frac{1}{2\pi\sigma_\mathrm{G}^2}e^{-\frac{i^2+j^2}{2\sigma_\mathrm{G}^2}})^2\sigma_\mathrm{coadd}^2,\\
    \sigma_\mathrm{smoothed}^2&\simeq\frac{\sigma_\mathrm{coadd}}{4\pi^2\sigma_\mathrm{G}^2}^2\int_{-\infty}^{\infty}\int_{-\infty}^{\infty}e^{-\frac{i^2}{\sigma_\mathrm{G}^2}}e^{-\frac{j^2}{\sigma_\mathrm{G}^2}}di\:dj, \label{eqn:std_1}\\
    \sigma_\mathrm{smoothed}&\simeq\frac{\sigma_\mathrm{coadd}}{2\sqrt{\pi}\sigma_\mathrm{G}}, \label{eqn:std_2}
\end{align}

where $\sigma_\mathrm{smoothed}$ is the standard deviation of the smoothed image, $\mathbf{im}_\mathrm{smoothed}$. Equation \ref{eqn:std_1} assumes that $\sigma_\mathrm{G}$ is large enough such that the square of the Gaussian is smooth and the image is large enough relative to the Gaussian size such that the sum of variances can be approximated as an integral from $-\infty$ to $\infty$ \citep[See the stack exchange post that prompted the analytical approximation used in Equation \ref{eqn:std_1};][]{niemitaloAnswerHowDoes2015}. We have checked that the first of these assumptions holds for $\sigma_\mathrm{G}=40$ pixels. The second assumption holds given our image length is $\sim60\sigma_\mathrm{G}$. The following uses Equation~\ref{eqn:std_2} to express $\sigma_\mathrm{subtracted}$, which is the standard deviation of $\mathbf{im}_\mathrm{subtracted}=\mathbf{im}_\mathrm{coadd}-\mathbf{im}_\mathrm{smoothed}$, in terms of $\sigma_\mathrm{coadd}$, $\sigma_\mathrm{G}$ and the covariance between $\mathbf{im}_\mathrm{coadd}$ and $\mathbf{im}_\mathrm{smoothed}$.

\begin{align}
    \sigma_\mathrm{subtracted} &= \sqrt{\sigma_\mathrm{smoothed}^2+\sigma_\mathrm{coadd}^2-2\mathrm{cov}(\mathbf{im}_\mathrm{coadd}\mathbf{im}_\mathrm{smoothed})},\\
    \sigma_\mathrm{subtracted} &\simeq\sqrt{\frac{\sigma_\mathrm{coadd}^2}{4\pi\sigma_\mathrm{G}^2}+\sigma_\mathrm{coadd}^2-2\mathrm{cov}(\mathbf{im}_\mathrm{coadd}\mathbf{im}_\mathrm{smoothed})},\\
    \sigma_\mathrm{subtracted}&\simeq\frac{\sigma_\mathrm{coadd}}{2\sqrt{\pi}\sigma_\mathrm{G}}\sqrt{4\pi\sigma_\mathrm{G}^2(1-\frac{2\mathrm{cov}(\mathbf{im}_\mathrm{coadd}\mathbf{im}_\mathrm{smoothed})}{\sigma_\mathrm{coadd}^2})+1},
\end{align}

We selected 400 pixels ($\sim4$ pkpc$^2$) as the minimum source area, as this covers the smallest sources in our sample but is not so small that parts of galaxies are detected as separate sources. We arrived at our relatively high detection threshold to reduce the detection of tidal features as separate sources. We present our tests of the deblending parameters in Appendix \ref{app:masking} to arrive at choices that lead to the best separation of host galaxies from nearby satellites.

We grow the combined hot+cold mask by dilating the segments in increments of 1 pixel using a circular footprint until the median background level matches the expected noise of the coadded image to within $\sigma_{\mathrm{coadd}}$ or up to a maximum of 20 pixels ($\sim2$ pkpc). The median background level is measured within as many $10^{\prime\prime}\times10^{\prime\prime}$ boxes fit within the image, allowing for a maximum of 30 per cent of the pixels within the box to be masked. Without a hard limit on the maximum size of the mask, the mask grower would not stop on cutouts of large groups and clusters, where there is a background of intra-group and intra-cluster light, until nearly every pixel in the image was masked. 

\subsubsection{Measuring galaxy shape and size}
\label{subsubsec:mge_model}

\begin{figure}
    \centering
    \includegraphics[width=\linewidth]{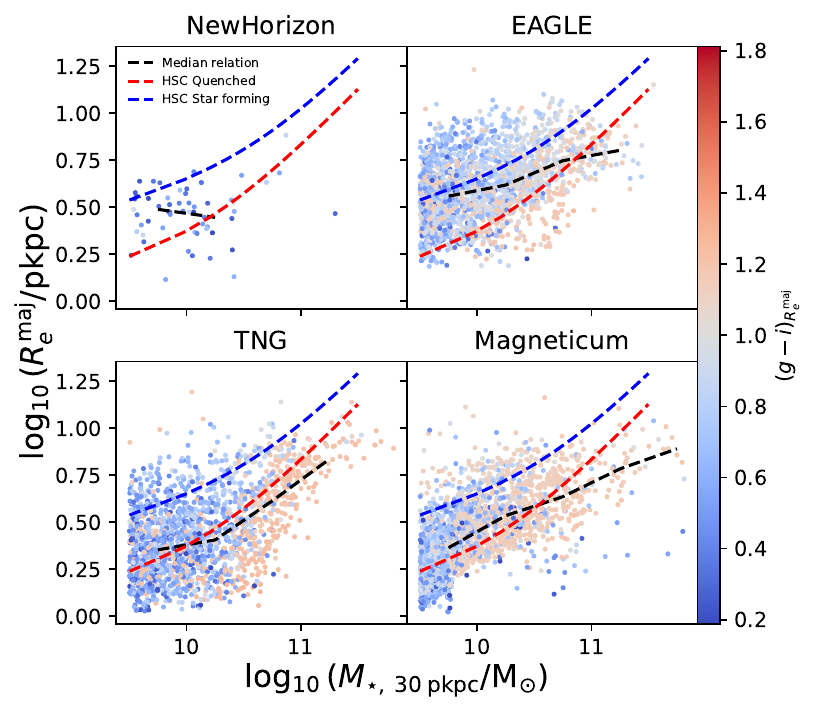}
    \caption{The size-stellar mass distribution for galaxies in each simulation. The full sample of galaxies for which we have reliable measurements out to 1 $R_{e}^\mathrm{maj}$ is indicated here. The sizes are given in proper kpc, and the stellar masses are provided in $\log_{10}(M_{\star\text{, 30 pkpc}}/\mathrm{M}_\odot)$. We see that qualitatively the distributions are similar, with the distribution for Magneticum having less scatter than EAGLE or TNG. The black dashed line is the median size-stellar mass relation for each simulation, binned in $\log_{10}(M_{\scriptstyle\star\mathrm{,\:30\:pkpc}}/$M$_{\scriptstyle\odot})$ with 0.5 dex wide bins ranging from 9.5 to 12, plotted only for bins with at least 10 galaxies. The blue and red dashed lines are empirical fit relations from the HSC-SSP survey \citep{kawinwanichakijHyperSuprimeCamSubaru2021} for star-forming and quiescent galaxies, respectively. Our size-mass relations are in reasonable agreement with those from HSC-SSP.}
    \label{fig:size_stellar_mass}
\end{figure}

We measure the shapes and sizes of galaxies in the masked \textit{gri}-coadd using the  Multi-Gaussian Expansion (MGE) method \citep{monnetModellingStellarIntensity1992,emsellemMultigaussianExpansionMethod1994} implemented in the \textsc{MGEfit} package \citep{cappellariEfficientMultiGaussianExpansion2002}. We use the MGE method as it provides an efficient and flexible way to model the extent of the galaxy, allowing for varying axial ratios and twisting isophotes. As we were not concerned with modelling asymmetries and features such as spiral arms, MGE models were sufficient for the task. We use the \texttt{find\_galaxy} function to measure the average ellipticity and position angle of the galaxy. This function finds regions of spatially connected pixels lying within the brightest 10 per cent of pixels in the image and then uses the first and second-order moments of the intensity distribution to calculate the galaxy centre, position angle and average ellipticity. The \texttt{find\_galaxy} function worked best on a smaller cutout of the original image, these cutouts were $150\mathrm{\:pixels}\times150\mathrm{\:pixels}=60^{\prime\prime}\times60^{\prime\prime}\simeq15\mathrm{\:pkpc}\times15\mathrm{\:pkpc}$, we use the 150 pixels $\times$ 150 pixels cutouts for the modelling of the galaxy with \texttt{MGEfit}. Given that our images are centred on the centre of the potential of the subhalo, we aim to find an object within 9 pixels ($\sim0.9$ pkpc) of the image centre. In instances where there is no object found within 9 pixels of the galaxy, we extend the search to include the brightest 50 per cent of pixels in the image, and if the algorithm fails to find the galaxy again, we flag the object and discard it from our sample. We discarded 5 galaxies from NewHorizon, 43 from TNG, 22 from EAGLE and 87 from Magneticum for this reason. Failure to identify the galaxy would usually occur in crowded fields where there was sufficient light from other sources to make it difficult to automatically detect the galaxy in the centre of the image.

We model each galaxy using multiple Gaussian components. For the modelling of the galaxy, there are options regularize the fit, which requires the broadest Gaussian component to have the roundest axial ratio while still fitting the observations and fixes the position angle of the Gaussian components \citep{cappellariEfficientMultiGaussianExpansion2002}. Regularized fits are used for dynamical modelling \citep[e.g.][]{scottSAURONProjectXIV2009}, which is not the aim of this study and unregularized fits have been found to provide the lowest $\chi^2$ values and yield the most realistic galaxy shapes \citep{deugenioSAMIGalaxySurvey2021}. We therefore use unregularized fits here. We allow for models to vary position angles with radius, as we found these further optimised our fits. We also convolve our fits with circular MGE models of the Hyper-Suprime Cam point spread function used in our mock images, to produce more accurate model galaxies.

Examples of fits to our galaxies can be seen in Fig.~\ref{fig:demo_size_measure_mask}. The fits reflect the data best, within the symmetric regions inside the galaxy, and get less accurate over the outskirts, which may exhibit tidal disruption and/or asymmetries. At this stage, we are using these models to measure galaxy size, so accurately modelling the outskirts is not necessary, as we are only trying to get a measurement of the overall extent and brightness of the galaxy. We use the MGE model and the average ellipticity ($\varepsilon$) and position angle ($\Psi$) of the galaxy from \texttt{find\_galaxy}, to measure the effective semi-major radius, $R_{e}^{\mathrm{maj}}$. The effective semi-major radius is half the semi-major axis of the elliptical aperture enclosing half the light from the MGE model of the galaxy. 

The galaxy size-stellar mass relation for the sample is shown in Fig.~\ref{fig:size_stellar_mass}. As in \citet{khalidCharacterizingTidalFeatures2024}, we use the 30 pkpc spherical aperture stellar mass, $M_{\star,\text{ 30 pkpc}}$. As discussed in \citet{schayeEAGLEProjectSimulating2015}, the 30 pkpc radius spherical aperture is comparable to the 2D Petrosian aperture often used in observational studies, making it a good choice for our aim to analyse the data in a manner directly comparable to observations. Fig.~\ref{fig:size_stellar_mass} shows that NewHorizon exhibits size values within the range covered by the other simulations. The smaller number of galaxies in this simulation is likely why we do not see any trends with stellar mass. EAGLE, TNG and Magneticum all show increasing galaxy size with increasing stellar mass. The median relations for each simulation span the range of values covered by the fits to the observed quenched and star-forming populations for galaxies at $z=0.3$ in HSC-SSP \citep{kawinwanichakijHyperSuprimeCamSubaru2021}. For $M_{\scriptstyle\star\mathrm{,\:30\:pkpc}}\gtrsim10^{11}\:\mathrm{M}_{\scriptstyle\odot}$, EAGLE, TNG and Magneticum exhibit a median size smaller than the relation for observed quenched HSC galaxies. While the medians are similar, TNG produces galaxy sizes smaller than EAGLE and Magneticum, particularly for bluer galaxies with $\log_{10}(M_{\scriptstyle\star\mathrm{,\:30\:pkpc}}/$M$_{\scriptstyle\odot})\lesssim 10.75$. In EAGLE, we see a tendency at a given mass for a bluer-coloured galaxy to have a larger size than a redder-coloured galaxy, whereas in TNG and Magneticum, red and blue galaxies appear to span similar ranges of sizes at a given stellar mass.

Given that MGE typically works best on elliptical galaxies \citep[e.g.][]{lablancheATLAS3DProjectXII2012,tahmasebzadehDeprojectionExternalBarred2021,dattathriDeprojectionStellarDynamical2024}, we tested how our measurements behave as a function of galaxy colour and shape in Appendix~\ref{app:size_measurement}. We find that the differences in measured galaxy sizes are driven primarily by their shape rather than their colour. Furthermore, we tested whether using the intrinsic simulation-measured galaxy sizes impacts our results, and find that they remain qualitatively unchanged.

\subsubsection{Measuring galaxy and galaxy outskirt colours}
\label{subsubsec:measure_colour}

We derive the intrinsic $g-i$ colours of our galaxies from the light in our mock image enclosed within the elliptical aperture (with an ellipticity $\varepsilon$ and a position angle $\Psi$) at $R_{e}^{\mathrm{maj}}$. The $g-i$ colour profiles are calculated using the light in apertures with the ellipticity and position angle fixed by that measured by \texttt{find\_galaxy}, of semi-major axis 0 - 1 $R_{e}^{\mathrm{maj}}$, 1 - 2 $R_{e}^{\mathrm{maj}}$, 2 - 3 $R_{e}^{\mathrm{maj}}$, 3 - 4 $R_{e}^{\mathrm{maj}}$ and 4 - 5 $R_{e}^{\mathrm{maj}}$. In cases where 5 $R_e^\mathrm{maj}$ was outside the $60^{\prime\prime}\times60^{\prime\prime}$ cutouts, we made a new cutout larger such that $5$ $R_e^\mathrm{maj}$ would fit within the cutout.



\subsection{Sample selection}
\label{subsec:measure_colour}

\begin{figure*}
    \centering
    \includegraphics[width=\linewidth]{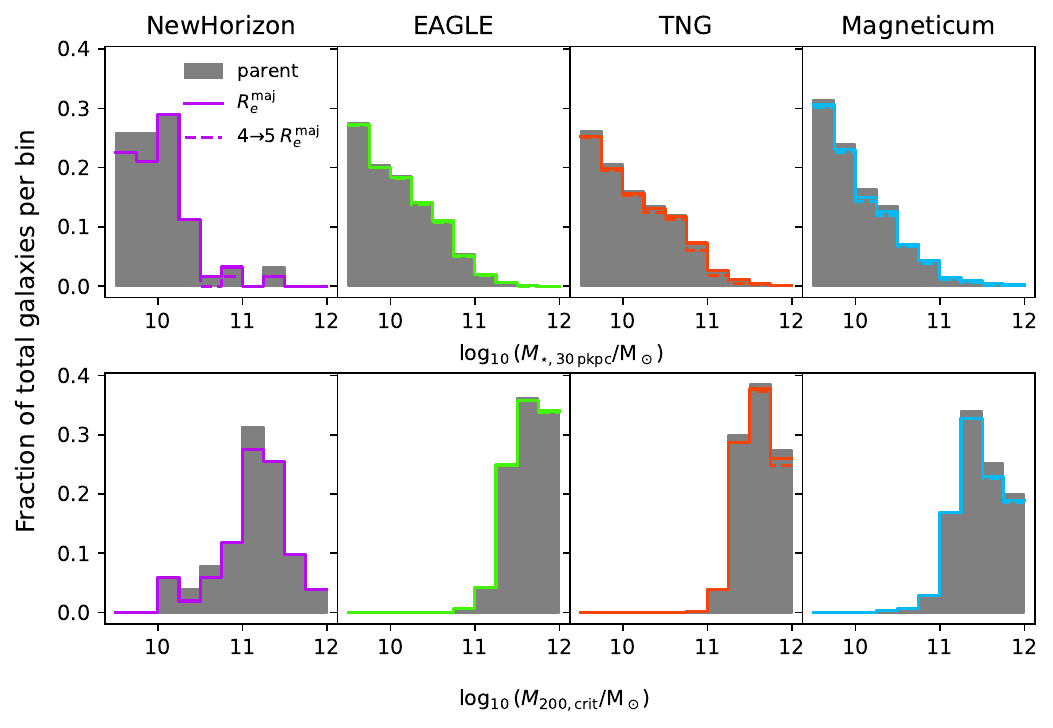
    }
    \caption{(Top panel) The stellar mass distributions for our samples from each simulation, measured in $\log_{10}(M_{\star\text{, 30 pkpc}}/\mathrm{M}_\odot)$. (Bottom panel) The halo mass distribution for our samples from each simulation, measured in $\log_{10}(M_{\text{200, crit}}/\mathrm{M}_\odot)$. We display the distributions in terms of the fraction of the total parent catalogue in each bin, i.e. counts in bin/total parent. We show the parent distribution of all galaxies in the tidal feature catalogue in grey. The solid lines show the distributions for the sample \aman{where our measurements of galaxy colour within $1$ $R_e^{\mathrm{maj}}$ meet the minimum size ($\geq10$ pixels) and maximum masking within aperture (50 per cent) criteria}, and the dashed lines show the distribution for the sample where \aman{the same criteria have been met} out to the aperture between $4$ and $5$ $R_e^{\mathrm{maj}}$. There are no significant systematic differences between any of the distributions.}
    \label{fig:stellar_halo_mass_distributions}
\end{figure*}

The catalogue of mock images and the measurements we have derived from them are summarised in Table \ref{tab:simulations}. Our catalogue of mock images with tidal feature classifications covers $62$ galaxies in NewHorizon, $1978$ galaxies in EAGLE, $1826$ galaxies in TNG and $1990$ galaxies in Magneticum. We run our masking and colour measurement code on each of these galaxies. We discard galaxies for which the \texttt{find\_galaxy} routine cannot locate the galaxy of interest (5, 43, 22 and 87 galaxies from NewHorizon, TNG, EAGLE and Magneticum, respectively). To ensure our colours are not impacted significantly by the softening length of the simulation, we only include galaxies with $R_e^\mathrm{maj}\geq2^{\prime\prime}$ (10 pixels $\sim1$ pkpc). This selection results in 1 galaxy being discarded from NewHorizon, 22 from EAGLE, 10 from TNG and 0 from Magneticum. To ensure we have enough pixels within any aperture for analysis, we allow for a maximum of $50$ per cent of any aperture to be masked. The resulting measurements of colour were made on a minimum of 443 pixels/aperture in NewHorizon, 539 pixels/aperture in EAGLE, 227 pixels/aperture in TNG and 233 pixels/aperture in Magneticum.

Table \ref{tab:simulations} also gives the number of measurements meeting the above criteria for the elliptical aperture out to $1$ $R_e^{\mathrm{maj}}$ for each simulation. These are the measurements we use in our analysis of galaxy colour, stellar mass and tidal feature fraction in Section \ref{subsec:col_mstar}. In this sample, we have 55 galaxies in NewHorizon, 1959 in EAGLE, 1773 in TNG and 1902 in Magneticum.

The colour profiles of galaxies differ between red sequence and blue cloud galaxies \citep[e.g.][]{millerColorGradientsHalfmass2023}. To limit the impact of the intrinsically different colour profiles of blue and red galaxies on our colour profile analysis, we divide the population into a red sequence and a blue cloud. In our analysis of colour profiles, we extend our measurements to 5 $R_e^\mathrm{maj}$ and apply the same masking threshold of 50 per cent as we did within $1\:R_e^\mathrm{maj}$. The number of galaxies remaining in the sample at each aperture between $1$ and $5\:R_e^{\mathrm{maj}}$ is given in Table \ref{tab:simulations}. Fig.~\ref{fig:stellar_halo_mass_distributions} illustrates the stellar and halo mass distributions for our final samples. For ease of illustration, we show only the distributions for the sample of colour and size measurements meeting the size and masking criteria out to 1 $R_e^{\mathrm{maj}}$ and those meeting the criteria out to the $4\to5$ $R_e^{\mathrm{maj}}$ aperture. We see that the distributions of the parent sample and for objects with reliable colours out to $4\to5$ $R_e^\mathrm{maj}$ are not significantly different from the parent distributions, yielding KS-test p-values of 1.00, 1.00, 0.78 and 0.97 for NewHorizon, EAGLE, TNG and Magneticum, respectively. For the $M_\mathrm{200,\: crit}$ distributions, the p-values are 1.00, 1.00, 0.96 and 0.84 for NewHorizon, EAGLE, TNG and Magneticum, respectively. While these differences are not statistically significant, we do observe the number of objects with valid measurements out to the $4\to5$ $R_e^\mathrm{maj}$ aperture drops with increasing stellar mass and at host halo masses corresponding to groups and clusters. This is expected as these higher stellar mass galaxies in crowded environments are more likely to have their outskirts impacted by masking from a neighbouring source and/or the intra-group/intra-cluster light \citep[e.g.][]{martinez-lombillaGalaxyMassAssembly2023,broughPreparingLowSurface2024}.

\begin{figure*}
    \centering
    \includegraphics[width=\linewidth]{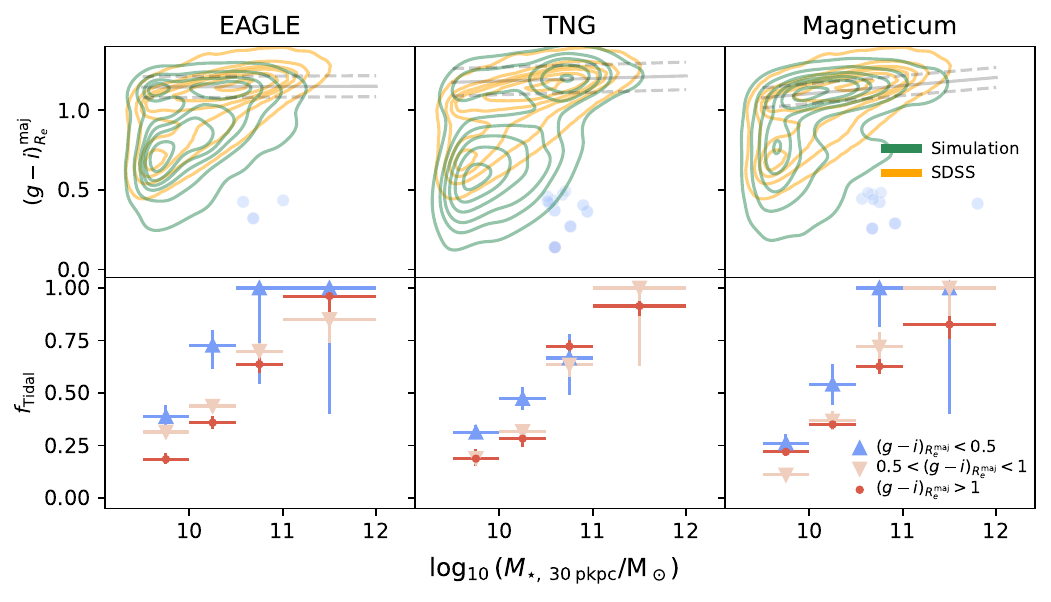}
    \caption{(Top panel) The $(g-i)_{R_e^\mathrm{maj}}$ colour and spherical aperture stellar mass contours for our simulations in green and the Galactic extinction corrected model $g-i$ colour and stellar masses from SDSS observations of galaxies at $z\leq0.035$ in orange, the lines show the 5 per cent, 20 per cent, 40 per cent, 60 per cent, 80 per cent and 95 per cent contours. We also show the galaxies with $M_{\star\mathrm{,\:30\:pkpc}}\geq10^{10.5}\:\mathrm{M}_\odot$ and $(g-i)_{R_e^\mathrm{maj}}<0.5$ as light blue points, as while this region lies outside the $95\%$ contours, these points do contribute to the tidal feature fractions shown. The solid grey line shows the linear fit to the red sequence for simulated galaxies, and the dashed grey lines show the selection of red sequence galaxies ($\pm 3\times$ the median distance of all non-blue cloud galaxies from the red sequence fit). Each simulation resolves a red sequence and a blue cloud of galaxies. There are quantitative differences in the shape and scatter of the relations across simulations and when compared to observations. The simulated blue clouds all scatter to bluer colours relative to observations, and the slopes of the red sequence in all simulations are all flatter than the slope for SDSS. (Bottom panel) Tidal feature fractions as a function of stellar mass for 4 stellar mass bins (bin edges $\log_{10}(M_{\star\mathrm{,\:30\:pkpc}}/$M$_{\scriptstyle\odot})=[9.5,\:10,\:10.5,\:11,\:12]$) and 3 colour bins (bin edges $(g-i)_{R_e^\mathrm{maj}}=[-0.02,\: 0.5,\: 1,\: 1.3]$), the blue points correspond to the bluest bin, the light pink points correspond to the middle colour bin and the red points correspond to the reddest bin. The error bars on the tidal feature fractions indicate the $1\sigma$ binomial uncertainties. The error bars in the stellar mass axis show the sizes of the bins. Tidal feature fractions increase with increasing stellar mass at all colours. The red sequence galaxies generally have lower tidal feature fractions than the blue cloud at all stellar masses.}
    \label{fig:col_mag_total}
\end{figure*}

As a preliminary division between the blue and red galaxies, we examine a $(g-i)_{R_e^\mathrm{maj}}$ histogram for each simulation and locate the local minimum between the two peaks for the red sequence and the blue cloud (see Fig.~\ref{fig:col_dist} in Appendix \ref{app:col_thresh}). The galaxies on the blue side of this threshold are allocated to the blue cloud, and the red side is linearly fit. NewHorizon does not have a sufficient number of `red' galaxies to resolve the red sequence. Therefore, we do not measure the threshold between the red and blue galaxies for this sample. The thresholds in intrinsic colour are $(g-i)_{R_e^\mathrm{maj}}=1.075,\: 1.025$ and $0.975$ for EAGLE, TNG and Magneticum respectively. The results of our linear fit of the red sequence are shown in the solid grey line in Fig.~\ref{fig:col_mag_total}. We compute the distances of all galaxies above the respective thresholds from the red sequence fit, and we then select all galaxies that are within 3 times the median distance of the fit relation to be part of the red sequence. This region is shown with the dashed grey lines. We select the region between the two dashed lines to be our sample of red sequence galaxies. We select the red sequence and blue cloud galaxies for EAGLE, TNG and Magneticum but not NewHorizon, as it does not have a sufficient number of `red' galaxies to resolve both populations. We give the colour distribution for NewHorizon in Appendix \ref{app:col_thresh}. We have 493 red sequence galaxies in EAGLE, 485  in TNG and 904 in Magneticum. We have 1448 blue cloud galaxies in EAGLE, 1246 in TNG and 905 in Magneticum. This selection leaves galaxies redder than the histogram-identified blue cloud threshold but not captured within the red sequence fit. We have 18 galaxies in EAGLE not assigned to either sample, 43 in TNG and 93 galaxies in Magneticum. This is the sample we use for our colour profile analysis in Section \ref{subsec:delta_col}.

\section{Results}
\label{sec:results}

\begin{table*}
    \centering
    \caption{The KS test results comparing the $(g-i)_{R_e^\mathrm{maj}}$ of galaxies without tidal features to galaxies hosting tidal features for each mass bin, $M_{\star,\:\mathrm{30,\:pkpc}}=[10^{9.5},\:10^{10},10^{10.5},\:10^{11},\:10^{12}]\:\mathrm{M}_\odot$. From left to right, the first value gives the critical $(g-i)_{R_e^\mathrm{maj}}$, where the test statistic is measured, the second value gives the test statistic, the third value gives the $p$-value, and the third and fourth values give the number of non-tidal feature $(conf.=0)$ and tidal feature $(conf.\geq2)$ host galaxies in the bin, respectively. A negative test statistic indicates that there is a greater proportion of tidal feature hosts than non-tidal feature hosts at colours bluer than the $(g-i)_{R_e^\mathrm{maj}}$ where the test statistic is measured. We bold negative test results which indicate that the two populations are not sampled from the same underlying distribution and the $p$-values are $<0.05$.}
    \begin{tabular}{l|l|l|l}
        \hline
         Stellar mass bin [M$_\odot$] & EAGLE & TNG & Magneticum\\
        \hline
        $[10^{9.5},\:10^{10}]$ & $0.79,\mathbf{\:-0.17,\:0.00013},\:584,\:234$ & $0.55,\:\mathbf{-0.15,\:0.0059},\:561,\:160$ & $0.97,\:0.14,\mathbf{\:0.0097},\:832,\:159$\\
        $[10^{10},\:10^{10.5}]$ & $0.91,\mathbf{\:-0.14,\:0.0058},\:339,\:241$ & $0.41,\:-0.13, 0.067,\:302,\:152$ & $1.11,\:\mathbf{-0.15, 0.0097},\:326,\:186$\\
        $[10^{10.5},\:10^{11}]$ & $0.91,\mathbf{\:-0.17},\:0.051,\:94,\:187$ & $0.92,\:0.14,\:0.14,\:90,\:207$ & $1.15,\mathbf{\:-0.29,\:0.00043},\:73,\:137$\\
        $[10^{11},\:10^{12}]$ & $0.97,\:0.40,\:0.49,\:4,\:41$ & $1.20,\:\mathbf{-0.20},\:0.94,\:6,\:63$ & $1.13,\:\mathbf{-0.35},\:0.31,\:8,\:49$\\
        \hline
    \end{tabular}
    \label{tab:ks_col_mag_total}
\end{table*}

\begin{figure}
    \centering
    \includegraphics[width=\linewidth]{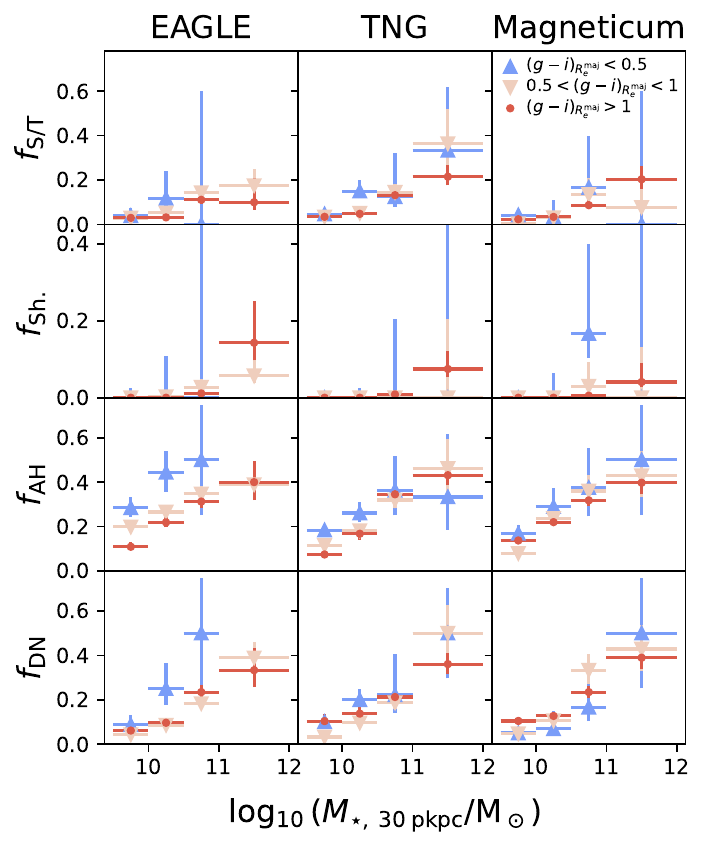}
    \caption{The tidal feature fraction as a function of spherical aperture stellar mass binned by colour for each simulation and tidal feature morphology. Top to bottom, the panels show fractions for: streams/tails, shells, asymmetric halos and double nuclei. The binning is identical to Fig.~\ref{fig:col_mag_total} and illustrated in the x-error bars. The y-error bars illustrate the $1\sigma$ binomial uncertainties on the tidal feature fractions. The trend for bluer galaxies to have a greater tidal feature fraction seen in Fig.~\ref{fig:col_mag_total} is driven primarily by the asymmetric halos.}
    \label{fig:col_mag_stf}
\end{figure}

\begin{figure}
    \centering
    \includegraphics[width=\linewidth]{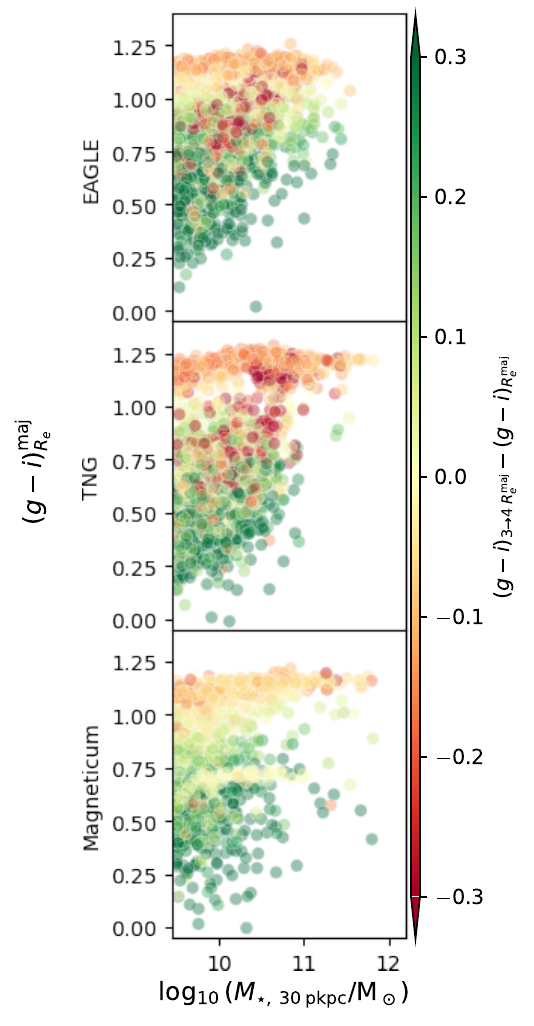}
    \caption{The $(g-i)_{R_e^\mathrm{maj}}$ colour and spherical aperture stellar mass for each galaxy in our simulations, the data points are coloured by $(g-i)_{3\to4\:R_e^\mathrm{maj}}-(g-i)_{R_e^\mathrm{maj}}$. Negative values indicate a negative slope between the inner effective semi-major axis and the $3\to4\:R_e^\mathrm{maj}$ apertures, and a positive value indicates a positive slope. We can see that red sequence galaxies have predominantly negative slopes.}
    \label{fig:col_mag_delta_col}
\end{figure}

In this Section, we present the relationships between tidal feature fraction, galaxy colour and the difference in colour between the galaxy centre and its outskirts. We compare the colours and colour profiles of galaxies with and without tidal features to provide insight into how any differences reflect the properties of the tidal features. In Section \ref{subsec:col_mstar}, we investigate the differences between galaxies with tidal features and galaxies without tidal features on the colour-stellar mass diagram. In Section \ref{subsec:delta_col}, we present a comparison of the radial colour profiles of 
tidal feature hosts and non-tidal feature hosts. 

We choose to use the confidence 2 and higher tidal features for this analysis. This is done to achieve a balance between the likelihood of correct classification and obtaining a significant sample of tidal feature galaxies to study. For our sample of galaxies without tidal features for our colour profile analysis, we selected galaxies with confidence 0 in every tidal feature category, indicating that no tidal feature was detected. This was done to construct as pure a sample of non-tidal feature galaxies as possible.

\subsection{Colour-stellar mass diagrams}
\label{subsec:col_mstar}

Fig.~\ref{fig:col_mag_total} shows the colour-stellar mass relations for EAGLE, TNG and Magneticum and tidal feature fractions as a function of stellar mass and colour. The top row of Fig.~\ref{fig:col_mag_total} compares the colour-stellar mass distributions for EAGLE, TNG and Magneticum with the colour-stellar mass distribution for the Sloan Digital Sky Survey \citep[e.g.][]{yorkSloanDigitalSky2000,aiharaEIGHTHDATARELEASE2011}. We do not include NewHorizon in this plot as it has too few data points for comparisons in colour-stellar mass. In Appendix \ref{app:col_thresh}, we show NewHorizon's $(g-i)_{R_e^\mathrm{maj}}$ distribution, which shows that blue cloud galaxies dominate the simulation. For SDSS we selected galaxies with $z\leq0.035$, as this produces a volume-limited sample complete to $M_\star\sim10^{9.5}$ \citep[][]{weigelStellarMassFunctions2016}. We plot their Galactic extinction-corrected model magnitudes \citep{schlegelMapsDustInfrared1998}, and stellar mass estimates based on the \citet{chenEvolutionMostMassive2012} method and \citet{bruzualStellarPopulationSynthesis2003} simple stellar populations. We transformed the \citet{chenEvolutionMostMassive2012} stellar masses from a Kroupa IMF \citep{kroupaVariationInitialMass2001} to Chabrier IMF \citep{chabrierGalacticStellarSubstellar2003} with a correction $\log_{10}(M_\mathrm{Chabrier}/\mathrm{M}_{\scriptstyle \odot})=\log_{10}(M_\mathrm{Kroupa}/\mathrm{M}_{\scriptstyle \odot})-0.05$, to align them with our simulations, which all assume a Chabrier IMF. We generally see that the simulations reproduce the red sequence and blue cloud structures on the colour-stellar mass diagram. Broadly, the distributions for the simulations and SDSS overlap. The slopes for the red sequence in all simulations are flatter than in SDSS. In all simulations, the blue cloud tends to scatter to bluer values than in SDSS. This difference is largest in TNG. Some of this difference could be due to the SDSS colours not being intrinsic, as they are not corrected for reddening due to dust intrinsic to the galaxy. The distributions of the red sequence and blue cloud galaxies vary between the simulations. EAGLE and Magneticum have similar colours in their red sequence, whereas TNG is shifted to redder colours. TNG has more high stellar mass members of the blue cloud, and the blue cloud in TNG is distributed to much bluer values than EAGLE or Magneticum.

The colour-stellar mass diagram allows us to investigate any relationships between host galaxy colour and the occurrence of visually-detected tidal features. We show how our tidal feature fraction as a function of stellar mass changes with colour in the bottom panels of Fig.~\ref{fig:col_mag_total}. We separate the population into three colour bins, ranging from -0.02 to 0.5, 0.5 to 1 and 1 to 1.3. The two bluer bins split the blue cloud, whereas the reddest bin spans the red sequence. We find across all simulations a systematic tendency for the bluest galaxies in the blue cloud to exhibit higher tidal feature fractions. We note that there are generally no substantial differences in the relationship between the tidal feature fraction-stellar mass relation followed by the red sequence galaxies and the redder galaxies in the blue cloud. 

Given that the trends in colour in Fig.~\ref{fig:col_mag_total} are not statistically strong, we further examine the result in a KS test \citep{hodgesSignificanceProbabilitySmirnov1958} of the colours $(g-i)_{R_e^\mathrm{maj}}$ of the galaxies with and without tidal features in each stellar mass bin. The two-sample KS test compares the cumulative distribution functions (CDFs) of the two samples. The test statistic is the largest difference between the CDFs of the two samples, and the p-value is the probability that this difference can arise if the two samples were drawn from the same distribution. Table~\ref{tab:ks_col_mag_total} shows our KS test results for the colour distributions for each stellar mass bin (bin edges $\log_{10}(M_{\star\mathrm{,\:30\:pkpc}}/\mathrm{M}_\odot)=[9.5,\:10,\:10.5,\:11,\:12]$). Negative KS test statistics indicate that at colours bluer than where the test statistic is measured, there are proportionately more galaxies with tidal features than without. We find that in EAGLE there is a clear and statistically significant ($>2\sigma$) shift to bluer colours for $M_{\star\mathrm{,\:30\:pkpc}}<10^{10.5}\:\mathrm{M}_\odot$. In TNG, this shift to bluer colours is only statistically significant in the lowest stellar mass bin. In Magneticum, the shift to bluer colours is statistically significant in the middle two mass bins ($10^{10}<M_{\star\mathrm{,\:30\:pkpc}}/\mathrm{M}_\odot<10^{11}$), however, we also see a statistically significant shift to redder colours for galaxies hosting tidal features in the lowest mass bin for Magneticum ($M_{\star\mathrm{,\:30\:pkpc}}<10^{10}\:\mathrm{M}_\odot$). The negative test statistics and some statistical significance support the systematic trends that we are seeing in our results, mainly that the tidal feature fraction is higher in the bluer population of galaxies as a function of stellar mass. However, the fact that this is not statistically significant across the board suggests that we should be cautious in our conclusions.

In Fig.~\ref{fig:col_mag_stf}, we investigate the relationship between tidal feature fraction and stellar mass for specific tidal feature morphologies, binned by galaxy colour. The number of each tidal feature in each of our stellar mass bins is given in Table \ref{tab:features_mstar_bin}. The third row of the figure highlights that it is primarily the occurrence of the most frequent features, asymmetric halos, that drive the trend of the bluest objects in our sample having higher tidal feature fractions (Fig.~\ref{fig:col_mag_total}). There are indications that galaxies hosting streams/tails and double nuclei also contribute to this. However, a larger sample is needed to show statistical significance. For the shells, we do see a hint of a preference for redder galaxies in the highest mass bins across all simulations. However, we cannot draw strong conclusions due to the small number of shells in our sample (12 in EAGLE, 7 in TNG and 5 in Magneticum).

\begin{table}
    \centering
    \caption{The total number of tidal features and each tidal feature morphology (stream/tail, shell, asymmetric halo and double nucleus) detected in each stellar mass bin. The bins are defined differently for the red sequence and the blue cloud due to the different stellar mass distributions of the populations.}
    \begin{tabular}{p{1.8cm}|c|c|c|c|c}
    \hline
         Stellar mass bin [M$_{\scriptstyle \odot}$] & Total features & S/T & Sh. & AH & DN\\
         \hline
         Red sequence & & & & &\\
         \hline
         EAGLE & & & & & \\
         $[10^{9.5}$, $10^{10}]$ & 34 & 7 & 0 & 16 & 11\\
         $[10^{10}$, $10^{10.5}]$ & 66 & 4 & 0 & 43 & 19\\
         $[10^{10.5}$, $10^{11}]$ & 110 & 15 & 1 & 57 & 37\\
         $\geq10^{11}$ & 32 & 4 & 2 & 13 & 13\\
         \hline
         TNG & & & & & \\
         $[10^{9.5}$, $10^{10}]$ & 16 & 2 & 0 & 6 & 8\\
         $[10^{10}$, $10^{10.5}]$ & 42 & 6 & 0 & 20 & 16\\
         $[10^{10.5}$, $10^{11}]$ & 177 & 25 & 2 & 96 & 54\\
         $\geq10^{11}$ & 72 & 8 & 5 & 31 & 28\\
         \hline
         Magneticum & & & & & \\
         $[10^{9.5}$, $10^{10}]$ & 79 & 5 & 0 & 43 & 31\\
         $[10^{10}$, $10^{10.5}]$ & 170 & 14 & 0 & 103 & 53\\
         $[10^{10.5}$, $10^{11}]$ & 138 & 15 & 1 & 75 & 47\\
         $\geq10^{11}$ & 46 & 9 & 1 & 18 & 18\\
         \hline
         Blue cloud & & & & &\\
         \hline
         EAGLE & & & & &\\
         $[10^{9.5}$, $10^{10}]$ & 212 & 23 & 0 & 200 & 40\\
         $[10^{10}$, $10^{10.5}]$ & 194 & 27 & 1 & 175 & 49\\
         $[10^{10.5}$, $10^{11}]$ & 109 & 28 & 5 & 90 & 42\\
         $\geq10^{11}$ & 20 & 5 & 3 & 16 & 16\\
         \hline
         TNG & & & & &\\
         $[10^{9.5}$, $10^{10}]$ & 142 & 26 & 0 & 111 & 43\\
         $[10^{10}$, $10^{10.5}]$ & 113 & 29 & 0 & 88 & 52\\
         $[10^{10.5}$, $10^{11}]$ & 52 & 16 & 0 & 45 & 22\\
         $\geq10^{11}$ & 2 & 1 & 0 & 2 & 2\\
         \hline
         Magneticum & & & & &\\
         $[10^{9.5}$, $10^{10}]$ & 88 & 6 & 0 & 72 & 36\\
         $[10^{10}$, $10^{10.5}]$ & 48 & 4 & 0 & 45 & 12\\
         $[10^{10.5}$, $10^{11}]$ & 25 & 6 & 2 & 18 & 15\\
         $\geq10^{10}$ & 7 & 0 & 0 & 6 & 5\\
         \hline
    \end{tabular}
    \label{tab:features_mstar_bin}
\end{table}

\subsection{Galaxy colour profiles}
\label{subsec:delta_col}
\begin{figure}
    \centering
    \includegraphics[width=\linewidth]{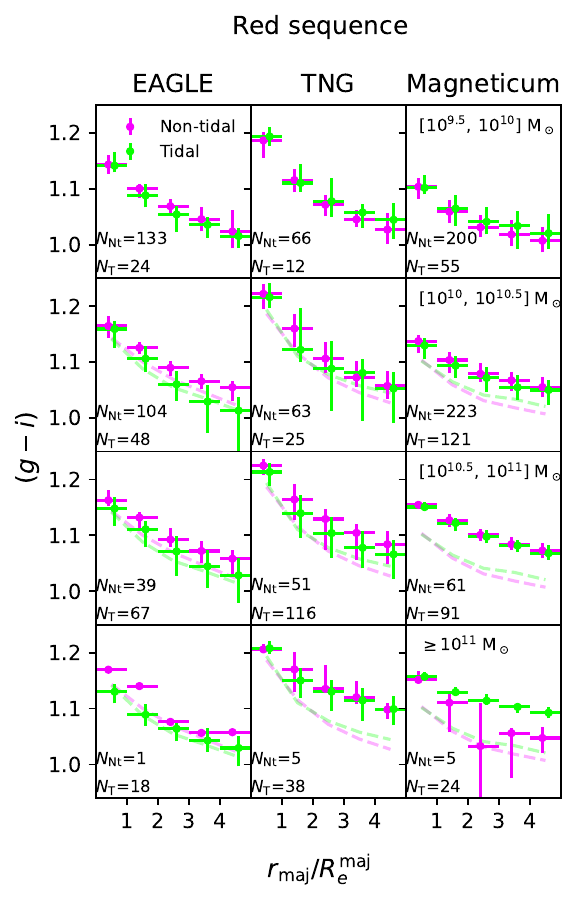}
    \caption{The distributions of the radial colour profiles for red sequence galaxies, given in terms of $(g-i)$ and the semi-major axis ($r_\mathrm{maj}/R_{e}^\mathrm{maj}$). The profiles are binned by stellar mass for galaxies hosting tidal features (lime green) and galaxies not hosting tidal features (pink), for each simulation. The profiles are measured using elliptical apertures spaced in integer values from $0\to5\:R_e^\mathrm{maj}$. Each data point shows the median colour in that aperture, and the error bars show the 25th and 75th percentiles. The number of non-tidal feature hosts is given in each panel by $N_\mathrm{Nt}$, and the number of tidal feature hosts is given by $N_\mathrm{T}$. The results for different stellar mass bins are shown with increasing stellar mass from top to bottom with bin edges $\log_{10}(M_{\scriptstyle\star\mathrm{,\:30\:pkpc}}/\mathrm{M}_{\scriptstyle \odot})=[9.5,\:10,\:10.5,\:11,\:12]$. The dashed lines in each panel show the profiles for tidal feature host galaxies and non-tidal feature host galaxies in the lowest stellar mass bin. They highlight that with increasing stellar mass, the colour profiles become redder at all radii across all simulations. In general, we see a tendency for galaxies hosting tidal features to have systematically bluer outskirts ($r_\mathrm{maj}> R_e^\mathrm{maj}$) than galaxies without tidal features for $10<\log_{10}(M_{\scriptstyle\star\mathrm{,\:30\:pkpc}}/\mathrm{M}_{\scriptstyle \odot})<11$. The colour profiles in Magneticum have significantly less scatter than those measured for EAGLE and TNG for $\log_{10}(M_{\scriptstyle\star\mathrm{,\:30\:pkpc}}/\mathrm{M}_{\scriptstyle \odot})\gtrsim10$. In the bottom panels, the sample of non-tidal feature galaxies drops significantly. Therefore, we expect the differences from the above panels are likely due to the small samples.}
    \label{fig:col_profile_rs}
\end{figure}

\begin{figure}
    \centering
    \includegraphics[width=\linewidth]{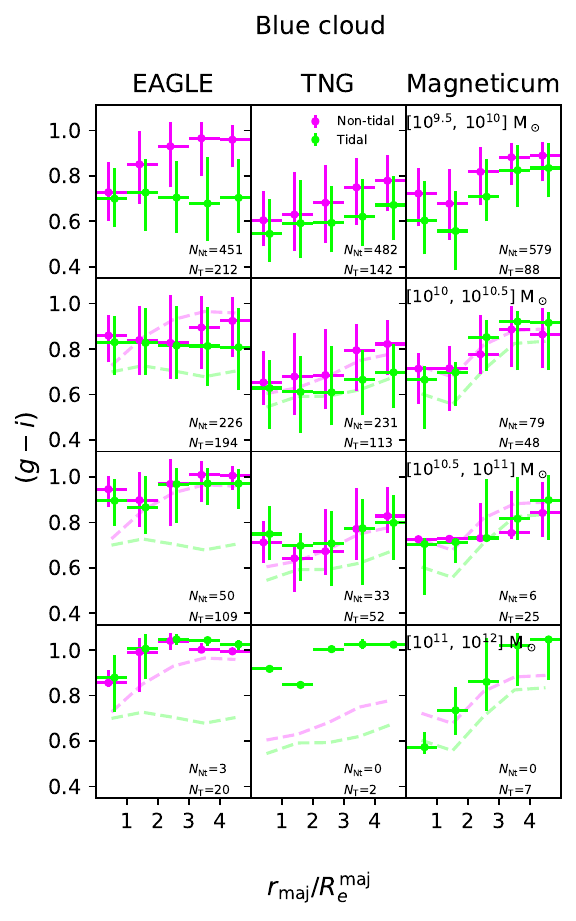}
    \caption{The distributions of the radial colour profiles for blue cloud galaxies, given in terms of $(g-i)$ and the semi-major axis ($r_\mathrm{maj}/R_{e}^\mathrm{maj}$). The binning and display of the distributions of colour profiles is identical to Fig.~\ref{fig:col_profile_rs}.}
    \label{fig:col_profile_bc}
\end{figure}

\begin{figure}
    \centering
    \includegraphics[width=\linewidth]{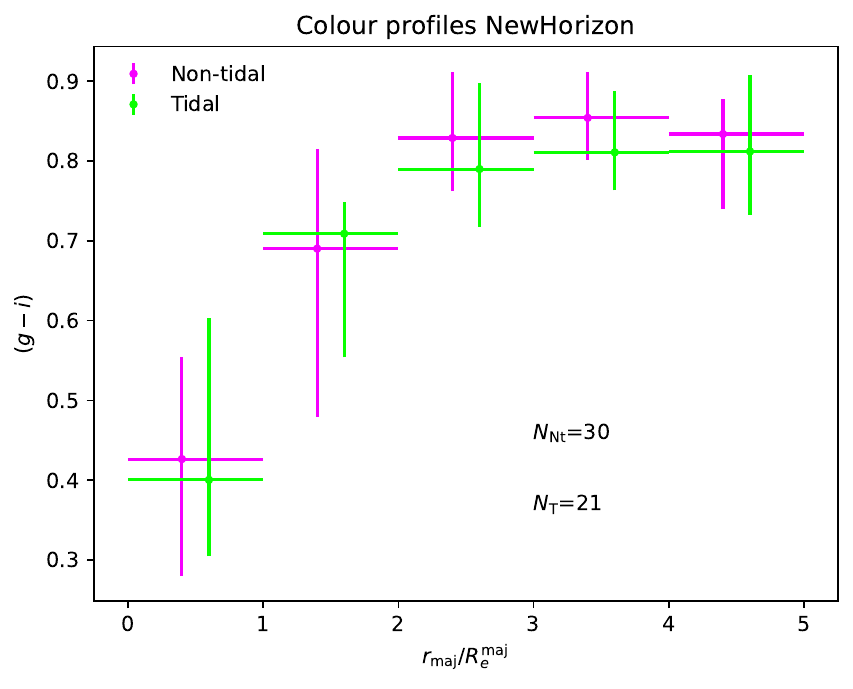}
    \caption{The distribution of the radial colour profiles for NewHorizon galaxies, given in terms of $(g-i)_{R_e^\mathrm{maj}}$ and the semi-major axis ($r_\mathrm{maj} [R_{e^\mathrm{maj}}]$). The binning and display of the distributions of colour profiles is identical to Fig.~\ref{fig:col_profile_rs}. The NewHorizon profiles only show a shift to bluer colours beyond 2 $R_e^\mathrm{maj}$. Binning by stellar mass is not applied here, due to the small sample of NewHorizon galaxies.}
    \label{fig:col_profile_nh}
\end{figure}

In this Section, we investigate whether tidal features imprint valuable information on galaxies' radial colour profiles. We measure the colour profiles using elliptical apertures as described in Section \ref{subsubsec:measure_colour}. In Fig.~\ref{fig:col_mag_delta_col}, we show that galaxies on the red sequence tend to have a negative gradient between their inner $R_e^\mathrm{maj}$ aperture and their $3\to4\:R_e^\mathrm{maj}$. There is a greater scatter in the colour gradients of the blue cloud population of galaxies, with the bluer galaxies tending to have a positive colour gradient. Due to the systematic differences between the red sequence and blue cloud galaxies, we consider the colour profiles of these populations separately.

\subsubsection{Red sequence galaxies}

In Fig.~\ref{fig:col_profile_rs}, we compare the colour profiles of galaxies hosting tidal features and the control sample of galaxies not hosting tidal features within the red sequence. Across all simulations, for both red sequence galaxies hosting tidal features and those without tidal features, we find that the radial profiles have a negative gradient. We note that Magneticum has the tightest distribution in its colour profiles.

For galaxies with $10^{10}<M_{\star\mathrm{,\:30\:pkpc}}/\mathrm{M}_\odot<10^{11}$, we see a tendency for colour profiles to be offset to bluer colours in tidal feature host galaxies compared to non-tidal feature galaxies in EAGLE and TNG, particularly beyond 1 $R_e^\mathrm{maj}$. Bootstrap resampling (without replacement) 100 random profiles from both the tidal feature host galaxies and non-tidal feature galaxies, we find the mean difference within $1\:R_e^\mathrm{maj}$ is $\Delta(g-i)=-0.0145\pm0.0002$, $-0.0070\pm0.0002$ and $0.0009\pm0.0003$ for EAGLE, TNG and Magneticum, respectively. Outside $1\:R_e^\mathrm{maj}$ the mean difference is $\Delta(g-i)=-0.0408\pm0.0005$, $-0.0215\pm0.0007$ and $0.0000\pm0.0004$ for EAGLE, TNG and Magneticum respectively. This illustrates that the colour profiles of tidal feature host galaxies in EAGLE and TNG are offset to slightly bluer colours within $1\:R_e^\mathrm{maj}$ compared to non-tidal feature host galaxies, and the magnitude of this deviation increases when considering the region outside $1\:R_e^\mathrm{maj}$. In Magneticum, while it is not the case over the entire mass range, we do see a tendency for tidal feature hosts to be offset to bluer colours beyond $1\:R_e^\mathrm{maj}$ for $10\leq\log_{10}(M_{\star\mathrm{,\:30\:pkpc}}/\mathrm{M}_\odot)\leq10.5$. As this is the bin with the most galaxies, it suggests that in Magneticum, the differences in the colour profiles from accreted material are so small that we may need a larger sample to detect them in the other stellar mass bins.

For EAGLE, the colour differences between the tidal feature and non-tidal feature hosting galaxies continue in the lowest stellar mass bin, whereas for TNG and Magneticum, we see a reversal of this trend with tidal feature hosts tending to be offset to redder colours. We note that the reversal of this trend coincides with a regime where red galaxies in TNG and Magneticum have systematically smaller sizes than red galaxies in EAGLE (Fig.~\ref{fig:size_stellar_mass}). This could result in radial trends being more poorly resolved in these simulations at these stellar masses ($M_{\star\mathrm{,\:30\:pkpc}}\leq10^{10}\:\mathrm{M}_{\scriptstyle \odot}$). In the highest mass bin ($M_{\star\mathrm{,\:30\:pkpc}}\geq10^{11}\:\mathrm{M}_{\scriptstyle \odot}$), the tendency of tidal feature host galaxies to be offset to bluer colours than non-tidal feature host galaxies at all radii continues for TNG and EAGLE. However, the results for the highest stellar mass bin are not statistically significant, as we do not have many non-tidal feature-hosting galaxies, 1 for EAGLE and 5 for TNG and Magneticum. We explored a variety of binning approaches to confirm that the results we present here for the colour profiles are robust to our choice of stellar mass and colour bins.

The dashed line in each of the stellar mass bins for $M_{\star\mathrm{,\:30\:pkpc}}>10^{10}\:\mathrm{M}_\odot$, shows the median colour profiles from the lowest stellar mass bin for comparison. In each of the stellar mass bins for $M_{\scriptstyle\star\mathrm{,\:30\:pkpc}}>10^{10}\:\mathrm{M}_{\scriptstyle\odot}$ we note that the profiles are generally redder at all radii than the profiles in the lowest stellar mass bin. This suggests that mass evolution impacts the red sequence profiles. However, it makes the galaxies redder at all radii instead of bluer. Therefore, the bluer profiles of tidal feature galaxies could not result from any stellar mass-related effect.

\subsubsection{Blue cloud galaxies}

Our results comparing the colour profiles between tidal feature hosting galaxies and non-tidal feature galaxies for the blue cloud galaxies are shown in Fig.~\ref{fig:col_profile_bc}. Compared to red sequence galaxies, the scatter in the colour profiles of the blue cloud galaxies is much larger. 

For $9.5<\log_{10}(M_{\star\mathrm{,\:30\:pkpc}}/\mathrm{M}_{\scriptstyle \odot})<10$, we see that tidal feature hosts tend to be offset to bluer colours than non-tidal feature hosts at all radii, across all simulations. EAGLE and TNG predict that the colour profiles for tidal feature host galaxies tend to be flatter than the colour profiles for non-tidal feature host galaxies for $\log_{10}(M_{\star\mathrm{,\:30\:pkpc}}/$M$_{\odot})\lesssim10.5$. In EAGLE for $10<\log_{10}(M_{\star\mathrm{,\:30\:pkpc}}/$M$_{\odot})<10.5$, the systematic offset to bluer colours for tidal feature hosts' colour profiles is only evident beyond $3\:R_e^\mathrm{maj}$, whereas, in TNG it is visible at all radii. In Magneticum, we observe the inverse of this effect, where the offset to bluer colours reduces in the outer radii with increasing stellar mass and only remains in the inner $1\:R_e^\mathrm{maj}$ for $\log_{10}(M_{\star\mathrm{,\:30\:pkpc}/\mathrm{M}_\odot}/\mathrm{M}_\odot)\gtrsim10$. For $10.5\lesssim\log_{10}(M_{\star\mathrm{,\:30\:pkpc}}/\mathrm{M}_\odot)\lesssim11$, only EAGLE continues to show any trend, with tidal feature host galaxy colour profiles, being systematically shifted to slightly bluer colours for $r_\mathrm{maj}<2\:R_e^\mathrm{maj}$ and $r_\mathrm{maj}>3\:R_e^\mathrm{maj}$. In the highest stellar mass bin, TNG and Magneticum do not have any blue cloud galaxies without tidal features, while EAGLE only has three and shows no systematic differences between the colour profiles of tidal feature and non-tidal feature galaxies. Given the subtle nature of the differences between the non-tidal feature and tidal feature host galaxies, we further test the statistical significance of these results in Section \ref{subsubsec:ks_color_profiles}.

Fig.~\ref{fig:col_profile_nh} shows our results comparing the colour profiles of tidal feature host galaxies and non-tidal feature host galaxies for the NewHorizon simulation. The NewHorizon sample does not include a substantial proportion of red galaxies, so we compare them to our consideration of the blue cloud galaxies. Due to the smaller sample, we have opted not to bin by stellar mass. We find that colour differences in the profiles are only evident beyond $2\:R_e^\mathrm{maj}$, where the profiles of tidal feature host galaxies tend to be subtly bluer than those of non-tidal feature host galaxies.

We also note that, given the large scatter in the colour profiles, there is no resolvable stellar mass evolution of these colour profiles. Therefore, stellar mass bias is unlikely to significantly impact our findings that tidal feature galaxies in the blue cloud tend to have bluer outskirts than non-tidal feature galaxies. 

\subsubsection{Statistical significance of colour profile results}
\label{subsubsec:ks_color_profiles}

\begin{figure}
    \centering
    \includegraphics[width=\linewidth]{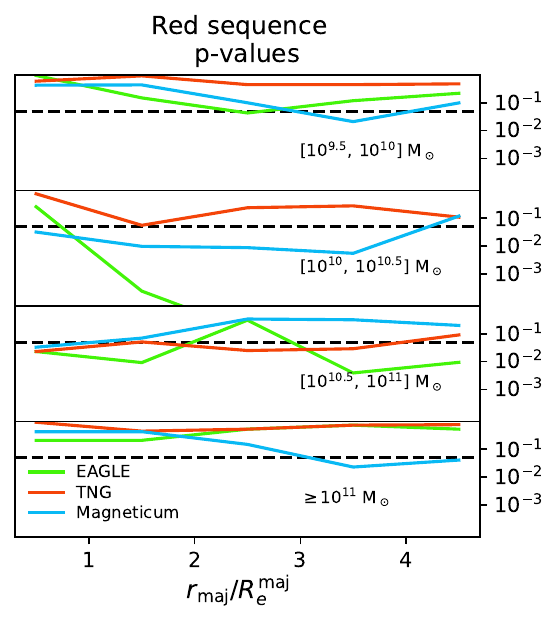}
    \caption{The p-values from the KS test for the colour distributions of galaxies in the red sequence with and without tidal features, in each of the radial and stellar mass bins in Fig.~\ref{fig:col_profile_rs}. The results for EAGLE are given in green, TNG in red and Magneticum in blue. The dashed line shows $p=0.05$ (reject null hypothesis with $2\sigma$ confidence). The $10^{10}<M_{\star\mathrm{,\:30\:pkpc}}/\mathrm{M}_\odot<10^{10.5}$ and $10^{10.5}<M_{\star\mathrm{,\:30\:pkpc}}/\mathrm{M}_\odot<10^{11}$ mass bins have the most points with $p<0.05$. For EAGLE and Magneticum, the most significant deviations occur at $r_\mathrm{maj}>R_e^\mathrm{maj}$. The large variance in the p-values and their generally high values suggests that no statistically significant systematic trends can be discerned with either radius or stellar mass.}
    \label{fig:ks_col_profile_rs}
\end{figure}

\begin{figure}
    \centering
    \includegraphics[width=\linewidth]{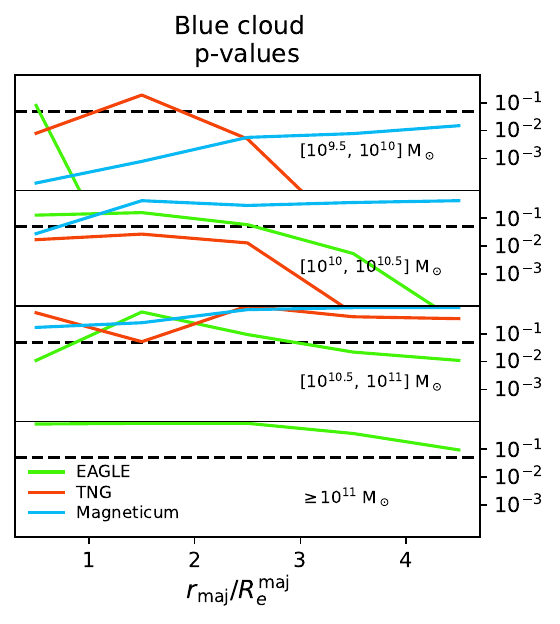}
    \caption{The p-values from the KS test for the colour distributions of galaxies in the red sequence with and without tidal features, in each of the radial and stellar mass bins in Fig.~\ref{fig:col_profile_bc}. The results are presented identically to Fig.~\ref{fig:ks_col_profile_rs}. Most of our statistically significant ($p<0.5$) results are in the $10^{9.5}<M_{\star\mathrm{,\:30\:pkpc}}/\mathrm{M}_\odot<10^{10}$ and $10^{10}<M_{\star\mathrm{,\:30\:pkpc}}/\mathrm{M}_\odot<10^{10.5}$ bins. The smallest p-values also tend to occur at larger radii, indicating that the significance of the shift of tidal feature host galaxies to bluer colours tends to be most significant at larger radii. Similarly to the results for the red sequence, the results for the blue cloud show a large variance in the p-values, which suggests that no statistically significant systematic trends with either radius or stellar mass can be discerned.}
    \label{fig:ks_color_profile_bc}
\end{figure}

Given the subtle nature of our results in Fig.~\ref{fig:col_profile_rs} and Fig.~\ref{fig:col_profile_bc}, we test the statistical significance of the shifts to bluer colours we are seeing in the tidal feature profiles using the KS test. The colour profiles for red sequence tidal feature host galaxies for $10^{10}<M_{\star\mathrm{,\:30\:pkpc}}/\mathrm{M}_\odot<10^{11}$ in Fig.~\ref{fig:col_profile_rs} and for blue cloud galaxies $10^{9.5}<M_{\star\mathrm{,\:30\:pkpc}}/\mathrm{M}_\odot<10^{10.5}$ in Fig.~\ref{fig:col_profile_bc}, tended to show deviations to bluer colours when compared to non-tidal feature host galaxies in the same stellar mass bin. Fig.~\ref{fig:ks_col_profile_rs} shows the KS test results comparing the colour profiles for tidal feature and non-tidal feature red sequence galaxies and Fig.~\ref{fig:ks_color_profile_bc} shows the same information for the blue cloud galaxies. The panels in both figures show the $p$-values for each radial bin for a given stellar mass range.

We find some bins with statistical significance ($p<0.05$) and negative test statistics for our simulations, particularly for $10^{10}<M_{\star\mathrm{,\:30\:pkpc}}/\mathrm{M}_\odot<10^{11}$ for the red sequence galaxies and $10^{9.5}<M_{\star\mathrm{,\:30\:pkpc}}/\mathrm{M}_\odot<10^{10.5}$ for the blue cloud galaxies. However, there remains significant variance in $p$-values across both radius and stellar mass bins, highlighting that no statistically significant systematic trends can be discerned from this sample.

\begin{table}
    \centering
    \caption{Summary of the KS test results comparing the colour profiles for galaxies with and without tidal features for both the red sequence and the blue cloud. We focus on the statistical significance achieved in the galaxy outskirts ($>R_e^\mathrm{maj}$). From left to right, the columns are stellar mass bin, total number of galaxies in the bin ($N_\mathrm{gal}=N_\mathrm{Nt}+N_\mathrm{T}$) $R_e^\mathrm{maj}$, and how many of the 4 radial bins beyond $R_e^\mathrm{maj}$ show significant $(p<0.05)$ results.}
    \begin{tabular}{c|c|c}
    \hline
    Stellar mass bin & $N_\mathrm{gal}$ & $N_\mathrm{sig}$ $(>R_e^\mathrm{maj})$\\
    $[\mathrm{M}_\odot]$ & & \\
    \hline
    \multicolumn{3}{c}{Red sequence}\\
    \hline
    \multicolumn{3}{c}{EAGLE}\\
    \hline
      $[10^{9.5},\:10^{10}]$ & 157 & 1\\
      $[10^{10},\:10^{10.5}]$ & 152 & 4\\
      $[10^{10.5},\:10^{11}]$ & 106 & 3\\
      $\geq10^{11}$ & 19 & 0\\
      \hline
      \multicolumn{3}{c}{TNG}\\
      \hline
      $[10^{9.5},\:10^{10}]$ & 78 & 0\\
      $[10^{10},\:10^{10.5}]$ & 88 & 0\\
      $[10^{10.5},\:10^{11}]$ & 167 & 2\\
      $\geq10^{11}$ & 43 & 0\\
      \hline
      \multicolumn{3}{c}{Magneticum}\\
      \hline
      $[10^{9.5},\:10^{10}]$ & 255 & 1\\
      $[10^{10},\:10^{10.5}]$ & 344 & 3\\
      $[10^{10.5},\:10^{11}]$ & 152 & 0\\
      $\geq10^{11}$ & 29 & 2\\
      \hline
      \multicolumn{3}{c}{Blue cloud}\\
      \hline
      \multicolumn{3}{c}{EAGLE}\\
      \hline
      $[10^{9.5},\:10^{10}]$ & 663 & 4\\
      $[10^{10},\:10^{10.5}]$ & 420 & 2\\
      $[10^{10.5},\:10^{11}]$ & 159 & 2\\
      $\geq10^{11}$ & 23 & 0\\
      \hline
      \multicolumn{3}{c}{TNG}\\
      \hline
      $[10^{9.5},\:10^{10}]$ & 624 & 3\\
      $[10^{10},\:10^{10.5}]$ & 344 & 4\\
      $[10^{10.5},\:10^{11}]$ & 85 & 0\\
      $\geq10^{11}$ & 2 &  \\
      \hline
      \multicolumn{3}{c}{Magneticum}\\
      \hline
      $[10^{9.5},\:10^{10}]$ & 667 & 4\\
      $[10^{10},\:10^{10.5}]$ & 127 & 0\\
      $[10^{10.5},\:10^{11}]$ & 31 & 0\\
      $\geq10^{11}$ & 7 & \\
      \hline
    \end{tabular}
    \label{tab:ks_col_profile}
\end{table}

Table~\ref{tab:ks_col_profile} summarizes the statistical significance achieved in the galaxy outskirts ($r_\mathrm{maj}>R_e^\mathrm{maj}$). We note that we tend to achieve statistical significance when there are $\gtrsim150$ galaxies within a stellar mass bin. This suggests that having samples of at least this size in each bin will help verify if the differences we see are statistically significant. The statistical significance of the results is also stronger when both the number of tidal feature host galaxies and the number of non-tidal feature host galaxies were sampled well, which is not the case for the lowest and highest stellar mass bins for the red sequence and the highest stellar mass bin for the blue cloud.

\section{Discussion}
\label{sec:discussion}

In this work, we have examined the relationships between tidal feature occurrence, galaxy colour and galaxy colour profiles in mock images of simulations. By studying galaxy colours and colour profiles across galaxies hosting tidal features, and not hosting tidal features, we investigate whether tidal features and their preceding interactions leave any informative imprints on the outskirts of the galaxy when viewed by optical imaging. In this Section, we place this work in context. In Section \ref{subsec:col_col_prof_discussion}, we review our results regarding the relationships between galaxy colours, colour profiles and tidal feature occurrence and see how they compare to observations and simulation results from the literature. In Section \ref{subsec:gas_frac_ssfr}, we investigate how the specific star formation rates of our galaxies could contextualize our trends with tidal feature fraction and colour. In Section \ref{subsec:mergers_subgrid}, we investigate the interplay between baryonic physics models in each simulation and the mergers that could lead to the results we present.

\subsection{The relationship between tidal features, colours and colour profiles}
\label{subsec:col_col_prof_discussion}


We investigate the galaxy-colour-stellar mass relation in Fig.~\ref{fig:col_mag_total}. The top panel compares our simulated data to observations from the SDSS survey \citep[e.g.][]{yorkSloanDigitalSky2000,stoughtonSloanDigitalSky2002,chenEvolutionMostMassive2012}. The top panel shows that all three simulations produced a red sequence and a blue cloud of galaxies. We noted quantitative differences in the distributions of both the red sequence and the blue cloud when compared to SDSS. All the simulations exhibit a blue cloud that extends to bluer colours than in SDSS and a red sequence that has a flatter gradient than in SDSS.  In EAGLE and TNG, blue cloud galaxies are shifted to higher stellar masses than in SDSS. In TNG red sequence galaxies of stellar masses $\leq10^{10.5}\:\mathrm{M}_{\scriptstyle \odot}$ are under-represented. 

To check the reliability of our measurements and further contextualize our observationally motivated predictions, we compare our results with previous analyses of observational and simulation data. The excess of high stellar mass blue cloud galaxies in EAGLE is consistent with the results of \citet{trayfordColoursLuminosities012015}, who also found the blue cloud of galaxies extended to higher stellar masses at $z=0.1$ in the EAGLE simulation we work with than in observations from the Galaxy and Mass Assembly survey \citep[GAMA;][]{driverGalaxyMassAssembly2011}. \citet{nelsonFirstResultsIllustrisTNG2018} which studied the evolution of galaxy colours in the colour-stellar mass diagram in the \textsc{TNG100-1} simulation. They found that their transition from a blue cloud-dominated population to a red sequence-dominated population occurs at too high a stellar mass compared to SDSS observations, which is consistent with the deficit of low stellar mass red galaxies in our TNG sample. \citet{nelsonFirstResultsIllustrisTNG2018} also found that the slope of the TNG red sequence is flatter than seen in observations, consistent with our Fig.~\ref{fig:col_mag_total}.

Our size-stellar mass relation (Fig.~\ref{fig:size_stellar_mass}) shows that TNG galaxies tend to have smaller sizes, particularly below $M_{\star\mathrm{,\:30\:pkpc}}\lesssim10^{10.6}\:\mathrm{M}_{\scriptstyle \odot}$, compared to other simulations. \citet{rodriguez-gomezOpticalMorphologiesGalaxies2019} studied the optical morphologies and colours of galaxies  with $M_\star>10^{9.5}\:\mathrm{M}_{\scriptstyle \odot}$ in \textsc{TNG100-1} using Pan-STARRS-like \citep[e.g.][]{chambersPanSTARRS1Surveys2019} mock observations. They also found that median sizes in their dust-free mock images had sizes that were smaller than the median for Pan-STARRS for $M_{\star\mathrm{,\:30\:pkpc}}\leq10^{10.6}\:\mathrm{M}_{\scriptstyle \odot}$, this is consistent with our findings. 

The bottom panels of Fig.~\ref{fig:col_mag_total}, and Fig.~\ref{fig:col_mag_stf} highlight that the likelihood of a galaxy exhibiting tidal features is not only a function of galaxy stellar mass \citep[e.g.][]{bilekCensusClassificationLowsurfacebrightness2020,martinPreparingLowSurface2022,desmonsGalaxyMassAssembly2023,khalidCharacterizingTidalFeatures2024,valenzuelaStreamComeTrue2024} but is also a function of galaxy colour at any given stellar mass in each of our simulations. All simulations tend to show that the intrinsically bluest galaxies, $(g-i)_{R_e^\mathrm{maj}}<0.5$, exhibit enhanced tidal feature fractions over redder galaxies at all stellar masses. While the trends are the same, we find qualitative and quantitative differences between the simulations regarding how the tidal feature fraction behaves as a function of colour. This is not surprising given the differences in how the subgrid physics processes for star formation and the heating of the gas through stellar and AGN feedback have been implemented in each, summarised in Section \ref{subsec:simulations}. This is reflected in the colour-stellar mass distributions, with the positions of the blue cloud and red sequence differing from simulation to simulation (Fig.~\ref{fig:col_mag_total}, top panel).

Observationally, \citet{atkinsonFAINTTIDALFEATURES2013} using the Canada France-Hawaii-Telescope (CFHT) Legacy Survey \citep{bouladeMegacamNewCanadaFranceHawaii2003}, found that tidal features are twice as prevalent around red galaxies than they are around blue galaxies for $0.04<z<0.2$ and $M_{\star}\geq10^{9.4}$ $M_\odot$ to a limiting surface brightness of 27.7 mag/arcsec$^{2}$ in the \textit{g} band. However, this is likely a consequence of the higher stellar masses of their red galaxies when compared to blue galaxies. \citet{kado-fongTidalFeatures0052018} did not find any significant relationship between the colours of galaxies and the occurrence of tidal features in their sample of 20,000 galaxies across a redshift range of $0.05<z<0.45$ in the HSC-SSP Wide survey.

There is, however, significant evidence for a correlation with galaxy colour and morphology and the prevalence of tidal features in samples of massive nearby galaxies, where the sample of tidal features is likely to be more complete. \citet{miskolcziTidalStreamsGalaxies2011} found their sample of nearby, edge-on late-type galaxies from SDSS DR7 \citep{yorkSloanDigitalSky2000,abazajianSEVENTHDATARELEASE2009} to have $f_\mathrm{Tidal}=0.19$, which is higher than $f_\mathrm{Tidal}=0.12$ found in the sample of 650 early-type galaxies in SDSS Stripe 82 coadded images \citep[$\sim27$ mag/arcsec$^2$; ][]{jiangSLOANDIGITALSKY2014} of \citet{yoonFrequencyTidalFeatures2020}, however, the edge-on inclination does increase the likelihood of detecting tidal features \citep{johnstonInterpretingDebrisSatellite2001}. \citet{yoonFrequencyTidalFeatures2020} also found that younger, more compact early-type galaxies and early-type galaxies with dust lanes were more likely to exhibit tidal features. More recently, in deeper CFHT imaging \footnote{The limiting surface brightness is $\mu_r\sim28.9$ mag/arcsec$^2$ for MATLAS \citep{ducATLAS3DProjectXXIX2015,ducMATLASDeepExploration2020,bilekCensusClassificationLowsurfacebrightness2020} and $28.3$ mag/arcsec$^2$ for the CFIS survey.}, \citet{solaCharacterizationLowSurface2022,solaLowSurfaceBrightness2025a} find differences in the occurrence rates and shapes of tidal tails with host galaxy morphology, tails being more common and thinner around late-type galaxies.

There are suggestions of trends of tidal feature fraction and colour in observations that could hint at the contribution of recent mergers to the formation of bluer galaxies. However, larger sample sizes and a broader analysis of tidal features around galaxies of all morphologies are required to better understand and test the predictions made in our work. While there are well-motivated reasons as to why we might expect bluer galaxies to have more tidal features in general, e.g. starbursts and AGN activity triggered by merging \citep[e.g.][]{hernquistTidalTriggeringStarbursts1989}, we still require a detailed study of tidal features across a statistically robust sample of galaxies covering all morphologies. Upcoming observations from LSST will provide the required depth and large sample sizes needed to extensively investigate trends between tidal feature fraction and galaxy colour.

It is valuable to discuss how our predictions might change if the LSST surface brightness limits are lower than expected, for example, due to the difficulty with sky subtraction for these depths \citep[e.g.][]{watkinsStrategiesOptimalSky2024}. The impact of surface brightness limits on tidal feature fractions was examined in detail in \citet{martinPreparingLowSurface2022} and in Appendix B of \citet{khalidCharacterizingTidalFeatures2024}. \citet{martinPreparingLowSurface2022} use the NewHorizon simulation to show that if the surface brightness limit is reduced from $30$ mag/arcsec$^2$ to $29$ mag/arcsec$^2$, the fraction of galaxies exhibiting tidal features at a given stellar mass will drop by $\sim0.1$ for $M_\star\lesssim10^{10.25}\:\mathrm{M}_\odot$ and $\sim0.08$ for $M_\star\sim10^{11}\:\mathrm{M}_\odot$ (see Fig.~14 in \citealt{martinPreparingLowSurface2022}). NewHorizon has a stellar mass resolution, a factor of $\sim100$ more than the EAGLE, TNG and Magneticum simulations used in this work. Appendix B of \citet{khalidCharacterizingTidalFeatures2024} shows how stellar mass resolution impacts the detectability of tidal features. In the toy model used, at 30.3 mag/arcsec$^2$, NewHorizon can resolve tidal features that are $10^6\:\mathrm{M}_\odot$, whereas the other simulations can only resolve tidal features from disrupted masses of $10^8\:\mathrm{M}_\odot$. A reduction in the surface brightness limit of $\sim1$ mag/arcsec$^2$, would result in a $\sim75$ per cent reduction in the area over which a tidal feature could be spread before it falls below the detection limit in NewHorizon, whereas, in EAGLE, TNG, and Magneticum, the reduction is on the order of $\sim28$ per cent for a tidally disrupted mass of $10^8\:\mathrm{M}_\odot$. Therefore, the predictions from \citet{martinPreparingLowSurface2022} provide an upper limit on the reduction in tidal feature fraction as a fraction of surface brightness for EAGLE, TNG and Magneticum.

It is also the case that light from bright parts of the PSF can interfere with the light from similarly bright parts of the galaxy. This could influence the detectability of tidal features. This effect becomes more significant as the size of the PSF increases with respect to the size of the galaxy. \citet{martinPreparingLowSurface2022} found that the impact of the PSF only begins to interfere with the visibility of tidal features significantly at $z=0.4$, far greater than the redshift of our galaxy of $z=0.025$. Therefore, the impact of the PSF on our results is negligible, and any differences between the model used here and the final LSST PSF are unlikely to significantly impact the predictions here.


\subsubsection{Colour profiles of red sequence galaxies}

Fig.~\ref{fig:col_profile_rs} and \ref{fig:col_profile_bc} show that the colour profiles we measured differed systematically between the red sequence and the blue cloud galaxies. We find that red sequence galaxies tend to have colour profiles with negative gradients and tight distributions, whereas blue cloud galaxies generally have positive gradients with a larger scatter. 

There are hints that there may be differences in the colour profiles of tidal feature host galaxies in the red sequence for $10^{10}<M_{\star\mathrm{,\:30\:pkpc}}/\mathrm{M}_\odot<10^{11}$, with tidal feature galaxies having colour profiles offset to slightly bluer colours, particularly in the outskirts, when compared to non-tidal feature galaxies. However, KS test results in Fig.~\ref{fig:ks_col_profile_rs} do not find any statistically significant trends within our sample. Larger samples will help in confirming if the differences in the colour profiles of tidal feature and non-tidal feature host galaxies are real. We nevertheless explore here the potential explanations for the offsets to bluer colours, particularly in the outskirts of these red sequence galaxies.

We compare our results to previous observational and simulation studies to check for consistency and contextualise our observationally motivated predictions. \citet{kado-fongTidalFeatures0052018} found that tidal features are generally bluer than their host galaxies for the early-type galaxies in their sample. Similarly, \citet{yoonImpactGalaxyMergers2023} studying the age and colour profiles of early-type galaxies within 1.5 effective radii in the Mapping Nearby Galaxies at Apache Point Observatory \citep[MANGA;][]{bundyOVERVIEWSDSSIVMaNGA2014} Integral Field Unit observations, found galaxies with tidal features in the corresponding deep SDSS Stripe 82 images \citep{fliriIACStripe822016} had lower metallicities and younger ages than galaxies without tidal features at all radii, for $M_\star\leq10^{10.6}\:\mathrm{M}_{\scriptstyle \odot}$ and, lower metallicities and younger ages within $1\:R_e$ and no differences beyond this radius for $M_\star\geq10^{10.6}\:\mathrm{M}_{\scriptstyle \odot}$. These observational results suggest that offsets to bluer colours in galaxy profiles could be caused by the tidal features themselves and merger-related processes that lead to lower metallicities and younger ages within one effective radius. The apparent offsets in our results could reflect qualitatively similar offsets in observational results.

We expect the colour profiles of our red sequence galaxies to on average reflect the two-phase model of initial in-situ star formation at $z\gtrsim2$ followed by accretion through predominantly minor/mini mergers since $z\lesssim2$ \citep[e.g.][]{naabMINORMERGERSSIZE2009,oserTWOPHASESGALAXY2010,martinPreparingLowSurface2022,remusAccretedNotAccreted2022}. Such mergers are expected to deposit more of their mass in the outskirts of galaxies \citep[e.g.][]{karademirOuterStellarHalos2019}, therefore, we expect there to be a radius in most of our red sequence galaxies beyond which accreted material dominates \citep[e.g.][]{pillepichFirstResultsIllustristng2018,davisonEAGLEsViewEx2020,remusAccretedNotAccreted2022}. Red sequence galaxies tend to be quenched and have older stellar populations \citep[e.g.][]{nelsonFirstResultsIllustrisTNG2018}, therefore, the colour gradients of these galaxies correlate more with metallicity than age as seen in observations \citep[e.g.][]{broughSpatiallyResolvedKinematics2007,tortoraColourStellarPopulation2010,santucciSAMIGalaxySurvey2020,yoonImpactGalaxyMergers2023} and simulations \citep[e.g.][Stoiber et al. in prep.]{tortoraStellarPopulationGradients2011}. From the metallicity-stellar mass relation \citep[e.g.][]{gallazziAgesMetallicitiesGalaxies2005,dolagDistributionEvolutionMetals2017,bellstedtGalaxyMassAssembly2021}, we expect that mini and minor mergers would deposit lower-metallicity stellar material that is bluer than the more massive host galaxy. Therefore, the offset to bluer colours in the galaxy outskirts exhibited by tidal feature hosts in Fig.~\ref{fig:col_profile_rs} could be driven by these galaxies having more accreted material from recent minor/mini mergers.

\subsubsection{Colour profiles of blue cloud galaxies}

Fig.~\ref{fig:col_profile_bc} shows the colour profiles for blue cloud galaxies. These show a large scatter in colour, suggesting a greater contribution from subgrid physics due to star formation, likely making stellar age a more significant contributor to the colour profiles \citep[e.g.][]{tortoraColourStellarPopulation2010,tortoraStellarPopulationGradients2011}. We find some evidence that the tidal feature host galaxies have outskirts offset to bluer colours than non-tidal feature galaxies in the blue cloud particularly for $10^{9.5}<M_{\star\mathrm{,\:30\:pkpc}}/\mathrm{M}_\odot<10^{10.5}$, however, the KS test results in Fig.~\ref{fig:ks_color_profile_bc} show that these differences are not statistically significant for our sample. We will likely need larger samples to confirm if there are any systematic differences in the colour profiles of tidal feature host galaxies and non-tidal feature host galaxies. We briefly compare with observational and simulation works and discuss the physics that could potentially cause an offset to bluer colours in the colour profiles.

The tidal feature host galaxy colour profiles being offset to bluer colours within $1$ to $2\:R_e^\mathrm{maj}$ could suggest merger-induced star formation at the centre of the galaxy \citep[e.g.][]{torreyMETALLICITYEVOLUTIONINTERACTING2012}. The qualitative differences in colour profiles across simulations are not unexpected, \citet{munMAGPISurveyRadial2024} found that while EAGLE, TNG and Magneticum agreed regarding radial star formation trends, they found varying degrees of suppression of star formation in the inner $1.5$ effective radii, which could be attributed to the different AGN feedback prescriptions.

Observationally \citet{gonzalez-perezColourGradientsSDSS2011} studied the colour gradients of SDSS galaxies $0.01<z<0.17$ \citep[][]{abazajianSEVENTHDATARELEASE2009} out to 2 half-light radii. They found the galaxies to have slightly redder cores than their outskirts, which is in agreement with our red sequence galaxy sample (Fig.~\ref{fig:col_profile_rs}) but in contrast with our results for the blue cloud (Fig.~\ref{fig:col_profile_bc}). Across all simulations, our measurements within $2\:R_e^\mathrm{maj}$ are on average flat or positive in non-interacting galaxies and flat or slightly negative for all but the lowest stellar mass bin for interacting galaxies. \citet{gonzalez-perezColourGradientsSDSS2011} also found that galaxies in interacting pairs were skewed towards having bluer cores, which is reflected in our results in Fig.~\ref{fig:col_profile_bc}. \citet{gonzalez-perezColourGradientsSDSS2011} found greater variance in the gradients of their late-type sample, which is consistent with the larger variance in gradients we see in the blue cloud across all simulations (Fig.~\ref{fig:col_mag_delta_col}). Differences between our results and \citet{gonzalez-perezColourGradientsSDSS2011} could be due to difficulty in detecting contribution from low-stellar mass merging galaxies, due to the brighter SDSS surface brightness limit and \citet{gonzalez-perezColourGradientsSDSS2011} measuring out to 2 half-light radii, whereas we measure out to $5\:R_e^\mathrm{maj}$.

\begin{figure}
    \centering
    \includegraphics[width=\linewidth]{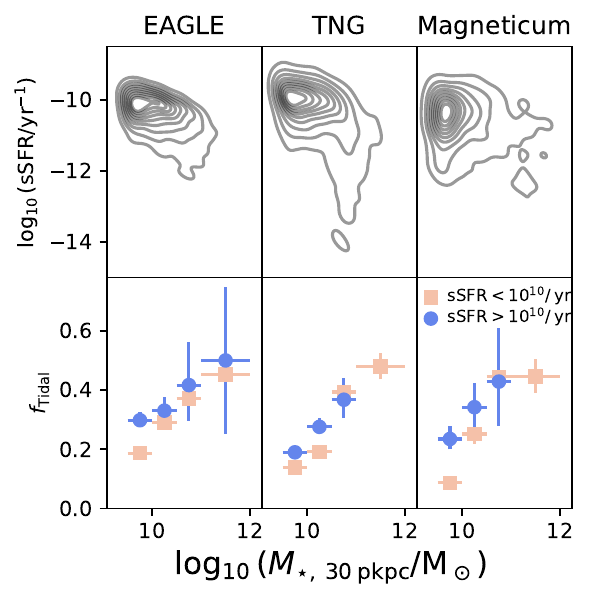}
    \caption{(Top panel) Specific star formation rate as a function of galaxy stellar mass for EAGLE, TNG and Magneticum. The contours are 10 per cent apart and range from 10 per cent to 90 per cent. (Bottom panel) Tidal feature fraction as a function of stellar mass and specific star formation rate. The galaxies are binned by the log of the specific star formation rate, with bin edges [-14, 10, -8.5], the blue points showing the fraction for galaxies with $\log_{10}(\mathrm{sSFR/yr^{-1}})$ the x-error bars show the sizes of the stellar mass bins, with bin edges $\log_{10}(M_{\star\mathrm{,\:30\:pkpc}}/\mathrm{M}_\odot)=$[9.5, 10, 10.5, 11, 12]. The y-error bars show the $1\sigma$ uncertainties on the tidal feature fractions. In the lowest two stellar mass bins, the galaxies with higher specific star formation rates have higher tidal feature fractions for both simulations. This is reflected in Fig.~\ref{fig:col_mag_total}, where the bluest galaxies in the same stellar mass bins have a higher tidal feature fraction than the redder galaxies.}
    \label{fig:tidal_feature_ssfr}
\end{figure}

\begin{figure*}
    \centering
    \includegraphics[width=\linewidth]{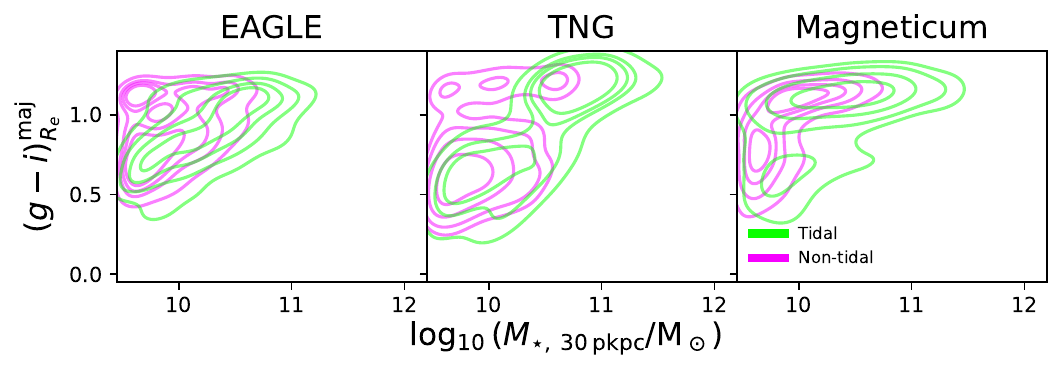}
    \caption{Comparing the distributions in intrinsic $(g-i)_{R_e^\mathrm{maj}}$ colour - spherical aperture stellar mass for galaxies hosting tidal features and galaxies without tidal features across our simulations. The distributions for tidal feature host galaxies are shown in lime green, and the distributions for non-tidal feature galaxies are shown in pink. The contours show the 20 per cent, 40 per cent, 60 per cent and 80 per cent confidence levels for the Gaussian kernel density estimate of the distributions. The under-representation of red sequence tidal feature host galaxies at the low stellar mass end in EAGLE and TNG could suggest that the merger-induced star formation and resultant stellar feedback is not sufficient to cause quenching. This could be due to starbursts not being resolved in cosmological simulations of this resolution.}
    \label{fig:col_mstar_tf_vs_ntf}
\end{figure*}

\subsection{How specific star formation rates relate to tidal features}
\label{subsec:gas_frac_ssfr}

The colours of galaxies reflect their stellar populations, therefore, we investigate the star formation rate for our sample of galaxies. Galaxies with bluer stellar populations are more likely to have higher star formation rates and could reflect star formation induced in a recent interaction \citep[e.g.][]{dimatteoStarFormationEfficiency2007,ohImpactGalaxyMergers2019}. In Fig.~\ref{fig:tidal_feature_ssfr}, we investigate the relationship between the specific star formation rate, and the tidal feature fraction for EAGLE, TNG and Magneticum. For TNG and Magneticum we have the instantaneous star formation rates measured by summing the star formation rates of all gas cells/particles in each galaxy, whereas for EAGLE as this measurement was not available for the snapshot used, we have estimated the star formation rate of each galaxy based on the star particles formed within the last 100 Myr. While the shapes of the star-forming main sequence and the quenched population of galaxies differ between the simulations, we do find that enhanced specific star formation rates correspond to enhanced tidal feature fractions in all three simulations. This is likely why we see enhanced tidal feature fractions in the bluest colour bin in Fig.~\ref{fig:col_mag_total}, as galaxies with higher specific star formation rates tend to be bluer.

\citet{ferreiraGalaxyEvolutionPostMerger2025} use deep learning-based predictions of time-since last merging event to measure star formation enhancement in galaxy mergers in SDSS. They find that star formation enhancements continue for $\sim1$ Gyr post coalescence. In a follow-up work using the same deep learning method, studying post-merger quenching in the Ultraviolet Near Infrared Optical Northern Survey (UNIONS), \citet{ellisonGalaxyEvolutionPostmerger2024a} find that the merging galaxy population is dominated by star formation up to $0.16$ Gyr post-merger and is dominated by post-starburst quenched galaxies beyond this time. Similarly, \citet{gordonLinkingEnhancedStar2025} also found enhanced star formation in galaxies with identified tidal features in CFHT and Dark Energy Camera Legacy Survey \citep[DECaLS;][]{deyOverviewDESILegacy2019} images. As tidal features are thought to last up to $\sim3$ Gyr post-merger \citep{jiLifetimeMergerFeatures2014,mancillasProbingMergerHistory2019,yoonFrequencyTidalFeatures2020,huangMassiveEarlyTypeGalaxies2022}, these results could indicate that our detected tidal features are tracing very recent mergers, or alternatively, that the AGN feedback, star formation and stellar feedback triggered by mergers \citep[e.g.][]{ellisonGalaxyMergersCan2022} in simulations are not quenching the galaxy populations on short enough timescales.

\subsection{Mergers and baryonic physics}
\label{subsec:mergers_subgrid}

The differences in both colours and colour profiles across our simulations highlight that the subgrid physics models are playing a role in our results. To see how much the mergers can play a role in these differences, we investigate how the distributions of galaxies hosting tidal features differ from those not hosting tidal features in the colour-stellar mass diagram in Fig.~\ref{fig:col_mstar_tf_vs_ntf}. In EAGLE and TNG, galaxies hosting tidal features are under-represented at the low-mass end of the red sequence, while still populating the blue cloud at these masses. In Magneticum, galaxies with tidal features are present on the red sequence and the blue cloud across all stellar masses. This reflects the higher tidal feature fractions for EAGLE and TNG galaxies around $(g-i)_{R_e^\mathrm{maj}}<0.5$ when compared to to $(g-i)_{R_e^\mathrm{maj}}>1$ for $M_{\star\mathrm{,\:30\:pkpc}}\lesssim10^{10.5}\mathrm{\:M}_\odot$, whereas, in Magneticum these differences are reduced and disappear for $M_{\star\mathrm{,\:30\:pkpc}}\lesssim10^{10}\mathrm{\:M}_\odot$ (Fig.~\ref{fig:col_mag_total}). The EAGLE and TNG results are consistent with the picture that merger-induced star formation and subsequent stellar feedback are insufficient to quench galaxies and place them on the red sequence \citep[e.g.][]{sparreZoomingMajorMergers2016}. In Magneticum, where the tidal feature fractions between red and blue galaxies are more similar, it could be that differences in their star formation models, such as allowing for a gas particle to form 4 generations of stars, could allow for a different distribution of feedback that still allows for quenching. It is also possible that environmental quenching due to processes such as galactic strangulation \citep{larsonEvolutionDiskGalaxies1980}, ram pressure stripping \citep[][]{gunnInfallMatterClusters1972,abadiRamPressureStripping1999}, tidal stripping \citep[][]{dekelGalacticHaloCuspcore2003,diemandFormationEvolutionGalaxy2007} or harassment \citep[][]{faroukiComputerSimulationsEnvironmental1981,mooreMorphologicalTransformationGalaxy1998} are playing a much stronger role in Magneticum rather than merger-induced processes. In EAGLE, TNG and Magneticum, tidal feature host galaxies have bluer colours (mean $(g-i)_{R_e^\mathrm{maj}}=0.72\pm0.01,\:0.58\pm0.01,\:0.62\pm0.02$, respectively) than non-tidal feature galaxies (mean $(g-i)_{R_e^\mathrm{maj}}=0.75\pm0.01,\:0.628\pm0.006,\:0.695\pm0.007$, respectively). However, in NewHorizon, where the initial gas mass resolution is $\sim10\times$ higher than the other simulations, we find no differences in the colours of tidal feature (mean $(g-i)_{R_e^\mathrm{maj}}=0.45\pm0.05$) and non-tidal feature (mean $(g-i)_{R_e^\mathrm{maj}}=0.45\pm0.04$) host galaxies, emphasising again the potential impacts of resolution.

In TNG it is known that galaxy quenching occurs primarily through AGN feedback when a black hole achieves a mass over $\gtrsim10^{8}\:\mathrm{M}_\odot$ \citep[][]{weinbergerSimulatingGalaxyFormation2017,nelsonFirstResultsIllustrisTNG2018} and mergers are not playing a significant role in quenching in TNG \citep[]{weinbergerSupermassiveBlackHoles2018}. This is consistent with our tidal feature population in TNG, for $M_{\star\mathrm{,\:30\:pkpc}}\lesssim10^{10}\:\mathrm{M}_{\scriptstyle \odot}$ being mostly blue cloud. Similar effects may be occurring for lower stellar masses in EAGLE, where stellar feedback from star formation triggered during merging and gas accretion may not be significant enough to form a quenched population of galaxies on the red sequence. The different modelling for SNe feedback for gas between TNG \citep[][]{vogelsbergerModelCosmologicalSimulations2013,pillepichSimulatingGalaxyFormation2018} and EAGLE \citep{dallavecchiaSimulatingGalacticOutflows2012} could be responsible for the differences in the stellar masses at which we start to see a population of red sequence galaxies hosting tidal features.

\section{Summary and Conclusions}
\label{sec:conclusion}

In this work, we use the catalogue of LSST-like mock observations from \citet{khalidCharacterizingTidalFeatures2024} for the NewHorizon, EAGLE, TNG and Magneticum cosmological-hydrodynamical simulations to investigate whether tidal features leave detectable imprints on the intrinsic $(g-i)_{R_e^\mathrm{maj}}$ colours and colour profiles of galaxies that LSST will observe. By comparing samples of galaxies hosting tidal features to those without, we investigate systematic differences in colours and colour profiles between the populations and find the following.

\begin{itemize}
    \item Bluer galaxies ($(g-i)_{R_e^\mathrm{maj}}<0.5$) tend to have higher tidal feature fractions than redder galaxies ($(g-i)_{R_e^\mathrm{maj}}>1$) across all galaxy stellar masses. This difference is seen most clearly in the galaxies hosting asymmetric halos, but is also visible in galaxies hosting streams/tails and double nuclei.
    \item The tidal feature fraction is enhanced in galaxies with high specific star formation rates. As the enhancement in specific star formation rate could be a result of the occurring/recent merger, it is likely that the correlation of enhanced specific star formation rate with an enhanced tidal feature fraction is at least partly responsible for the trend we observe with colour.
    \item We do not find statistically significant systematic differences between the colour profiles of tidal feature and non-tidal feature galaxy colour profiles. We do see some evidence that tidal feature galaxies have colour profiles offset to bluer colours compared to their non-tidal feature counterparts, particularly at larger radii for $10^{10}<M_{\star\mathrm{,\:30\:pkpc}}/\mathrm{M}_\odot<10^{11}$ red sequence galaxies and $10^{9.5}<M_{\star\mathrm{,\:30\:pkpc}}/\mathrm{M}_\odot<10^{10.5}$ blue cloud galaxies. Larger samples of $\gtrsim150$ galaxies per stellar mass bin will help confirm if these differences are real.
    \item Our results indicate that studying the colours of galaxies with and without tidal features provides a useful probe of recent accretion events in red sequence and blue cloud galaxies. 
    \item The differences in the colours of tidal feature host galaxies across simulations highlight the need for LSST observations to further constrain models of galaxy evolution.
    \item Our observationally-motivated approach allows direct comparison to future LSST observations.
\end{itemize}

\section*{Acknowledgements}
We thank the reviewer for their helpful comments and suggestions. We thank Jesse van de Sande for helpful discussions regarding the use of \texttt{MGE}. We thank Elizabeth Iles for a helpful discussion regarding the estimation of star formation rates in simulations. We thank Annette Ferguson for helpful feedback. We thank Pierre-Alain Duc for helpful feedback. AK acknowledges the support of the Astronomical Society of Australia Student Travel Grant. SB acknowledges funding support from the Australian Research Council through a Discovery Project DP190101943. This research includes computations using the computational cluster Katana supported by Research Technology Services at UNSW Sydney. This work was performed on the OzSTAR national facility at Swinburne University of Technology. The OzSTAR program receives funding in part from the Astronomy National Collaborative Research Infrastructure Strategy (NCRIS) allocation provided by the Australian Government, and from the Victorian Higher Education State Investment Fund (VHESIF) provided by the Victorian Government. The \textsc{NewHorizon} simulation was undertaken with HPC resources of CINES under the allocations  c2016047637, A0020407637 and A0070402192 by Genci, KSC-2017-G2-0003 by KISTI, and as a “Grand Challenge” project granted by GENCI on the AMD Rome extension of the Joliot Curie supercomputer at TGCC. A large data transfer was supported by KREONET which is managed and operated by KISTI. We acknowledge the Virgo Consortium for making their simulation data available. The EAGLE simulations were performed using the DiRAC-2 facility at Durham, managed by the ICC, and the PRACE facility Curie based in France at TGCC, CEA, Bruyères-le-Châtel. TNG100 was run on the HazelHen Cray XC40 system at the High Performance Computing Center Stuttgart as part of project GCS-ILLU of the Gauss Centre for Supercomputing (GCS). Ancillary and test runs of the IllustrisTNG project were also run on the Stampede supercomputer at TACC/XSEDE (allocation AST140063), at the Hydra and Draco supercomputers at the Max Planck Computing and Data Facility, and on the MIT/Harvard computing facilities supported by FAS and MIT MKI. The {\it Magneticum} simulations were performed at the Leibniz-Rechenzentrum with CPU time assigned to the Project {\it pr83li}. This work was supported by the Deutsche Forschungsgemeinschaft (DFG, German Research Foundation) under Germany's Excellence Strategy - EXC-2094 - 390783311. Parts of this research were supported by the Australian Research Council Centre of Excellence for All Sky Astrophysics in 3 Dimensions (ASTRO 3D), through project number CE170100013.

Software: \textsc{Astropy} \citep{astropycollaborationAstropyCommunityPython2013,astropycollaborationAstropyProjectBuilding2018,astropycollaborationAstropyProjectSustaining2022}, \textsc{Numpy} \citep{harrisArrayProgrammingNumPy2020a}, \textsc{Pandas} \citep{mckinney-proc-scipy-2010,thepandasdevelopmentteamPandasdevPandasPandas2023}, \textsc{Photutils} \citep{larry_bradley_2023_8248020}, \textsc{Scipy} \citep{virtanenSciPy10Fundamental2020}.
\section*{Data Availability}

\textsc{NewHorizon} data may be requested from \url{https://new.horizon-simulation.org/data.html}. \textsc{eagle} data used in this work are publicly available at \url{http://icc.dur.ac.uk/Eagle/}. The \textsc{IllustrisTNG} data used in this work are publicly available at \url{http://www.tng-project.org}. \textsc{Magneticum Pathfinder} data are partially available at \url{https://c2papcosmosim.uc.lrz.de/} \citep{ragagninWebPortalHydrodynamical2017}, with
larger data sets on request.
 



\bibliographystyle{mnras}
\bibliography{references} 

@inproceedings{2001cghr.confE..64H,
  title = {Modelling the {{UV}}/x-Ray Cosmic Background with {{CUBA}}},
  booktitle = {Clusters of Galaxies and the High Redshift Universe Observed in X-Rays},
  author = {Haardt, F. and Madau, P.},
  editor = {Neumann, D. M. and Tran, J. T. V.},
  year = {2001},
  month = jan,
  number = {64},
  eprint = {astro-ph/0106018},
  pages = {64},
  doi = {10.48550/arXiv.astro-ph/0106018},
  archiveprefix = {arXiv}
}

@article{abadiRamPressureStripping1999,
  title = {Ram Pressure Stripping of Spiral Galaxies in Clusters},
  author = {Abadi, Mario G. and Moore, Ben and Bower, Richard G.},
  year = {1999},
  month = oct,
  journal = {MNRAS},
  volume = {308},
  number = {4},
  pages = {947--954},
  issn = {0035-8711},
  doi = {10.1046/j.1365-8711.1999.02715.x},
  urldate = {2025-01-20}
}

@article{abazajianSEVENTHDATARELEASE2009,
  title = {{{THE SEVENTH DATA RELEASE OF THE SLOAN DIGITAL SKY SURVEY}}},
  author = {Abazajian, Kevork N. and {Adelman-McCarthy}, Jennifer K. and Ag{\"u}eros, Marcel A. and Allam, Sahar S. and Prieto, Carlos Allende and An, Deokkeun and Anderson, Kurt S. J. and Anderson, Scott F. and Annis, James and Bahcall, Neta A. and {Bailer-Jones}, C. A. L. and Barentine, J. C. and Bassett, Bruce A. and Becker, Andrew C. and Beers, Timothy C. and Bell, Eric F. and Belokurov, Vasily and Berlind, Andreas A. and Berman, Eileen F. and Bernardi, Mariangela and Bickerton, Steven J. and Bizyaev, Dmitry and Blakeslee, John P. and Blanton, Michael R. and Bochanski, John J. and Boroski, William N. and Brewington, Howard J. and Brinchmann, Jarle and Brinkmann, J. and Brunner, Robert J. and Budav{\'a}ri, Tam{\'a}s and Carey, Larry N. and Carliles, Samuel and Carr, Michael A. and Castander, Francisco J. and Cinabro, David and Connolly, A. J. and Csabai, Istv{\'a}n and Cunha, Carlos E. and Czarapata, Paul C. and Davenport, James R. A. and de Haas, Ernst and Dilday, Ben and Doi, Mamoru and Eisenstein, Daniel J. and Evans, Michael L. and Evans, N. W. and Fan, Xiaohui and Friedman, Scott D. and Frieman, Joshua A. and Fukugita, Masataka and G{\"a}nsicke, Boris T. and Gates, Evalyn and Gillespie, Bruce and Gilmore, G. and Gonzalez, Belinda and Gonzalez, Carlos F. and Grebel, Eva K. and Gunn, James E. and Gy{\"o}ry, Zsuzsanna and Hall, Patrick B. and Harding, Paul and Harris, Frederick H. and Harvanek, Michael and Hawley, Suzanne L. and Hayes, Jeffrey J. E. and Heckman, Timothy M. and Hendry, John S. and Hennessy, Gregory S. and Hindsley, Robert B. and Hoblitt, J. and Hogan, Craig J. and Hogg, David W. and Holtzman, Jon A. and Hyde, Joseph B. and Ichikawa, Shin-ichi and Ichikawa, Takashi and Im, Myungshin and Ivezi{\'c}, {\v Z}eljko and Jester, Sebastian and Jiang, Linhua and Johnson, Jennifer A. and Jorgensen, Anders M. and Juri{\'c}, Mario and Kent, Stephen M. and Kessler, R. and Kleinman, S. J. and Knapp, G. R. and Konishi, Kohki and Kron, Richard G. and Krzesinski, Jurek and Kuropatkin, Nikolay and Lampeitl, Hubert and Lebedeva, Svetlana and Lee, Myung Gyoon and Lee, Young Sun and Leger, R. French and L{\'e}pine, S{\'e}bastien and Li, Nolan and Lima, Marcos and Lin, Huan and Long, Daniel C. and Loomis, Craig P. and Loveday, Jon and Lupton, Robert H. and Magnier, Eugene and Malanushenko, Olena and Malanushenko, Viktor and Mandelbaum, Rachel and Margon, Bruce and Marriner, John P. and {Mart{\'i}nez-Delgado}, David and Matsubara, Takahiko and McGehee, Peregrine M. and McKay, Timothy A. and Meiksin, Avery and Morrison, Heather L. and Mullally, Fergal and Munn, Jeffrey A. and Murphy, Tara and Nash, Thomas and Nebot, Ada and Neilsen, Eric H. and Newberg, Heidi Jo and Newman, Peter R. and Nichol, Robert C. and Nicinski, Tom and {Nieto-Santisteban}, Maria and Nitta, Atsuko and Okamura, Sadanori and Oravetz, Daniel J. and Ostriker, Jeremiah P. and Owen, Russell and Padmanabhan, Nikhil and Pan, Kaike and Park, Changbom and Pauls, George and Peoples, John and Percival, Will J. and Pier, Jeffrey R. and Pope, Adrian C. and Pourbaix, Dimitri and Price, Paul A. and Purger, Norbert and Quinn, Thomas and Raddick, M. Jordan and Fiorentin, Paola Re and Richards, Gordon T. and Richmond, Michael W. and Riess, Adam G. and Rix, Hans-Walter and Rockosi, Constance M. and Sako, Masao and Schlegel, David J. and Schneider, Donald P. and Scholz, Ralf-Dieter and Schreiber, Matthias R. and Schwope, Axel D. and Seljak, Uro{\v s} and Sesar, Branimir and Sheldon, Erin and Shimasaku, Kazu and Sibley, Valena C. and Simmons, A. E. and Sivarani, Thirupathi and Smith, J. Allyn and Smith, Martin C. and Smol{\v c}i{\'c}, Vernesa and Snedden, Stephanie A. and Stebbins, Albert and Steinmetz, Matthias and Stoughton, Chris and Strauss, Michael A. and SubbaRao, Mark and Suto, Yasushi and Szalay, Alexander S. and Szapudi, Istv{\'a}n and Szkody, Paula and Tanaka, Masayuki and Tegmark, Max and Teodoro, Luis F. A. and Thakar, Aniruddha R. and Tremonti, Christy A. and Tucker, Douglas L. and Uomoto, Alan and Berk, Daniel E. Vanden and Vandenberg, Jan and Vidrih, S. and Vogeley, Michael S. and Voges, Wolfgang and Vogt, Nicole P. and Wadadekar, Yogesh and Watters, Shannon and Weinberg, David H. and West, Andrew A. and White, Simon D. M. and Wilhite, Brian C. and Wonders, Alainna C. and Yanny, Brian and Yocum, D. R. and York, Donald G. and Zehavi, Idit and Zibetti, Stefano and Zucker, Daniel B.},
  year = {2009},
  month = may,
  journal = {ApJS},
  volume = {182},
  number = {2},
  pages = {543},
  publisher = {The American Astronomical Society},
  issn = {0067-0049},
  doi = {10.1088/0067-0049/182/2/543},
  urldate = {2024-11-26},
  langid = {english}
}

@article{adePlanck2013Results2014,
  title = {Planck 2013 Results. {{XVI}}. {{Cosmological}} Parameters},
  author = {Ade, P. a. R. and Aghanim, N. and {Armitage-Caplan}, C. and Arnaud, M. and Ashdown, M. and {Atrio-Barandela}, F. and Aumont, J. and Baccigalupi, C. and Banday, A. J. and Barreiro, R. B. and Bartlett, J. G. and Battaner, E. and Benabed, K. and Beno{\^i}t, A. and {Benoit-L{\'e}vy}, A. and Bernard, J.-P. and Bersanelli, M. and Bielewicz, P. and Bobin, J. and Bock, J. J. and Bonaldi, A. and Bond, J. R. and Borrill, J. and Bouchet, F. R. and Bridges, M. and Bucher, M. and Burigana, C. and Butler, R. C. and Calabrese, E. and Cappellini, B. and Cardoso, J.-F. and Catalano, A. and Challinor, A. and Chamballu, A. and Chary, R.-R. and Chen, X. and Chiang, H. C. and Chiang, L.-Y. and Christensen, P. R. and Church, S. and Clements, D. L. and Colombi, S. and Colombo, L. P. L. and Couchot, F. and Coulais, A. and Crill, B. P. and Curto, A. and Cuttaia, F. and Danese, L. and Davies, R. D. and Davis, R. J. and de Bernardis, P. and de Rosa, A. and de Zotti, G. and Delabrouille, J. and Delouis, J.-M. and D{\'e}sert, F.-X. and Dickinson, C. and Diego, J. M. and Dolag, K. and Dole, H. and Donzelli, S. and Dor{\'e}, O. and Douspis, M. and Dunkley, J. and Dupac, X. and Efstathiou, G. and Elsner, F. and En{\ss}lin, T. A. and Eriksen, H. K. and Finelli, F. and Forni, O. and Frailis, M. and Fraisse, A. A. and Franceschi, E. and Gaier, T. C. and Galeotta, S. and Galli, S. and Ganga, K. and Giard, M. and Giardino, G. and {Giraud-H{\'e}raud}, Y. and Gjerl{\o}w, E. and {Gonz{\'a}lez-Nuevo}, J. and G{\'o}rski, K. M. and Gratton, S. and Gregorio, A. and Gruppuso, A. and Gudmundsson, J. E. and Haissinski, J. and Hamann, J. and Hansen, F. K. and Hanson, D. and Harrison, D. and {Henrot-Versill{\'e}}, S. and {Hern{\'a}ndez-Monteagudo}, C. and Herranz, D. and Hildebrandt, S. R. and Hivon, E. and Hobson, M. and Holmes, W. A. and Hornstrup, A. and Hou, Z. and Hovest, W. and Huffenberger, K. M. and Jaffe, A. H. and Jaffe, T. R. and Jewell, J. and Jones, W. C. and Juvela, M. and Keih{\"a}nen, E. and Keskitalo, R. and Kisner, T. S. and Kneissl, R. and Knoche, J. and Knox, L. and Kunz, M. and {Kurki-Suonio}, H. and Lagache, G. and L{\"a}hteenm{\"a}ki, A. and Lamarre, J.-M. and Lasenby, A. and Lattanzi, M. and Laureijs, R. J. and Lawrence, C. R. and Leach, S. and Leahy, J. P. and Leonardi, R. and {Le{\'o}n-Tavares}, J. and Lesgourgues, J. and Lewis, A. and Liguori, M. and Lilje, P. B. and {Linden-V{\o}rnle}, M. and {L{\'o}pez-Caniego}, M. and Lubin, P. M. and {Mac{\'i}as-P{\'e}rez}, J. F. and Maffei, B. and Maino, D. and Mandolesi, N. and Maris, M. and Marshall, D. J. and Martin, P. G. and {Mart{\'i}nez-Gonz{\'a}lez}, E. and Masi, S. and Massardi, M. and Matarrese, S. and Matthai, F. and Mazzotta, P. and Meinhold, P. R. and Melchiorri, A. and Melin, J.-B. and Mendes, L. and Menegoni, E. and Mennella, A. and Migliaccio, M. and Millea, M. and Mitra, S. and {Miville-Desch{\^e}nes}, M.-A. and Moneti, A. and Montier, L. and Morgante, G. and Mortlock, D. and Moss, A. and Munshi, D. and Murphy, J. A. and Naselsky, P. and Nati, F. and Natoli, P. and Netterfield, C. B. and {N{\o}rgaard-Nielsen}, H. U. and Noviello, F. and Novikov, D. and Novikov, I. and O'Dwyer, I. J. and Osborne, S. and Oxborrow, C. A. and Paci, F. and Pagano, L. and Pajot, F. and Paladini, R. and Paoletti, D. and Partridge, B. and Pasian, F. and Patanchon, G. and Pearson, D. and Pearson, T. J. and Peiris, H. V. and Perdereau, O. and Perotto, L. and Perrotta, F. and Pettorino, V. and Piacentini, F. and Piat, M. and Pierpaoli, E. and Pietrobon, D. and Plaszczynski, S. and Platania, P. and Pointecouteau, E. and Polenta, G. and Ponthieu, N. and Popa, L. and Poutanen, T. and Pratt, G. W. and Pr{\'e}zeau, G. and Prunet, S. and Puget, J.-L. and Rachen, J. P. and Reach, W. T. and Rebolo, R. and Reinecke, M. and Remazeilles, M. and Renault, C. and Ricciardi, S. and Riller, T. and Ristorcelli, I. and Rocha, G. and Rosset, C. and Roudier, G. and {Rowan-Robinson}, M. and {Rubi{\~n}o-Mart{\'i}n}, J. A. and Rusholme, B. and Sandri, M. and Santos, D. and Savelainen, M. and Savini, G. and Scott, D. and Seiffert, M. D. and Shellard, E. P. S. and Spencer, L. D. and Starck, J.-L. and Stolyarov, V. and Stompor, R. and Sudiwala, R. and Sunyaev, R. and Sureau, F. and Sutton, D. and {Suur-Uski}, A.-S. and Sygnet, J.-F. and Tauber, J. A. and Tavagnacco, D. and Terenzi, L. and Toffolatti, L. and Tomasi, M. and Tristram, M. and Tucci, M. and Tuovinen, J. and T{\"u}rler, M. and Umana, G. and Valenziano, L. and Valiviita, J. and Tent, B. Van and Vielva, P. and Villa, F. and Vittorio, N. and Wade, L. A. and Wandelt, B. D. and Wehus, I. K. and White, M. and White, S. D. M. and Wilkinson, A. and Yvon, D. and Zacchei, A. and Zonca, A.},
  year = {2014},
  month = nov,
  journal = {A\&A},
  volume = {571},
  pages = {A16},
  publisher = {EDP Sciences},
  issn = {0004-6361, 1432-0746},
  doi = {10.1051/0004-6361/201321591},
  urldate = {2024-04-30},
  copyright = {{\copyright} ESO, 2014},
  langid = {english}
}

@article{adePlanck2015Results2016,
  title = {Planck 2015 Results - {{XIII}}. {{Cosmological}} Parameters},
  author = {Ade, P. a. R. and Aghanim, N. and Arnaud, M. and Ashdown, M. and Aumont, J. and Baccigalupi, C. and Banday, A. J. and Barreiro, R. B. and Bartlett, J. G. and Bartolo, N. and Battaner, E. and Battye, R. and Benabed, K. and Beno{\^i}t, A. and {Benoit-L{\'e}vy}, A. and Bernard, J.-P. and Bersanelli, M. and Bielewicz, P. and Bock, J. J. and Bonaldi, A. and Bonavera, L. and Bond, J. R. and Borrill, J. and Bouchet, F. R. and Boulanger, F. and Bucher, M. and Burigana, C. and Butler, R. C. and Calabrese, E. and Cardoso, J.-F. and Catalano, A. and Challinor, A. and Chamballu, A. and Chary, R.-R. and Chiang, H. C. and Chluba, J. and Christensen, P. R. and Church, S. and Clements, D. L. and Colombi, S. and Colombo, L. P. L. and Combet, C. and Coulais, A. and Crill, B. P. and Curto, A. and Cuttaia, F. and Danese, L. and Davies, R. D. and Davis, R. J. and de Bernardis, P. and de Rosa, A. and de Zotti, G. and Delabrouille, J. and D{\'e}sert, F.-X. and Valentino, E. Di and Dickinson, C. and Diego, J. M. and Dolag, K. and Dole, H. and Donzelli, S. and Dor{\'e}, O. and Douspis, M. and Ducout, A. and Dunkley, J. and Dupac, X. and Efstathiou, G. and Elsner, F. and En{\ss}lin, T. A. and Eriksen, H. K. and Farhang, M. and Fergusson, J. and Finelli, F. and Forni, O. and Frailis, M. and Fraisse, A. A. and Franceschi, E. and Frejsel, A. and Galeotta, S. and Galli, S. and Ganga, K. and Gauthier, C. and Gerbino, M. and Ghosh, T. and Giard, M. and {Giraud-H{\'e}raud}, Y. and Giusarma, E. and Gjerl{\o}w, E. and {Gonz{\'a}lez-Nuevo}, J. and G{\'o}rski, K. M. and Gratton, S. and Gregorio, A. and Gruppuso, A. and Gudmundsson, J. E. and Hamann, J. and Hansen, F. K. and Hanson, D. and Harrison, D. L. and Helou, G. and {Henrot-Versill{\'e}}, S. and {Hern{\'a}ndez-Monteagudo}, C. and Herranz, D. and Hildebrandt, S. R. and Hivon, E. and Hobson, M. and Holmes, W. A. and Hornstrup, A. and Hovest, W. and Huang, Z. and Huffenberger, K. M. and Hurier, G. and Jaffe, A. H. and Jaffe, T. R. and Jones, W. C. and Juvela, M. and Keih{\"a}nen, E. and Keskitalo, R. and Kisner, T. S. and Kneissl, R. and Knoche, J. and Knox, L. and Kunz, M. and {Kurki-Suonio}, H. and Lagache, G. and L{\"a}hteenm{\"a}ki, A. and Lamarre, J.-M. and Lasenby, A. and Lattanzi, M. and Lawrence, C. R. and Leahy, J. P. and Leonardi, R. and Lesgourgues, J. and Levrier, F. and Lewis, A. and Liguori, M. and Lilje, P. B. and {Linden-V{\o}rnle}, M. and {L{\'o}pez-Caniego}, M. and Lubin, P. M. and {Mac{\'i}as-P{\'e}rez}, J. F. and Maggio, G. and Maino, D. and Mandolesi, N. and Mangilli, A. and Marchini, A. and Maris, M. and Martin, P. G. and Martinelli, M. and {Mart{\'i}nez-Gonz{\'a}lez}, E. and Masi, S. and Matarrese, S. and McGehee, P. and Meinhold, P. R. and Melchiorri, A. and Melin, J.-B. and Mendes, L. and Mennella, A. and Migliaccio, M. and Millea, M. and Mitra, S. and {Miville-Desch{\^e}nes}, M.-A. and Moneti, A. and Montier, L. and Morgante, G. and Mortlock, D. and Moss, A. and Munshi, D. and Murphy, J. A. and Naselsky, P. and Nati, F. and Natoli, P. and Netterfield, C. B. and {N{\o}rgaard-Nielsen}, H. U. and Noviello, F. and Novikov, D. and Novikov, I. and Oxborrow, C. A. and Paci, F. and Pagano, L. and Pajot, F. and Paladini, R. and Paoletti, D. and Partridge, B. and Pasian, F. and Patanchon, G. and Pearson, T. J. and Perdereau, O. and Perotto, L. and Perrotta, F. and Pettorino, V. and Piacentini, F. and Piat, M. and Pierpaoli, E. and Pietrobon, D. and Plaszczynski, S. and Pointecouteau, E. and Polenta, G. and Popa, L. and Pratt, G. W. and Pr{\'e}zeau, G. and Prunet, S. and Puget, J.-L. and Rachen, J. P. and Reach, W. T. and Rebolo, R. and Reinecke, M. and Remazeilles, M. and Renault, C. and Renzi, A. and Ristorcelli, I. and Rocha, G. and Rosset, C. and Rossetti, M. and Roudier, G. and {d'Orfeuil}, B. Rouill{\'e} and {Rowan-Robinson}, M. and {Rubi{\~n}o-Mart{\'i}n}, J. A. and Rusholme, B. and Said, N. and Salvatelli, V. and Salvati, L. and Sandri, M. and Santos, D. and Savelainen, M. and Savini, G. and Scott, D. and Seiffert, M. D. and Serra, P. and Shellard, E. P. S. and Spencer, L. D. and Spinelli, M. and Stolyarov, V. and Stompor, R. and Sudiwala, R. and Sunyaev, R. and Sutton, D. and {Suur-Uski}, A.-S. and Sygnet, J.-F. and Tauber, J. A. and Terenzi, L. and Toffolatti, L. and Tomasi, M. and Tristram, M. and Trombetti, T. and Tucci, M. and Tuovinen, J. and T{\"u}rler, M. and Umana, G. and Valenziano, L. and Valiviita, J. and Tent, F. Van and Vielva, P. and Villa, F. and Wade, L. A. and Wandelt, B. D. and Wehus, I. K. and White, M. and White, S. D. M. and Wilkinson, A. and Yvon, D. and Zacchei, A. and Zonca, A.},
  year = {2016},
  month = oct,
  journal = {A\&A},
  volume = {594},
  pages = {A13},
  publisher = {EDP Sciences},
  issn = {0004-6361, 1432-0746},
  doi = {10.1051/0004-6361/201525830},
  urldate = {2024-04-29},
  copyright = {{\copyright} ESO, 2016},
  langid = {english}
}

@article{agertzFormationDiscGalaxies2011,
  title = {The Formation of Disc Galaxies in a {{$\Lambda$CDM}} Universe},
  author = {Agertz, Oscar and Teyssier, Romain and Moore, Ben},
  year = {2011},
  month = jan,
  journal = {MNRAS},
  volume = {410},
  pages = {1391--1408},
  issn = {0035-8711},
  doi = {10.1111/j.1365-2966.2010.17530.x},
  urldate = {2023-05-02}
}

@article{aiharaEIGHTHDATARELEASE2011,
  title = {{{THE EIGHTH DATA RELEASE OF THE SLOAN DIGITAL SKY SURVEY}}: {{FIRST DATA FROM SDSS-III}}},
  shorttitle = {{{THE EIGHTH DATA RELEASE OF THE SLOAN DIGITAL SKY SURVEY}}},
  author = {Aihara, Hiroaki and Prieto, Carlos Allende and An, Deokkeun and Anderson, Scott F. and Aubourg, {\'E}ric and Balbinot, Eduardo and Beers, Timothy C. and Berlind, Andreas A. and Bickerton, Steven J. and Bizyaev, Dmitry and Blanton, Michael R. and Bochanski, John J. and Bolton, Adam S. and Bovy, Jo and Brandt, W. N. and Brinkmann, J. and Brown, Peter J. and Brownstein, Joel R. and Busca, Nicolas G. and Campbell, Heather and Carr, Michael A. and Chen, Yanmei and Chiappini, Cristina and Comparat, Johan and Connolly, Natalia and Cortes, Marina and Croft, Rupert A. C. and Cuesta, Antonio J. and da Costa, Luiz N. and Davenport, James R. A. and Dawson, Kyle and Dhital, Saurav and Ealet, Anne and Ebelke, Garrett L. and Edmondson, Edward M. and Eisenstein, Daniel J. and Escoffier, Stephanie and Esposito, Massimiliano and Evans, Michael L. and Fan, Xiaohui and Castell{\'a}, Bruno Femen{\'i}a and {Font-Ribera}, Andreu and Frinchaboy, Peter M. and Ge, Jian and Gillespie, Bruce A. and Gilmore, G. and Hern{\'a}ndez, Jonay I. Gonz{\'a}lez and Gott, J. Richard and Gould, Andrew and Grebel, Eva K. and Gunn, James E. and Hamilton, Jean-Christophe and Harding, Paul and Harris, David W. and Hawley, Suzanne L. and Hearty, Frederick R. and Ho, Shirley and Hogg, David W. and Holtzman, Jon A. and Honscheid, Klaus and Inada, Naohisa and Ivans, Inese I. and Jiang, Linhua and Johnson, Jennifer A. and Jordan, Cathy and Jordan, Wendell P. and Kazin, Eyal A. and Kirkby, David and Klaene, Mark A. and Knapp, G. R. and Kneib, Jean-Paul and Kochanek, C. S. and Koesterke, Lars and Kollmeier, Juna A. and Kron, Richard G. and Lampeitl, Hubert and Lang, Dustin and Goff, Jean-Marc Le and Lee, Young Sun and Lin, Yen-Ting and Long, Daniel C. and Loomis, Craig P. and Lucatello, Sara and Lundgren, Britt and Lupton, Robert H. and Ma, Zhibo and MacDonald, Nicholas and Mahadevan, Suvrath and Maia, Marcio A. G. and Makler, Martin and Malanushenko, Elena and Malanushenko, Viktor and Mandelbaum, Rachel and Maraston, Claudia and Margala, Daniel and Masters, Karen L. and McBride, Cameron K. and McGehee, Peregrine M. and McGreer, Ian D. and M{\'e}nard, Brice and {Miralda-Escud{\'e}}, Jordi and Morrison, Heather L. and Mullally, F. and Muna, Demitri and Munn, Jeffrey A. and Murayama, Hitoshi and Myers, Adam D. and Naugle, Tracy and Neto, Angelo Fausti and Nguyen, Duy Cuong and Nichol, Robert C. and O'Connell, Robert W. and Ogando, Ricardo L. C. and Olmstead, Matthew D. and Oravetz, Daniel J. and Padmanabhan, Nikhil and {Palanque-Delabrouille}, Nathalie and Pan, Kaike and Pandey, Parul and P{\^a}ris, Isabelle and Percival, Will J. and Petitjean, Patrick and Pfaffenberger, Robert and Pforr, Janine and Phleps, Stefanie and Pichon, Christophe and Pieri, Matthew M. and Prada, Francisco and {Price-Whelan}, Adrian M. and Raddick, M. Jordan and Ramos, Beatriz H. F. and Reyl{\'e}, C{\'e}line and Rich, James and Richards, Gordon T. and Rix, Hans-Walter and Robin, Annie C. and {Rocha-Pinto}, Helio J. and Rockosi, Constance M. and Roe, Natalie A. and Rollinde, Emmanuel and Ross, Ashley J. and Ross, Nicholas P. and Rossetto, Bruno M. and S{\'a}nchez, Ariel G. and Sayres, Conor and Schlegel, David J. and Schlesinger, Katharine J. and Schmidt, Sarah J. and Schneider, Donald P. and Sheldon, Erin and Shu, Yiping and Simmerer, Jennifer and Simmons, Audrey E. and Sivarani, Thirupathi and Snedden, Stephanie A. and Sobeck, Jennifer S. and Steinmetz, Matthias and Strauss, Michael A. and Szalay, Alexander S. and Tanaka, Masayuki and Thakar, Aniruddha R. and Thomas, Daniel and Tinker, Jeremy L. and Tofflemire, Benjamin M. and Tojeiro, Rita and Tremonti, Christy A. and Vandenberg, Jan and Maga{\~n}a, M. Vargas and Verde, Licia and Vogt, Nicole P. and Wake, David A. and Wang, Ji and Weaver, Benjamin A. and Weinberg, David H. and White, Martin and White, Simon D. M. and Yanny, Brian and Yasuda, Naoki and Yeche, Christophe and Zehavi, Idit},
  year = {2011},
  month = mar,
  journal = {ApJS},
  volume = {193},
  number = {2},
  pages = {29},
  publisher = {The American Astronomical Society},
  issn = {0067-0049},
  doi = {10.1088/0067-0049/193/2/29},
  urldate = {2024-11-28},
  langid = {english}
}

@article{amoriscoContributionsAccretedStellar2017,
  title = {Contributions to the Accreted Stellar Halo: An Atlas of Stellar Deposition},
  shorttitle = {Contributions to the Accreted Stellar Halo},
  author = {Amorisco, N. C.},
  year = {2017},
  month = jan,
  journal = {MNRAS},
  volume = {464},
  number = {3},
  pages = {2882--2895},
  issn = {0035-8711},
  doi = {10.1093/mnras/stw2229},
  urldate = {2025-04-08}
}

@article{amoriscoFeathersBifurcationsShells2015,
  title = {On Feathers, Bifurcations and Shells: The Dynamics of Tidal Streams across the Mass Scale},
  shorttitle = {On Feathers, Bifurcations and Shells},
  author = {Amorisco, N. C.},
  year = {2015},
  month = jun,
  journal = {MNRAS},
  volume = {450},
  number = {1},
  pages = {575--591},
  issn = {0035-8711},
  doi = {10.1093/mnras/stv648},
  urldate = {2022-06-11}
}

@misc{arthAnisotropicThermalConduction2017,
  title = {Anisotropic Thermal Conduction in Galaxy Clusters with {{MHD}} in {{Gadget}}},
  author = {Arth, Alexander and Dolag, Klaus and Beck, Alexander M. and Petkova, Margarita and Lesch, Harald},
  year = {2017},
  month = sep,
  number = {arXiv:1412.6533},
  eprint = {1412.6533},
  primaryclass = {astro-ph},
  publisher = {arXiv},
  doi = {10.48550/arXiv.1412.6533},
  urldate = {2024-03-22},
  archiveprefix = {arXiv}
}

@article{astropycollaborationAstropyCommunityPython2013,
  title = {Astropy: {{A}} Community {{Python}} Package for Astronomy},
  shorttitle = {Astropy},
  author = {{Astropy Collaboration} and Robitaille, Thomas P. and Tollerud, Erik J. and Greenfield, Perry and Droettboom, Michael and Bray, Erik and Aldcroft, Tom and Davis, Matt and Ginsburg, Adam and {Price-Whelan}, Adrian M. and Kerzendorf, Wolfgang E. and Conley, Alexander and Crighton, Neil and Barbary, Kyle and Muna, Demitri and Ferguson, Henry and Grollier, Fr{\'e}d{\'e}ric and Parikh, Madhura M. and Nair, Prasanth H. and Unther, Hans M. and Deil, Christoph and Woillez, Julien and Conseil, Simon and Kramer, Roban and Turner, James E. H. and Singer, Leo and Fox, Ryan and Weaver, Benjamin A. and Zabalza, Victor and Edwards, Zachary I. and Azalee Bostroem, K. and Burke, D. J. and Casey, Andrew R. and Crawford, Steven M. and Dencheva, Nadia and Ely, Justin and Jenness, Tim and Labrie, Kathleen and Lim, Pey Lian and Pierfederici, Francesco and Pontzen, Andrew and Ptak, Andy and Refsdal, Brian and Servillat, Mathieu and Streicher, Ole},
  year = {2013},
  month = oct,
  journal = {A\&A},
  volume = {558},
  pages = {A33},
  issn = {0004-6361},
  doi = {10.1051/0004-6361/201322068},
  urldate = {2023-09-29}
}

@article{astropycollaborationAstropyProjectBuilding2018,
  title = {The {{Astropy Project}}: {{Building}} an {{Open-science Project}} and {{Status}} of the v2.0 {{Core Package}}},
  shorttitle = {The {{Astropy Project}}},
  author = {{Astropy Collaboration} and {Price-Whelan}, A. M. and Sip{\H o}cz, B. M. and G{\"u}nther, H. M. and Lim, P. L. and Crawford, S. M. and Conseil, S. and Shupe, D. L. and Craig, M. W. and Dencheva, N. and Ginsburg, A. and VanderPlas, J. T. and Bradley, L. D. and {P{\'e}rez-Su{\'a}rez}, D. and {de Val-Borro}, M. and Aldcroft, T. L. and Cruz, K. L. and Robitaille, T. P. and Tollerud, E. J. and Ardelean, C. and Babej, T. and Bach, Y. P. and Bachetti, M. and Bakanov, A. V. and Bamford, S. P. and Barentsen, G. and Barmby, P. and Baumbach, A. and Berry, K. L. and Biscani, F. and Boquien, M. and Bostroem, K. A. and Bouma, L. G. and Brammer, G. B. and Bray, E. M. and Breytenbach, H. and Buddelmeijer, H. and Burke, D. J. and Calderone, G. and Cano Rodr{\'i}guez, J. L. and Cara, M. and Cardoso, J. V. M. and Cheedella, S. and Copin, Y. and Corrales, L. and Crichton, D. and D'Avella, D. and Deil, C. and Depagne, {\'E}. and Dietrich, J. P. and Donath, A. and Droettboom, M. and Earl, N. and Erben, T. and Fabbro, S. and Ferreira, L. A. and Finethy, T. and Fox, R. T. and Garrison, L. H. and Gibbons, S. L. J. and Goldstein, D. A. and Gommers, R. and Greco, J. P. and Greenfield, P. and Groener, A. M. and Grollier, F. and Hagen, A. and Hirst, P. and Homeier, D. and Horton, A. J. and Hosseinzadeh, G. and Hu, L. and Hunkeler, J. S. and Ivezi{\'c}, {\v Z}. and Jain, A. and Jenness, T. and Kanarek, G. and Kendrew, S. and Kern, N. S. and Kerzendorf, W. E. and Khvalko, A. and King, J. and Kirkby, D. and Kulkarni, A. M. and Kumar, A. and Lee, A. and Lenz, D. and Littlefair, S. P. and Ma, Z. and Macleod, D. M. and Mastropietro, M. and McCully, C. and Montagnac, S. and Morris, B. M. and Mueller, M. and Mumford, S. J. and Muna, D. and Murphy, N. A. and Nelson, S. and Nguyen, G. H. and Ninan, J. P. and N{\"o}the, M. and Ogaz, S. and Oh, S. and Parejko, J. K. and Parley, N. and Pascual, S. and Patil, R. and Patil, A. A. and Plunkett, A. L. and Prochaska, J. X. and Rastogi, T. and Reddy Janga, V. and Sabater, J. and Sakurikar, P. and Seifert, M. and Sherbert, L. E. and {Sherwood-Taylor}, H. and Shih, A. Y. and Sick, J. and Silbiger, M. T. and Singanamalla, S. and Singer, L. P. and Sladen, P. H. and Sooley, K. A. and Sornarajah, S. and Streicher, O. and Teuben, P. and Thomas, S. W. and Tremblay, G. R. and Turner, J. E. H. and Terr{\'o}n, V. and {van Kerkwijk}, M. H. and {de la Vega}, A. and Watkins, L. L. and Weaver, B. A. and Whitmore, J. B. and Woillez, J. and Zabalza, V. and {Astropy Contributors}},
  year = {2018},
  month = sep,
  journal = {AJ},
  volume = {156},
  pages = {123},
  issn = {0004-6256},
  doi = {10.3847/1538-3881/aabc4f},
  urldate = {2023-09-29}
}

@article{astropycollaborationAstropyProjectSustaining2022,
  title = {The {{Astropy Project}}: {{Sustaining}} and {{Growing}} a {{Community-oriented Open-source Project}} and the {{Latest Major Release}} (v5.0) of the {{Core Package}}},
  shorttitle = {The {{Astropy Project}}},
  author = {{Astropy Collaboration} and {Price-Whelan}, Adrian M. and Lim, Pey Lian and Earl, Nicholas and Starkman, Nathaniel and Bradley, Larry and Shupe, David L. and Patil, Aarya A. and Corrales, Lia and Brasseur, C. E. and N{\"o}the, Maximilian and Donath, Axel and Tollerud, Erik and Morris, Brett M. and Ginsburg, Adam and Vaher, Eero and Weaver, Benjamin A. and Tocknell, James and Jamieson, William and {van Kerkwijk}, Marten H. and Robitaille, Thomas P. and Merry, Bruce and Bachetti, Matteo and G{\"u}nther, H. Moritz and Aldcroft, Thomas L. and {Alvarado-Montes}, Jaime A. and Archibald, Anne M. and B{\'o}di, Attila and Bapat, Shreyas and Barentsen, Geert and Baz{\'a}n, Juanjo and Biswas, Manish and Boquien, M{\'e}d{\'e}ric and Burke, D. J. and Cara, Daria and Cara, Mihai and Conroy, Kyle E. and Conseil, Simon and Craig, Matthew W. and Cross, Robert M. and Cruz, Kelle L. and D'Eugenio, Francesco and Dencheva, Nadia and Devillepoix, Hadrien A. R. and Dietrich, J{\"o}rg P. and Eigenbrot, Arthur Davis and Erben, Thomas and Ferreira, Leonardo and {Foreman-Mackey}, Daniel and Fox, Ryan and Freij, Nabil and Garg, Suyog and Geda, Robel and Glattly, Lauren and Gondhalekar, Yash and Gordon, Karl D. and Grant, David and Greenfield, Perry and Groener, Austen M. and Guest, Steve and Gurovich, Sebastian and Handberg, Rasmus and Hart, Akeem and {Hatfield-Dodds}, Zac and Homeier, Derek and Hosseinzadeh, Griffin and Jenness, Tim and Jones, Craig K. and Joseph, Prajwel and Kalmbach, J. Bryce and Karamehmetoglu, Emir and Ka{\l}uszy{\'n}ski, Miko{\l}aj and Kelley, Michael S. P. and Kern, Nicholas and Kerzendorf, Wolfgang E. and Koch, Eric W. and Kulumani, Shankar and Lee, Antony and Ly, Chun and Ma, Zhiyuan and MacBride, Conor and Maljaars, Jakob M. and Muna, Demitri and Murphy, N. A. and Norman, Henrik and O'Steen, Richard and Oman, Kyle A. and Pacifici, Camilla and Pascual, Sergio and {Pascual-Granado}, J. and Patil, Rohit R. and Perren, Gabriel I. and Pickering, Timothy E. and Rastogi, Tanuj and Roulston, Benjamin R. and Ryan, Daniel F. and Rykoff, Eli S. and Sabater, Jose and Sakurikar, Parikshit and Salgado, Jes{\'u}s and Sanghi, Aniket and Saunders, Nicholas and Savchenko, Volodymyr and Schwardt, Ludwig and {Seifert-Eckert}, Michael and Shih, Albert Y. and Jain, Anany Shrey and Shukla, Gyanendra and Sick, Jonathan and Simpson, Chris and Singanamalla, Sudheesh and Singer, Leo P. and Singhal, Jaladh and Sinha, Manodeep and Sip{\H o}cz, Brigitta M. and Spitler, Lee R. and Stansby, David and Streicher, Ole and {\v S}umak, Jani and Swinbank, John D. and Taranu, Dan S. and Tewary, Nikita and Tremblay, Grant R. and {de Val-Borro}, Miguel and Van Kooten, Samuel J. and Vasovi{\'c}, Zlatan and Verma, Shresth and {de Miranda Cardoso}, Jos{\'e} Vin{\'i}cius and Williams, Peter K. G. and Wilson, Tom J. and Winkel, Benjamin and {Wood-Vasey}, W. M. and Xue, Rui and Yoachim, Peter and Zhang, Chen and Zonca, Andrea and {Astropy Project Contributors}},
  year = {2022},
  month = aug,
  journal = {ApJ},
  volume = {935},
  pages = {167},
  issn = {0004-637X},
  doi = {10.3847/1538-4357/ac7c74},
  urldate = {2023-09-29}
}

@article{atkinsonFAINTTIDALFEATURES2013,
  title = {{{FAINT TIDAL FEATURES IN GALAXIES WITHIN THE CANADA-FRANCE-HAWAII TELESCOPE LEGACY SURVEY WIDE FIELDS}}},
  author = {Atkinson, Adam M and Abraham, Roberto G and Ferguson, Annette M N},
  year = {2013},
  journal = {ApJ},
  volume = {765},
  number = {1},
  pages = {28--28},
  publisher = {American Astronomical Society},
  doi = {10.1088/0004-637x/765/1/28}
}

@article{banksGalaxyMassAssembly2021,
  title = {Galaxy {{And Mass Assembly}} ({{GAMA}}): {{The Merging Potential}} of {{Brightest Group Galaxies}}},
  shorttitle = {Galaxy {{And Mass Assembly}} ({{GAMA}})},
  author = {Banks, K. and Brough, S. and Holwerda, B. W. and Hopkins, A. M. and {L{\'o}pez-S{\'a}nchez}, {\'A} R. and Phillipps, S. and Pimbblet, K. A. and Robotham, A. S. G.},
  year = {2021},
  month = oct,
  journal = {ApJ},
  volume = {921},
  number = {1},
  pages = {47},
  publisher = {American Astronomical Society},
  issn = {0004-637X},
  doi = {10.3847/1538-4357/ac1c0a},
  urldate = {2022-06-10},
  langid = {english}
}

@article{barnesEncountersDiskHalo1988,
  title = {Encounters of Disk/Halo Galaxies},
  author = {Barnes, Joshua E.},
  year = {1988},
  month = aug,
  journal = {ApJ},
  volume = {331},
  pages = {699},
  issn = {0004-637X, 1538-4357},
  doi = {10.1086/166593},
  urldate = {2025-01-10},
  langid = {english}
}

@misc{bazkiaeiBrightStarSubtraction2024,
  title = {Bright {{Star Subtraction Pipeline}} for {{LSST}}: {{Progress Review}}},
  shorttitle = {Bright {{Star Subtraction Pipeline}} for {{LSST}}},
  author = {Bazkiaei, Amir E. and Kelvin, Lee S. and Brough, Sarah and O'Toole, Simon J. and Watkins, Aaron and Schmitz, Morgen A.},
  year = {2024},
  month = apr,
  number = {arXiv:2404.04802},
  eprint = {2404.04802},
  primaryclass = {astro-ph},
  publisher = {arXiv},
  doi = {10.48550/arXiv.2404.04802},
  urldate = {2025-04-22},
  archiveprefix = {arXiv}
}

@article{bellstedtGalaxyMassAssembly2021,
  title = {Galaxy and Mass Assembly ({{GAMA}}): The Inferred Mass--Metallicity Relation from z~= 0 to 3.5 via Forensic {{SED}} Fitting},
  shorttitle = {Galaxy and Mass Assembly ({{GAMA}})},
  author = {Bellstedt, Sabine and Robotham, Aaron S G and Driver, Simon P and Thorne, Jessica E and Davies, Luke J M and Holwerda, Benne W and Hopkins, Andrew M and {Lara-Lopez}, Maritza A and {L{\'o}pez-S{\'a}nchez}, {\'A}ngel R and Phillipps, Steven},
  year = {2021},
  month = may,
  journal = {MNRAS},
  volume = {503},
  number = {3},
  pages = {3309--3325},
  issn = {0035-8711},
  doi = {10.1093/mnras/stab550},
  urldate = {2024-11-24}
}

@article{bilekCensusClassificationLowsurfacebrightness2020,
  title = {Census and Classification of Low-Surface-Brightness Structures in Nearby Early-Type Galaxies from the {{MATLAS}} Survey},
  author = {B{\'i}lek, Michal and Duc, Pierre-Alain and Cuillandre, Jean-Charles and Gwyn, Stephen and Cappellari, Michele and Bekaert, David V. and Bonfini, Paolo and Bitsakis, Theodoros and Paudel, Sanjaya and Krajnovi{\'c}, Davor and Durrell, Patrick R. and Marleau, Francine},
  year = {2020},
  month = oct,
  journal = {MNRAS},
  volume = {498},
  pages = {2138--2166},
  issn = {0035-8711},
  doi = {10.1093/mnras/staa2248},
  urldate = {2022-06-09}
}

@article{boothCosmologicalSimulationsGrowth2009,
  title = {Cosmological Simulations of the Growth of Supermassive Black Holes and Feedback from Active Galactic Nuclei: Method and Tests},
  shorttitle = {Cosmological Simulations of the Growth of Supermassive Black Holes and Feedback from Active Galactic Nuclei},
  author = {Booth, C. M. and Schaye, Joop},
  year = {2009},
  month = sep,
  journal = {MNRAS},
  volume = {398},
  pages = {53--74},
  issn = {0035-8711},
  doi = {10.1111/j.1365-2966.2009.15043.x},
  urldate = {2023-04-11}
}

@inproceedings{bouladeMegacamNewCanadaFranceHawaii2003,
  title = {Megacam: The New {{Canada-France-Hawaii Telescope}} Wide-Field Imaging Camera},
  shorttitle = {Megacam},
  booktitle = {Instrument {{Design}} and {{Performance}} for {{Optical}}/{{Infrared Ground-based Telescopes}}},
  author = {Boulade, Olivier and Charlot, Xavier and Abbon, P. and Aune, Stephan and Borgeaud, Pierre and Carton, Pierre-Henri and Carty, M. and Costa, J. Da and Deschamps, H. and Desforge, D. and Eppelle, Dominique and Gallais, Pascal and Gosset, L. and Granelli, Remy and Gros, Michel and de Kat, Jean and Loiseau, Denis and Ritou, J.-L. and Rousse, Jean Yves and Starzynski, Pierre and Vignal, Nicolas and Vigroux, Laurent G.},
  year = {2003},
  month = mar,
  volume = {4841},
  pages = {72--81},
  publisher = {SPIE},
  doi = {10.1117/12.459890},
  urldate = {2024-11-26}
}

@article{broughPreparingLowSurface2024,
  title = {Preparing for Low Surface Brightness Science with the {{Vera C}}. {{Rubin Observatory}}: A Comparison of Observable and Simulated Intracluster Light Fractions},
  shorttitle = {Preparing for Low Surface Brightness Science with the {{Vera C}}. {{Rubin Observatory}}},
  author = {Brough, Sarah and Ahad, Syeda Lammim and Bah{\'e}, Yannick M and Ellien, Ama{\"e}l and Gonzalez, Anthony H and {Jim{\'e}nez-Teja}, Yolanda and Kimmig, Lucas C and Martin, Garreth and {Mart{\'i}nez-Lombilla}, Cristina and Montes, Mireia and Pillepich, Annalisa and Ragusa, Rossella and Remus, Rhea-Silvia and Collins, Chris A and Knapen, Johan H and Mihos, J Christopher},
  year = {2024},
  month = feb,
  journal = {MNRAS},
  volume = {528},
  number = {1},
  pages = {771--795},
  issn = {0035-8711},
  doi = {10.1093/mnras/stad3810},
  urldate = {2024-04-05}
}

@article{broughSpatiallyResolvedKinematics2007,
  title = {Spatially Resolved Kinematics and Stellar Populations of Brightest Cluster and Group Galaxies},
  author = {Brough, S. and Proctor, Robert and Forbes, Duncan A. and Couch, Warrick J. and Collins, C. A. and Burke, D. J. and Mann, R. G.},
  year = {2007},
  month = jun,
  journal = {MNRAS},
  volume = {378},
  number = {4},
  pages = {1507--1530},
  issn = {0035-8711},
  doi = {10.1111/j.1365-2966.2007.11900.x},
  urldate = {2024-11-26}
}

@article{broughVeraRubinObservatory2020,
  title = {The {{Vera Rubin Observatory Legacy Survey}} of {{Space}} and {{Time}} and the {{Low Surface Brightness Universe}}},
  author = {Brough, Sarah and Collins, Chris and Demarco, Ricardo and Ferguson, Henry C and Galaz, Gaspar and Holwerda, Benne and {Martinez-Lombilla}, Cristina and Mihos, Chris and Montes, Mireia},
  year = {2020},
  month = jan,
  journal = {arXiv e-prints},
  pages = {arXiv:2001.11067-arXiv:2001.11067}
}

@article{bruzualStellarPopulationSynthesis2003,
  title = {Stellar Population Synthesis at the Resolution of 2003},
  author = {Bruzual, G. and Charlot, S.},
  year = {2003},
  month = oct,
  journal = {MNRAS},
  volume = {344},
  number = {4},
  pages = {1000--1028},
  issn = {0035-8711, 1365-2966},
  doi = {10.1046/j.1365-8711.2003.06897.x},
  urldate = {2022-06-09},
  langid = {english}
}

@article{bundyOVERVIEWSDSSIVMaNGA2014,
  title = {{{OVERVIEW OF THE SDSS-IV MaNGA SURVEY}}: {{MAPPING NEARBY GALAXIES AT APACHE POINT OBSERVATORY}}},
  shorttitle = {{{OVERVIEW OF THE SDSS-IV MaNGA SURVEY}}},
  author = {Bundy, Kevin and Bershady, Matthew A. and Law, David R. and Yan, Renbin and Drory, Niv and MacDonald, Nicholas and Wake, David A. and Cherinka, Brian and {S{\'a}nchez-Gallego}, Jos{\'e} R. and Weijmans, Anne-Marie and Thomas, Daniel and Tremonti, Christy and Masters, Karen and Coccato, Lodovico and {Diamond-Stanic}, Aleksandar M. and {Arag{\'o}n-Salamanca}, Alfonso and {Avila-Reese}, Vladimir and Badenes, Carles and {Falc{\'o}n-Barroso}, J{\'e}sus and Belfiore, Francesco and Bizyaev, Dmitry and Blanc, Guillermo A. and {Bland-Hawthorn}, Joss and Blanton, Michael R. and Brownstein, Joel R. and Byler, Nell and Cappellari, Michele and Conroy, Charlie and Dutton, Aaron A. and Emsellem, Eric and Etherington, James and Frinchaboy, Peter M. and Fu, Hai and Gunn, James E. and Harding, Paul and Johnston, Evelyn J. and Kauffmann, Guinevere and Kinemuchi, Karen and Klaene, Mark A. and Knapen, Johan H. and Leauthaud, Alexie and Li, Cheng and Lin, Lihwai and Maiolino, Roberto and Malanushenko, Viktor and Malanushenko, Elena and Mao, Shude and Maraston, Claudia and McDermid, Richard M. and Merrifield, Michael R. and Nichol, Robert C. and Oravetz, Daniel and Pan, Kaike and Parejko, John K. and Sanchez, Sebastian F. and Schlegel, David and Simmons, Audrey and Steele, Oliver and Steinmetz, Matthias and Thanjavur, Karun and Thompson, Benjamin A. and Tinker, Jeremy L. and van den Bosch, Remco C. E. and Westfall, Kyle B. and Wilkinson, David and Wright, Shelley and Xiao, Ting and Zhang, Kai},
  year = {2014},
  month = dec,
  journal = {ApJ},
  volume = {798},
  number = {1},
  pages = {7},
  publisher = {The American Astronomical Society},
  issn = {0004-637X},
  doi = {10.1088/0004-637X/798/1/7},
  urldate = {2024-11-24},
  langid = {english}
}

@article{cameronEstimationConfidenceIntervals2011,
  title = {On the {{Estimation}} of {{Confidence Intervals}} for {{Binomial Population Proportions}} in {{Astronomy}}: {{The Simplicity}} and {{Superiority}} of the {{Bayesian Approach}}},
  shorttitle = {On the {{Estimation}} of {{Confidence Intervals}} for {{Binomial Population Proportions}} in {{Astronomy}}},
  author = {Cameron, Ewan},
  year = {2011},
  month = jun,
  journal = {PASA},
  volume = {28},
  pages = {128--139},
  issn = {1323-3580},
  doi = {10.1071/AS10046},
  urldate = {2023-03-02}
}

@article{cannarozzoContributionSituEx2023,
  title = {The Contribution of in Situ and Ex Situ Star Formation in Early-Type Galaxies: {{MaNGA}} versus {{IllustrisTNG}}},
  shorttitle = {The Contribution of in Situ and Ex Situ Star Formation in Early-Type Galaxies},
  author = {Cannarozzo, Carlo and Leauthaud, Alexie and Oyarz{\'u}n, Grecco A and Nipoti, Carlo and Diemer, Benedikt and Huang, Song and {Rodriguez-Gomez}, Vicente and Sonnenfeld, Alessandro and Bundy, Kevin},
  year = {2023},
  month = apr,
  journal = {MNRAS},
  volume = {520},
  number = {4},
  pages = {5651--5670},
  issn = {0035-8711},
  doi = {10.1093/mnras/stac3023},
  urldate = {2024-02-01}
}

@article{cappellariEfficientMultiGaussianExpansion2002,
  title = {Efficient Multi-{{Gaussian}} Expansion of Galaxies},
  author = {Cappellari, Michele},
  year = {2002},
  month = jun,
  journal = {MNRAS},
  volume = {333},
  number = {2},
  pages = {400--410},
  issn = {0035-8711},
  doi = {10.1046/j.1365-8711.2002.05412.x},
  urldate = {2024-07-25}
}

@article{chabrierGalacticStellarSubstellar2003,
  title = {Galactic {{Stellar}} and {{Substellar Initial Mass Function}}},
  author = {Chabrier, Gilles},
  year = {2003},
  month = jul,
  journal = {PASP},
  volume = {115},
  pages = {763--795},
  issn = {0004-6280},
  doi = {10.1086/376392},
  urldate = {2023-08-02}
}

@article{chamberlainPhysicallyMotivatedFramework2024,
  title = {A {{Physically Motivated Framework}} to {{Compare Pair Fractions}} of {{Isolated Low-}} and {{High-mass Galaxies}} across {{Cosmic Time}}},
  author = {Chamberlain, Katie and Besla, Gurtina and Patel, Ekta and {Rodriguez-Gomez}, Vicente and Torrey, Paul and Martin, Garreth and Johnson, Kelsey and Kallivayalil, Nitya and Patton, David and Pearson, Sarah and Privon, George and Stierwalt, Sabrina},
  year = {2024},
  month = feb,
  journal = {ApJ},
  volume = {962},
  number = {2},
  pages = {162},
  publisher = {The American Astronomical Society},
  issn = {0004-637X},
  doi = {10.3847/1538-4357/ad19d0},
  urldate = {2024-12-19},
  langid = {english}
}

@misc{chambersPanSTARRS1Surveys2019,
  title = {The {{Pan-STARRS1 Surveys}}},
  author = {Chambers, K. C. and Magnier, E. A. and Metcalfe, N. and Flewelling, H. A. and Huber, M. E. and Waters, C. Z. and Denneau, L. and Draper, P. W. and Farrow, D. and Finkbeiner, D. P. and Holmberg, C. and Koppenhoefer, J. and Price, P. A. and Rest, A. and Saglia, R. P. and Schlafly, E. F. and Smartt, S. J. and Sweeney, W. and Wainscoat, R. J. and Burgett, W. S. and Chastel, S. and Grav, T. and Heasley, J. N. and Hodapp, K. W. and Jedicke, R. and Kaiser, N. and Kudritzki, R.-P. and Luppino, G. A. and Lupton, R. H. and Monet, D. G. and Morgan, J. S. and Onaka, P. M. and Shiao, B. and Stubbs, C. W. and Tonry, J. L. and White, R. and Ba{\~n}ados, E. and Bell, E. F. and Bender, R. and Bernard, E. J. and Boegner, M. and Boffi, F. and Botticella, M. T. and Calamida, A. and Casertano, S. and Chen, W.-P. and Chen, X. and Cole, S. and Deacon, N. and Frenk, C. and Fitzsimmons, A. and Gezari, S. and Gibbs, V. and Goessl, C. and Goggia, T. and Gourgue, R. and Goldman, B. and Grant, P. and Grebel, E. K. and Hambly, N. C. and Hasinger, G. and Heavens, A. F. and Heckman, T. M. and Henderson, R. and Henning, T. and Holman, M. and Hopp, U. and Ip, W.-H. and Isani, S. and Jackson, M. and Keyes, C. D. and Koekemoer, A. M. and Kotak, R. and Le, D. and Liska, D. and Long, K. S. and Lucey, J. R. and Liu, M. and Martin, N. F. and Masci, G. and McLean, B. and Mindel, E. and Misra, P. and Morganson, E. and Murphy, D. N. A. and Obaika, A. and Narayan, G. and {Nieto-Santisteban}, M. A. and Norberg, P. and Peacock, J. A. and Pier, E. A. and Postman, M. and Primak, N. and Rae, C. and Rai, A. and Riess, A. and Riffeser, A. and Rix, H. W. and R{\"o}ser, S. and Russel, R. and Rutz, L. and Schilbach, E. and Schultz, A. S. B. and Scolnic, D. and Strolger, L. and Szalay, A. and Seitz, S. and Small, E. and Smith, K. W. and Soderblom, D. R. and Taylor, P. and Thomson, R. and Taylor, A. N. and Thakar, A. R. and Thiel, J. and Thilker, D. and Unger, D. and Urata, Y. and Valenti, J. and Wagner, J. and Walder, T. and Walter, F. and Watters, S. P. and Werner, S. and {Wood-Vasey}, W. M. and Wyse, R.},
  year = {2019},
  month = jan,
  number = {arXiv:1612.05560},
  eprint = {1612.05560},
  publisher = {arXiv},
  doi = {10.48550/arXiv.1612.05560},
  urldate = {2024-11-25},
  archiveprefix = {arXiv}
}

@article{chenEvolutionMostMassive2012,
  title = {Evolution of the Most Massive Galaxies to z= 0.6 -- {{I}}. {{A}} New Method for Physical Parameter Estimation},
  author = {Chen, Yan-Mei and Kauffmann, Guinevere and Tremonti, Christy A. and White, Simon and Heckman, Timothy M. and Kova{\v c}, Katarina and Bundy, Kevin and Chisholm, John and Maraston, Claudia and Schneider, Donald P. and Bolton, Adam S. and Weaver, Benjamin A. and Brinkmann, Jon},
  year = {2012},
  month = mar,
  journal = {MNRAS},
  volume = {421},
  number = {1},
  pages = {314--332},
  issn = {0035-8711},
  doi = {10.1111/j.1365-2966.2011.20306.x},
  urldate = {2024-11-22}
}

@article{crainEAGLESimulationsGalaxy2015,
  title = {The {{EAGLE}} Simulations of Galaxy Formation: Calibration of Subgrid Physics and Model Variations},
  author = {Crain, Robert A and Schaye, Joop and Bower, Richard G and Furlong, Michelle and Schaller, Matthieu and Theuns, Tom and Dalla Vecchia, Claudio and Frenk, Carlos S and McCarthy, Ian G and Helly, John C and Jenkins, Adrian and {Rosas-Guevara}, Yetli M and White, Simon D.{\textasciitilde}M. and Trayford, James W},
  year = {2015},
  month = jun,
  journal = {MNRAS},
  volume = {450},
  number = {2},
  pages = {1937--1961},
  doi = {10.1093/mnras/stv725}
}

@article{dalgarnoHeatingIonizationHI1972,
  title = {Heating and {{Ionization}} of {{HI Regions}}},
  author = {Dalgarno, A. and McCray, R. A.},
  year = {1972},
  month = jan,
  journal = {ARA\&A},
  volume = {10},
  pages = {375},
  issn = {0066-4146},
  doi = {10.1146/annurev.aa.10.090172.002111},
  urldate = {2023-08-02}
}

@article{dallavecchiaSimulatingGalacticOutflows2012,
  title = {Simulating Galactic Outflows with Thermal Supernova Feedback},
  author = {Dalla Vecchia, Claudio and Schaye, Joop},
  year = {2012},
  month = oct,
  journal = {MNRAS},
  volume = {426},
  pages = {140--158},
  issn = {0035-8711},
  doi = {10.1111/j.1365-2966.2012.21704.x},
  urldate = {2023-08-02}
}

@article{dattathriDeprojectionStellarDynamical2024,
  title = {Deprojection and Stellar Dynamical Modelling of Boxy/Peanut Bars in Edge-on Discs},
  author = {Dattathri, Shashank and Valluri, Monica and Vasiliev, Eugene and Wheeler, Vance and Erwin, Peter},
  year = {2024},
  month = may,
  journal = {MNRAS},
  volume = {530},
  number = {1},
  pages = {1195--1217},
  issn = {0035-8711},
  doi = {10.1093/mnras/stae802},
  urldate = {2025-07-08}
}

@article{davisonEAGLEsViewEx2020,
  title = {An {{EAGLE}}'s View of Ex Situ Galaxy Growth},
  author = {Davison, Thomas A. and Norris, Mark A. and Pfeffer, Joel L. and Davies, Jonathan J. and Crain, Robert A.},
  year = {2020},
  month = sep,
  journal = {MNRAS},
  volume = {497},
  pages = {81--93},
  issn = {0035-8711},
  doi = {10.1093/mnras/staa1816},
  urldate = {2024-02-01}
}

@article{dekelGalacticHaloCuspcore2003,
  title = {Galactic Halo Cusp-Core: Tidal Compression in Mergers},
  shorttitle = {Galactic Halo Cusp-Core},
  author = {Dekel, Avishai and Devor, Jonathan and Hetzroni, Guy},
  year = {2003},
  month = may,
  journal = {MNRAS},
  volume = {341},
  number = {1},
  pages = {326--342},
  issn = {0035-8711},
  doi = {10.1046/j.1365-8711.2003.06432.x},
  urldate = {2025-01-20}
}

@article{desmonsGalaxyMassAssembly2023,
  title = {Galaxy and Mass Assembly ({{GAMA}}): Comparing Visually and Spectroscopically Identified Galaxy Merger Samples},
  shorttitle = {Galaxy and Mass Assembly ({{GAMA}})},
  author = {Desmons, Alice and Brough, Sarah and {Mart{\'i}nez-Lombilla}, Cristina and De~Propris, Roberto and Holwerda, Benne and {L{\'o}pez-S{\'a}nchez}, {\'A}ngel R},
  year = {2023},
  month = aug,
  journal = {MNRAS},
  volume = {523},
  number = {3},
  pages = {4381--4393},
  issn = {0035-8711},
  doi = {10.1093/mnras/stad1639},
  urldate = {2024-03-12}
}

@article{deugenioSAMIGalaxySurvey2021,
  title = {The {{SAMI Galaxy Survey}}: Stellar Population and Structural Trends across the {{Fundamental Plane}}},
  shorttitle = {The {{SAMI Galaxy Survey}}},
  author = {D'Eugenio, Francesco and Colless, Matthew and Scott, Nicholas and {van~der~Wel}, Arjen and Davies, Roger L and {van~de~Sande}, Jesse and Sweet, Sarah M and Oh, Sree and Groves, Brent and Sharp, Rob and Owers, Matt S and {Bland-Hawthorn}, Joss and Croom, Scott M and Brough, Sarah and Bryant, Julia J and Goodwin, Michael and Lawrence, Jon S and Lorente, Nuria P F and Richards, Samuel N},
  year = {2021},
  month = jul,
  journal = {MNRAS},
  volume = {504},
  number = {4},
  pages = {5098--5130},
  issn = {0035-8711},
  doi = {10.1093/mnras/stab1146},
  urldate = {2024-09-10}
}

@article{deyOverviewDESILegacy2019,
  title = {Overview of the {{DESI Legacy Imaging Surveys}}},
  author = {Dey, Arjun and Schlegel, David J. and Lang, Dustin and Blum, Robert and Burleigh, Kaylan and Fan, Xiaohui and Findlay, Joseph R. and Finkbeiner, Doug and Herrera, David and Juneau, St{\'e}phanie and Landriau, Martin and Levi, Michael and McGreer, Ian and Meisner, Aaron and Myers, Adam D. and Moustakas, John and Nugent, Peter and Patej, Anna and Schlafly, Edward F. and Walker, Alistair R. and Valdes, Francisco and Weaver, Benjamin A. and Y{\`e}che, Christophe and Zou, Hu and Zhou, Xu and Abareshi, Behzad and Abbott, T. M. C. and Abolfathi, Bela and Aguilera, C. and Alam, Shadab and Allen, Lori and Alvarez, A. and Annis, James and Ansarinejad, Behzad and Aubert, Marie and Beechert, Jacqueline and Bell, Eric F. and BenZvi, Segev Y. and Beutler, Florian and Bielby, Richard M. and Bolton, Adam S. and Brice{\~n}o, C{\'e}sar and {Buckley-Geer}, Elizabeth J. and Butler, Karen and Calamida, Annalisa and Carlberg, Raymond G. and Carter, Paul and Casas, Ricard and Castander, Francisco J. and Choi, Yumi and Comparat, Johan and Cukanovaite, Elena and Delubac, Timoth{\'e}e and DeVries, Kaitlin and Dey, Sharmila and Dhungana, Govinda and Dickinson, Mark and Ding, Zhejie and Donaldson, John B. and Duan, Yutong and Duckworth, Christopher J. and Eftekharzadeh, Sarah and Eisenstein, Daniel J. and Etourneau, Thomas and Fagrelius, Parker A. and Farihi, Jay and Fitzpatrick, Mike and {Font-Ribera}, Andreu and Fulmer, Leah and G{\"a}nsicke, Boris T. and Gaztanaga, Enrique and George, Koshy and Gerdes, David W. and A Gontcho, Satya Gontcho and Gorgoni, Claudio and Green, Gregory and Guy, Julien and Harmer, Diane and Hernandez, M. and Honscheid, Klaus and Huang, Lijuan (Wendy) and James, David J. and Jannuzi, Buell T. and Jiang, Linhua and Joyce, Richard and Karcher, Armin and Karkar, Sonia and Kehoe, Robert and Kneib, Jean-Paul and {Kueter-Young}, Andrea and Lan, Ting-Wen and Lauer, Tod R. and Guillou, Laurent Le and Van Suu, Auguste Le and Lee, Jae Hyeon and Lesser, Michael and Levasseur, Laurence Perreault and Li, Ting S. and Mann, Justin L. and Marshall, Robert and {Mart{\'i}nez-V{\'a}zquez}, C. E. and Martini, Paul and {du Mas des Bourboux}, H{\'e}lion and McManus, Sean and Meier, Tobias Gabriel and M{\'e}nard, Brice and Metcalfe, Nigel and {Mu{\~n}oz-Guti{\'e}rrez}, Andrea and Najita, Joan and Napier, Kevin and Narayan, Gautham and Newman, Jeffrey A. and Nie, Jundan and Nord, Brian and Norman, Dara J. and Olsen, Knut A. G. and Paat, Anthony and {Palanque-Delabrouille}, Nathalie and Peng, Xiyan and Poppett, Claire L. and Poremba, Megan R. and Prakash, Abhishek and Rabinowitz, David and Raichoor, Anand and Rezaie, Mehdi and Robertson, A. N. and Roe, Natalie A. and Ross, Ashley J. and Ross, Nicholas P. and Rudnick, Gregory and Gaines, Sasha and Saha, Abhijit and S{\'a}nchez, F. Javier and Savary, Elodie and Schweiker, Heidi and Scott, Adam and Seo, Hee-Jong and Shan, Huanyuan and Silva, David R. and Slepian, Zachary and Soto, Christian and Sprayberry, David and Staten, Ryan and Stillman, Coley M. and Stupak, Robert J. and Summers, David L. and Tie, Suk Sien and Tirado, H. and {Vargas-Maga{\~n}a}, Mariana and Vivas, A. Katherina and Wechsler, Risa H. and Williams, Doug and Yang, Jinyi and Yang, Qian and Yapici, Tolga and Zaritsky, Dennis and Zenteno, A. and Zhang, Kai and Zhang, Tianmeng and Zhou, Rongpu and Zhou, Zhimin},
  year = {2019},
  month = apr,
  journal = {AJ},
  volume = {157},
  number = {5},
  pages = {168},
  publisher = {The American Astronomical Society},
  issn = {1538-3881},
  doi = {10.3847/1538-3881/ab089d},
  urldate = {2025-06-10},
  langid = {english}
}

@article{diemandFormationEvolutionGalaxy2007,
  title = {Formation and {{Evolution}} of {{Galaxy Dark Matter Halos}} and {{Their Substructure}}},
  author = {Diemand, J{\"u}rg and Kuhlen, Michael and Madau, Piero},
  year = {2007},
  month = oct,
  journal = {ApJ},
  volume = {667},
  number = {2},
  pages = {859},
  publisher = {IOP Publishing},
  issn = {0004-637X},
  doi = {10.1086/520573},
  urldate = {2025-01-20},
  langid = {english}
}

@article{dimatteoEnergyInputQuasars2005,
  title = {Energy Input from Quasars Regulates the Growth and Activity of Black Holes and Their Host Galaxies},
  author = {Di Matteo, Tiziana and Springel, Volker and Hernquist, Lars},
  year = {2005},
  month = feb,
  journal = {Nature},
  volume = {433},
  number = {7026},
  pages = {604--607},
  publisher = {Nature Publishing Group},
  issn = {1476-4687},
  doi = {10.1038/nature03335},
  urldate = {2025-01-09},
  copyright = {2005 Macmillan Magazines Ltd.},
  langid = {english}
}

@article{dimatteoStarFormationEfficiency2007,
  title = {Star Formation Efficiency in Galaxy Interactions and Mergers: A Statistical Study},
  shorttitle = {Star Formation Efficiency in Galaxy Interactions and Mergers},
  author = {Di Matteo, P. and Combes, F. and Melchior, A.-L. and Semelin, B.},
  year = {2007},
  month = jun,
  journal = {A\&A},
  volume = {468},
  number = {1},
  pages = {61--81},
  issn = {0004-6361, 1432-0746},
  doi = {10.1051/0004-6361:20066959},
  urldate = {2022-06-09}
}

@article{dolagDistributionEvolutionMetals2017,
  title = {Distribution and {{Evolution}} of {{Metals}} in the {{Magneticum Simulations}}},
  author = {Dolag, Klaus and Mevius, Emilio and Remus, Rhea-Silvia},
  year = {2017},
  month = aug,
  journal = {Galaxies},
  volume = {5},
  pages = {35},
  doi = {10.3390/galaxies5030035},
  urldate = {2023-08-02}
}

@misc{dolagEncyclopediaMagneticumScaling2025,
  title = {Encyclopedia {{Magneticum}}: {{Scaling Relations}} from {{Cosmic Dawn}} to {{Present Day}}},
  shorttitle = {Encyclopedia {{Magneticum}}},
  author = {Dolag, Klaus and Remus, Rhea-Silvia and Valenzuela, Lucas M. and Kimmig, Lucas C. and Seidel, Benjamin and Fortune, Silvio and Stoiber, Johannes and Ivleva, Anna and Hoffmann, Tadziu and Biffi, Veronica and Marini, Ilaria and Popesso, Paola and {Vladutescu-Zopp}, Stephan},
  year = {2025},
  month = apr,
  number = {arXiv:2504.01061},
  eprint = {2504.01061},
  primaryclass = {astro-ph},
  publisher = {arXiv},
  doi = {10.48550/arXiv.2504.01061},
  urldate = {2025-04-28},
  archiveprefix = {arXiv}
}

@article{dolagThermalConductionSimulated2004,
  title = {Thermal {{Conduction}} in {{Simulated Galaxy Clusters}}},
  author = {Dolag, K. and Jubelgas, M. and Springel, V. and Borgani, S. and Rasia, E.},
  year = {2004},
  month = apr,
  journal = {ApJ},
  volume = {606},
  number = {2},
  pages = {L97},
  publisher = {IOP Publishing},
  issn = {0004-637X},
  doi = {10.1086/420966},
  urldate = {2024-03-22},
  langid = {english}
}

@article{driverGalaxyMassAssembly2011,
  title = {Galaxy and {{Mass Assembly}} ({{GAMA}}): Survey Diagnostics and Core Data Release},
  author = {Driver, S.{\textasciitilde}P. and Hill, D.{\textasciitilde}T. and Kelvin, L.{\textasciitilde}S. and Robotham, A.{\textasciitilde}S.{\textasciitilde}G. and Liske, J and Norberg, P and Baldry, I.{\textasciitilde}K. and Bamford, S.{\textasciitilde}P. and Hopkins, A.{\textasciitilde}M. and Loveday, J and Peacock, J.{\textasciitilde}A. and Andrae, E and {Bland-Hawthorn}, J and Brough, S and Brown, M.{\textasciitilde}J.{\textasciitilde}I. and Cameron, E and Ching, J.{\textasciitilde}H.{\textasciitilde}Y. and Colless, M and Conselice, C.{\textasciitilde}J. and Croom, S.{\textasciitilde}M. and Cross, N.{\textasciitilde}J.{\textasciitilde}G. and {de Propris}, R and Dye, S and Drinkwater, M.{\textasciitilde}J. and Ellis, S and Graham, Alister W and Grootes, M.{\textasciitilde}W. and Gunawardhana, M and Jones, D.{\textasciitilde}H. and {van Kampen}, E and Maraston, C and Nichol, R.{\textasciitilde}C. and Parkinson, H.{\textasciitilde}R. and Phillipps, S and Pimbblet, K and Popescu, C.{\textasciitilde}C. and Prescott, M and Roseboom, I.{\textasciitilde}G. and Sadler, E.{\textasciitilde}M. and Sansom, A.{\textasciitilde}E. and Sharp, R.{\textasciitilde}G. and Smith, D.{\textasciitilde}J.{\textasciitilde}B. and Taylor, E and Thomas, D and Tuffs, R.{\textasciitilde}J. and Wijesinghe, D and Dunne, L and Frenk, C.{\textasciitilde}S. and Jarvis, M.{\textasciitilde}J. and Madore, B.{\textasciitilde}F. and Meyer, M.{\textasciitilde}J. and Seibert, M and {Staveley-Smith}, L and Sutherland, W.{\textasciitilde}J. and Warren, S.{\textasciitilde}J.},
  year = {2011},
  month = may,
  journal = {MNRAS},
  volume = {413},
  number = {2},
  pages = {971--995},
  doi = {10.1111/j.1365-2966.2010.18188.x}
}

@article{duboisDancingDarkGalactic2014,
  title = {Dancing in the Dark: {{Galactic}} Properties Trace Spin Swings along the Cosmic Web},
  author = {Dubois, Y. and Pichon, C. and Welker, C. and Le Borgne, D. and Devriendt, J. and Laigle, C. and Codis, S. and Pogosyan, D. and Arnouts, S. and Benabed, K. and Bertin, E. and Blaizot, J. and Bouchet, F. and Cardoso, J. F. and Colombi, S. and De Lapparent, V. and Desjacques, V. and Gavazzi, R. and Kassin, S. and Kimm, T. and McCracken, H. and Milliard, B. and Peirani, S. and Prunet, S. and Rouberol, S. and Silk, J. and Slyz, A. and Sousbie, T. and Teyssier, R. and Tresse, L. and Treyer, M. and Vibert, D. and Volonteri, M.},
  year = {2014},
  journal = {MNRAS},
  volume = {444},
  number = {2},
  pages = {1453--1468},
  publisher = {Oxford University Press},
  doi = {10.1093/mnras/stu1227}
}

@article{duboisHORIZONAGNSimulationMorphological2016,
  title = {The {{HORIZON-AGN}} Simulation: Morphological Diversity of Galaxies Promoted by {{AGN}} Feedback},
  shorttitle = {The {{HORIZON-AGN}} Simulation},
  author = {Dubois, Yohan and Peirani, S{\'e}bastien and Pichon, Christophe and Devriendt, Julien and Gavazzi, Rapha{\"e}l and Welker, Charlotte and Volonteri, Marta},
  year = {2016},
  month = dec,
  journal = {MNRAS},
  volume = {463},
  pages = {3948--3964},
  issn = {0035-8711},
  doi = {10.1093/mnras/stw2265},
  urldate = {2024-02-01}
}

@article{duboisIntroducingNEWHORIZONSimulation2021,
  title = {Introducing the {{NEWHORIZON}} Simulation: {{Galaxy}} Properties with Resolved Internal Dynamics across Cosmic Time},
  shorttitle = {Introducing the {{NEWHORIZON}} Simulation},
  author = {Dubois, Yohan and Beckmann, Ricarda and Bournaud, Fr{\'e}d{\'e}ric and Choi, Hoseung and Devriendt, Julien and Jackson, Ryan and Kaviraj, Sugata and Kimm, Taysun and Kraljic, Katarina and Laigle, Clotilde and Martin, Garreth and Park, Min-Jung and Peirani, S{\'e}bastien and Pichon, Christophe and Volonteri, Marta and Yi, Sukyoung K.},
  year = {2021},
  month = jul,
  journal = {A\&A},
  volume = {651},
  pages = {A109},
  issn = {0004-6361, 1432-0746},
  doi = {10.1051/0004-6361/202039429},
  urldate = {2022-05-14},
  langid = {english}
}

@article{duboisJetregulatedCoolingCatastrophe2010,
  title = {Jet-Regulated Cooling Catastrophe},
  author = {Dubois, Yohan and Devriendt, Julien and Slyz, Adrianne and Teyssier, Romain},
  year = {2010},
  month = dec,
  journal = {MNRAS},
  volume = {409},
  pages = {985--1001},
  issn = {0035-8711},
  doi = {10.1111/j.1365-2966.2010.17338.x},
  urldate = {2023-08-02}
}

@article{ducATLAS3DProjectXXIX2015,
  title = {The {{ATLAS3D}} Project -- {{XXIX}}. {{The}} New Look of Early-Type Galaxies and Surrounding Fields Disclosed by Extremely Deep Optical Images},
  author = {Duc, Pierre-Alain and Cuillandre, Jean-Charles and Karabal, Emin and Cappellari, Michele and Alatalo, Katherine and Blitz, Leo and Bournaud, Fr{\'e}d{\'e}ric and Bureau, Martin and Crocker, Alison F. and Davies, Roger L. and Davis, Timothy A. and {de Zeeuw}, P. T. and Emsellem, Eric and Khochfar, Sadegh and Krajnovi{\'c}, Davor and Kuntschner, Harald and McDermid, Richard M. and {Michel-Dansac}, Leo and Morganti, Raffaella and Naab, Thorsten and Oosterloo, Tom and Paudel, Sanjaya and Sarzi, Marc and Scott, Nicholas and Serra, Paolo and Weijmans, Anne-Marie and Young, Lisa M.},
  year = {2015},
  month = jan,
  journal = {MNRAS},
  volume = {446},
  number = {1},
  pages = {120--143},
  issn = {0035-8711},
  doi = {10.1093/mnras/stu2019},
  urldate = {2025-05-13}
}

@misc{ducMATLASDeepExploration2020,
  title = {{{MATLAS}}: A Deep Exploration of the Surroundings of Massive Early-Type Galaxies},
  shorttitle = {{{MATLAS}}},
  author = {Duc, Pierre-Alain},
  year = {2020},
  month = jul,
  number = {arXiv:2007.13874},
  eprint = {2007.13874},
  publisher = {arXiv},
  doi = {10.48550/arXiv.2007.13874},
  urldate = {2024-11-24},
  archiveprefix = {arXiv}
}

@article{eisertERGOMLInferringAssembly2023,
  title = {{{ERGO-ML I}}: Inferring the Assembly Histories of {{IllustrisTNG}} Galaxies from Integral Observable Properties via Invertible Neural Networks},
  shorttitle = {{{ERGO-ML I}}},
  author = {Eisert, Lukas and Pillepich, Annalisa and Nelson, Dylan and Klessen, Ralf S and {Huertas-Company}, Marc and {Rodriguez-Gomez}, Vicente},
  year = {2023},
  month = feb,
  journal = {MNRAS},
  volume = {519},
  number = {2},
  pages = {2199--2223},
  issn = {0035-8711},
  doi = {10.1093/mnras/stac3295},
  urldate = {2024-10-10}
}

@article{ellisonGalaxyEvolutionPostmerger2024a,
  title = {Galaxy Evolution in the Post-Merger Regime. {{II}} -- {{Post-merger}} Quenching Peaks within 500 {{Myr}} of Coalescence},
  author = {Ellison, Sara and Ferreira, Leonardo and Wild, Vivienne and Wilkinson, Scott and Rowlands, Kate and Patton, David R.},
  year = {2024},
  month = dec,
  journal = {The Open Journal of Astrophysics},
  volume = {7},
  publisher = {Maynooth Academic Publishing},
  doi = {10.33232/001c.127779},
  urldate = {2025-09-29},
  langid = {english}
}

@article{ellisonGalaxyMergersCan2022,
  title = {Galaxy Mergers Can Rapidly Shut down Star Formation},
  author = {Ellison, Sara L and Wilkinson, Scott and Woo, Joanna and Leung, Ho-Hin and Wild, Vivienne and Bickley, Robert W and Patton, David R and Quai, Salvatore and Gwyn, Stephen},
  year = {2022},
  month = nov,
  journal = {MNRAS},
  volume = {517},
  number = {1},
  pages = {L92-L96},
  issn = {1745-3925},
  doi = {10.1093/mnrasl/slac109},
  urldate = {2025-01-20}
}

@article{emsellemMultigaussianExpansionMethod1994,
  title = {The Multi-Gaussian Expansion Method: A Tool for Building Realistic Photometric and Kinematical Models of Stellar Systems {{I}}. {{The}} Formalism},
  shorttitle = {The Multi-Gaussian Expansion Method},
  author = {Emsellem, E. and Monnet, G. and Bacon, R.},
  year = {1994},
  month = may,
  journal = {A\&A},
  volume = {285},
  pages = {723--738},
  issn = {0004-6361},
  urldate = {2025-04-07}
}

@article{fabjanSimulatingEffectActive2010,
  title = {Simulating the Effect of Active Galactic Nuclei Feedback on the Metal Enrichment of Galaxy Clusters},
  author = {Fabjan, D. and Borgani, S. and Tornatore, L. and Saro, A. and Murante, G. and Dolag, K.},
  year = {2010},
  month = jan,
  journal = {MNRAS},
  volume = {401},
  pages = {1670--1690},
  issn = {0035-8711},
  doi = {10.1111/j.1365-2966.2009.15794.x},
  urldate = {2023-08-02}
}

@article{fallFormationRotationDisc1980,
  title = {Formation and Rotation of Disc Galaxies with Haloes.},
  author = {Fall, S.{\textasciitilde}M. and Efstathiou, G},
  year = {1980},
  month = oct,
  journal = {MNRAS},
  volume = {193},
  pages = {189--206},
  doi = {10.1093/mnras/193.2.189}
}

@article{faroukiComputerSimulationsEnvironmental1981,
  title = {Computer Simulations of Environmental Influences on Galaxy Evolution in Dense Clusters. {{II}} - {{Rapid}} Tidal Encounters},
  author = {Farouki, R. and Shapiro, S. L.},
  year = {1981},
  month = jan,
  journal = {ApJ},
  volume = {243},
  pages = {32--41},
  publisher = {IOP},
  issn = {0004-637X},
  doi = {10.1086/158563},
  urldate = {2025-01-20}
}

@article{faucher-giguereNewCalculationIonizing2009,
  title = {A {{New Calculation}} of the {{Ionizing Background Spectrum}} and the {{Effects}} of {{He II Reionization}}},
  author = {{Faucher-Gigu{\`e}re}, Claude-Andr{\'e} and Lidz, Adam and Zaldarriaga, Matias and Hernquist, Lars},
  year = {2009},
  month = oct,
  journal = {ApJ},
  volume = {703},
  pages = {1416--1443},
  issn = {0004-637X},
  doi = {10.1088/0004-637X/703/2/1416},
  urldate = {2023-08-02}
}

@article{ferlandCLOUDY90Numerical1998,
  title = {{{CLOUDY}} 90: {{Numerical Simulation}} of {{Plasmas}} and {{Their Spectra}}},
  shorttitle = {{{CLOUDY}} 90},
  author = {Ferland, G. J. and Korista, K. T. and Verner, D. A. and Ferguson, J. W. and Kingdon, J. B. and Verner, E. M.},
  year = {1998},
  month = jul,
  journal = {PASP},
  volume = {110},
  pages = {761--778},
  issn = {0004-6280},
  doi = {10.1086/316190},
  urldate = {2023-08-02}
}

@article{ferreiraGalaxyEvolutionPostMerger2025,
  title = {Galaxy Evolution in the {{Post-Merger Regime}} -- {{I}}. {{Most}} Merger-Induced in Situ Stellar Mass Growth Happens Post-Coalescence},
  author = {Ferreira, Leonardo and Ellison, Sara L and Patton, David R and {Byrne-Mamahit}, Shoshannah and Wilkinson, Scott and Bickley, Robert and Conselice, Christopher J and Bottrell, Connor},
  year = {2025},
  month = mar,
  journal = {Mon Not R Astron Soc Lett},
  volume = {538},
  number = {1},
  pages = {L31-L36},
  issn = {1745-3925},
  doi = {10.1093/mnrasl/slaf004},
  urldate = {2025-09-29}
}

@article{fliriIACStripe822016,
  title = {The {{IAC Stripe}} 82 {{Legacy Project}}: A Wide-Area Survey for Faint Surface Brightness Astronomy},
  shorttitle = {The {{IAC Stripe}} 82 {{Legacy Project}}},
  author = {Fliri, J{\"u}rgen and Trujillo, Ignacio},
  year = {2016},
  month = feb,
  journal = {MNRAS},
  volume = {456},
  number = {2},
  pages = {1359--1373},
  issn = {0035-8711},
  doi = {10.1093/mnras/stv2686},
  urldate = {2024-11-24}
}

@article{gallazziAgesMetallicitiesGalaxies2005,
  title = {The Ages and Metallicities of Galaxies in the Local Universe},
  author = {Gallazzi, Anna and Charlot, St{\'e}phane and Brinchmann, Jarle and White, Simon D. M. and Tremonti, Christy A.},
  year = {2005},
  month = sep,
  journal = {Monthly Notices of the Royal Astronomical Society},
  volume = {362},
  number = {1},
  pages = {41--58},
  issn = {0035-8711},
  doi = {10.1111/j.1365-2966.2005.09321.x},
  urldate = {2024-12-04}
}

@article{genelSizeEvolutionStarforming2018,
  title = {The Size Evolution of Star-Forming and Quenched Galaxies in the {{IllustrisTNG}} Simulation},
  author = {Genel, Shy and Nelson, Dylan and Pillepich, Annalisa and Springel, Volker and Pakmor, R{\"u}diger and Weinberger, Rainer and Hernquist, Lars and Naiman, Jill and Vogelsberger, Mark and Marinacci, Federico and Torrey, Paul},
  year = {2018},
  month = mar,
  journal = {MNRAS},
  volume = {474},
  number = {3},
  pages = {3976--3996},
  issn = {0035-8711},
  doi = {10.1093/mnras/stx3078},
  urldate = {2024-09-04}
}

@article{gonzalez-perezColourGradientsSDSS2011,
  title = {Colour Gradients within {{SDSS DR7}} Galaxies: Hints of Recent Evolution},
  shorttitle = {Colour Gradients within {{SDSS DR7}} Galaxies},
  author = {{Gonzalez-Perez}, V. and Castander, F. J. and Kauffmann, G.},
  year = {2011},
  month = feb,
  journal = {MNRAS},
  volume = {411},
  number = {2},
  pages = {1151--1166},
  issn = {0035-8711},
  doi = {10.1111/j.1365-2966.2010.17744.x},
  urldate = {2024-11-26}
}

@misc{gordonLinkingEnhancedStar2025,
  title = {Linking Enhanced Star Formation and Quenching to Faint Tidal Features in Galaxies},
  author = {Gordon, Alexander J. and Ferguson, Annette M. N. and Mann, Robert G. and Wild, Vivienne},
  year = {2025},
  month = jul,
  number = {arXiv:2507.21050},
  eprint = {2507.21050},
  primaryclass = {astro-ph},
  publisher = {arXiv},
  doi = {10.48550/arXiv.2507.21050},
  urldate = {2025-07-29},
  archiveprefix = {arXiv}
}

@article{gunnInfallMatterClusters1972,
  title = {On the {{Infall}} of {{Matter Into Clusters}} of {{Galaxies}} and {{Some Effects}} on {{Their Evolution}}},
  author = {Gunn, James E. and Gott, III, J. Richard},
  year = {1972},
  month = aug,
  journal = {ApJ},
  volume = {176},
  pages = {1},
  publisher = {IOP},
  issn = {0004-637X},
  doi = {10.1086/151605},
  urldate = {2025-01-20}
}

@article{harrisArrayProgrammingNumPy2020a,
  title = {Array Programming with {{NumPy}}},
  author = {Harris, Charles R. and Millman, K. Jarrod and {van der Walt}, St{\'e}fan J. and Gommers, Ralf and Virtanen, Pauli and Cournapeau, David and Wieser, Eric and Taylor, Julian and Berg, Sebastian and Smith, Nathaniel J. and Kern, Robert and Picus, Matti and Hoyer, Stephan and {van Kerkwijk}, Marten H. and Brett, Matthew and Haldane, Allan and {del R{\'i}o}, Jaime Fern{\'a}ndez and Wiebe, Mark and Peterson, Pearu and {G{\'e}rard-Marchant}, Pierre and Sheppard, Kevin and Reddy, Tyler and Weckesser, Warren and Abbasi, Hameer and Gohlke, Christoph and Oliphant, Travis E.},
  year = {2020},
  month = sep,
  journal = {Nature},
  volume = {585},
  number = {7825},
  pages = {357--362},
  publisher = {Nature Publishing Group},
  issn = {1476-4687},
  doi = {10.1038/s41586-020-2649-2},
  urldate = {2023-09-29},
  copyright = {2020 The Author(s)},
  langid = {english}
}

@article{hendelTidalDebrisMorphology2015,
  title = {Tidal Debris Morphology and the Orbits of Satellite Galaxies},
  author = {Hendel, David and Johnston, Kathryn V},
  year = {2015},
  month = dec,
  journal = {MNRAS},
  volume = {454},
  number = {3},
  pages = {2472--2485},
  doi = {10.1093/mnras/stv2035}
}

@article{hennebelleAnalyticalStarFormation2011,
  title = {Analytical {{Star Formation Rate}} from {{Gravoturbulent Fragmentation}}},
  author = {Hennebelle, Patrick and Chabrier, Gilles},
  year = {2011},
  month = dec,
  journal = {ApJ},
  volume = {743},
  pages = {L29},
  issn = {0004-637X},
  doi = {10.1088/2041-8205/743/2/L29},
  urldate = {2023-08-02}
}

@article{hernquistOriginKinematicSubsystems1991,
  title = {Origin of Kinematic Subsystems in Elliptical Galaxies},
  author = {Hernquist, Lars and Barnes, Joshua E.},
  year = {1991},
  month = nov,
  journal = {Nature},
  volume = {354},
  number = {6350},
  pages = {210--212},
  publisher = {Nature Publishing Group},
  issn = {1476-4687},
  doi = {10.1038/354210a0},
  urldate = {2025-01-09},
  copyright = {1991 Springer Nature Limited},
  langid = {english}
}

@article{hernquistTidalTriggeringStarbursts1989,
  title = {Tidal Triggering of Starbursts and Nuclear Activity in Galaxies},
  author = {Hernquist, Lars},
  year = {1989},
  month = aug,
  journal = {Nature},
  volume = {340},
  number = {6236},
  pages = {687--691},
  publisher = {Nature Publishing Group},
  issn = {1476-4687},
  doi = {10.1038/340687a0},
  urldate = {2025-04-27},
  copyright = {1989 Springer Nature Limited},
  langid = {english}
}

@article{hirschmannCosmologicalSimulationsBlack2014,
  title = {Cosmological Simulations of Black Hole Growth: {{AGN}} Luminosities and Downsizing},
  shorttitle = {Cosmological Simulations of Black Hole Growth},
  author = {Hirschmann, Michaela and Dolag, Klaus and Saro, Alexandro and Bachmann, Lisa and Borgani, Stefano and Burkert, Andreas},
  year = {2014},
  month = aug,
  journal = {MNRAS},
  volume = {442},
  pages = {2304--2324},
  issn = {0035-8711},
  doi = {10.1093/mnras/stu1023},
  urldate = {2022-04-12}
}

@article{hodgesSignificanceProbabilitySmirnov1958,
  title = {The Significance Probability of the Smirnov Two-Sample Test},
  author = {Hodges, J L},
  year = {1958},
  journal = {Arkiv f{\"o}r Matematik},
  volume = {3},
  pages = {469--486},
  doi = {10.1007/BF02589501}
}

@article{huangMassiveEarlytypeGalaxies2022,
  title = {Massive {{Early-type Galaxies}} in the {{HSC-SSP}}: {{Flux Fraction}} of {{Tidal Features}} and {{Merger Rates}}},
  shorttitle = {Massive {{Early-type Galaxies}} in the {{HSC-SSP}}},
  author = {Huang, Qifeng and Fan, Lulu},
  year = {2022},
  month = sep,
  journal = {ApJS},
  volume = {262},
  number = {2},
  pages = {39},
  publisher = {The American Astronomical Society},
  issn = {0067-0049},
  doi = {10.3847/1538-4365/ac85b1},
  urldate = {2024-04-29},
  langid = {english}
}

@article{ivezicLSSTScienceDrivers2019a,
  title = {{{LSST}}: {{From Science Drivers}} to {{Reference Design}} and {{Anticipated Data Products}}},
  shorttitle = {{{LSST}}},
  author = {Ivezi{\'c}, {\v Z}eljko and Kahn, Steven M. and Tyson, J. Anthony and Abel, Bob and Acosta, Emily and Allsman, Robyn and Alonso, David and AlSayyad, Yusra and Anderson, Scott F. and Andrew, John and Angel, James Roger P. and Angeli, George Z. and Ansari, Reza and Antilogus, Pierre and Araujo, Constanza and Armstrong, Robert and Arndt, Kirk T. and Astier, Pierre and Aubourg, {\'E}ric and Auza, Nicole and Axelrod, Tim S. and Bard, Deborah J. and Barr, Jeff D. and Barrau, Aurelian and Bartlett, James G. and Bauer, Amanda E. and Bauman, Brian J. and Baumont, Sylvain and Bechtol, Ellen and Bechtol, Keith and Becker, Andrew C. and Becla, Jacek and Beldica, Cristina and Bellavia, Steve and Bianco, Federica B. and Biswas, Rahul and Blanc, Guillaume and Blazek, Jonathan and Blandford, Roger D. and Bloom, Josh S. and Bogart, Joanne and Bond, Tim W. and Booth, Michael T. and Borgland, Anders W. and Borne, Kirk and Bosch, James F. and Boutigny, Dominique and Brackett, Craig A. and Bradshaw, Andrew and Brandt, William Nielsen and Brown, Michael E. and Bullock, James S. and Burchat, Patricia and Burke, David L. and Cagnoli, Gianpietro and Calabrese, Daniel and Callahan, Shawn and Callen, Alice L. and Carlin, Jeffrey L. and Carlson, Erin L. and Chandrasekharan, Srinivasan and {Charles-Emerson}, Glenaver and Chesley, Steve and Cheu, Elliott C. and Chiang, Hsin-Fang and Chiang, James and Chirino, Carol and Chow, Derek and Ciardi, David R. and Claver, Charles F. and {Cohen-Tanugi}, Johann and Cockrum, Joseph J. and Coles, Rebecca and Connolly, Andrew J. and Cook, Kem H. and Cooray, Asantha and Covey, Kevin R. and Cribbs, Chris and Cui, Wei and Cutri, Roc and Daly, Philip N. and Daniel, Scott F. and Daruich, Felipe and Daubard, Guillaume and Daues, Greg and Dawson, William and Delgado, Francisco and Dellapenna, Alfred and de Peyster, Robert and de {Val-Borro}, Miguel and Digel, Seth W. and Doherty, Peter and Dubois, Richard and {Dubois-Felsmann}, Gregory P. and Durech, Josef and Economou, Frossie and Eifler, Tim and Eracleous, Michael and Emmons, Benjamin L. and Neto, Angelo Fausti and Ferguson, Henry and Figueroa, Enrique and {Fisher-Levine}, Merlin and Focke, Warren and Foss, Michael D. and Frank, James and Freemon, Michael D. and Gangler, Emmanuel and Gawiser, Eric and Geary, John C. and Gee, Perry and Geha, Marla and Gessner, Charles J. B. and Gibson, Robert R. and Gilmore, D. Kirk and Glanzman, Thomas and Glick, William and Goldina, Tatiana and Goldstein, Daniel A. and Goodenow, Iain and Graham, Melissa L. and Gressler, William J. and Gris, Philippe and Guy, Leanne P. and Guyonnet, Augustin and Haller, Gunther and Harris, Ron and Hascall, Patrick A. and Haupt, Justine and Hernandez, Fabio and Herrmann, Sven and Hileman, Edward and Hoblitt, Joshua and Hodgson, John A. and Hogan, Craig and Howard, James D. and Huang, Dajun and Huffer, Michael E. and Ingraham, Patrick and Innes, Walter R. and Jacoby, Suzanne H. and Jain, Bhuvnesh and Jammes, Fabrice and Jee, M. James and Jenness, Tim and Jernigan, Garrett and Jevremovi{\'c}, Darko and Johns, Kenneth and Johnson, Anthony S. and Johnson, Margaret W. G. and Jones, R. Lynne and {Juramy-Gilles}, Claire and Juri{\'c}, Mario and Kalirai, Jason S. and Kallivayalil, Nitya J. and Kalmbach, Bryce and Kantor, Jeffrey P. and Karst, Pierre and Kasliwal, Mansi M. and Kelly, Heather and Kessler, Richard and Kinnison, Veronica and Kirkby, David and Knox, Lloyd and Kotov, Ivan V. and Krabbendam, Victor L. and Krughoff, K. Simon and Kub{\'a}nek, Petr and Kuczewski, John and Kulkarni, Shri and Ku, John and Kurita, Nadine R. and Lage, Craig S. and Lambert, Ron and Lange, Travis and Langton, J. Brian and Guillou, Laurent Le and Levine, Deborah and Liang, Ming and Lim, Kian-Tat and Lintott, Chris J. and Long, Kevin E. and Lopez, Margaux and Lotz, Paul J. and Lupton, Robert H. and Lust, Nate B. and MacArthur, Lauren A. and Mahabal, Ashish and Mandelbaum, Rachel and Markiewicz, Thomas W. and Marsh, Darren S. and Marshall, Philip J. and Marshall, Stuart and May, Morgan and McKercher, Robert and McQueen, Michelle and Meyers, Joshua and Migliore, Myriam and Miller, Michelle and Mills, David J. and Miraval, Connor and Moeyens, Joachim and Moolekamp, Fred E. and Monet, David G. and Moniez, Marc and Monkewitz, Serge and Montgomery, Christopher and Morrison, Christopher B. and Mueller, Fritz and Muller, Gary P. and Arancibia, Freddy Mu{\~n}oz and Neill, Douglas R. and Newbry, Scott P. and Nief, Jean-Yves and Nomerotski, Andrei and Nordby, Martin and O'Connor, Paul and Oliver, John and Olivier, Scot S. and Olsen, Knut and O'Mullane, William and Ortiz, Sandra and Osier, Shawn and Owen, Russell E. and Pain, Reynald and Palecek, Paul E. and Parejko, John K. and Parsons, James B. and Pease, Nathan M. and Peterson, J. Matt and Peterson, John R. and Petravick, Donald L. and Petrick, M. E. Libby and Petry, Cathy E. and Pierfederici, Francesco and Pietrowicz, Stephen and Pike, Rob and Pinto, Philip A. and Plante, Raymond and Plate, Stephen and Plutchak, Joel P. and Price, Paul A. and Prouza, Michael and Radeka, Veljko and Rajagopal, Jayadev and Rasmussen, Andrew P. and Regnault, Nicolas and Reil, Kevin A. and Reiss, David J. and Reuter, Michael A. and Ridgway, Stephen T. and Riot, Vincent J. and Ritz, Steve and Robinson, Sean and Roby, William and Roodman, Aaron and Rosing, Wayne and Roucelle, Cecille and Rumore, Matthew R. and Russo, Stefano and Saha, Abhijit and Sassolas, Benoit and Schalk, Terry L. and Schellart, Pim and Schindler, Rafe H. and Schmidt, Samuel and Schneider, Donald P. and Schneider, Michael D. and Schoening, William and Schumacher, German and Schwamb, Megan E. and Sebag, Jacques and Selvy, Brian and Sembroski, Glenn H. and Seppala, Lynn G. and Serio, Andrew and Serrano, Eduardo and Shaw, Richard A. and Shipsey, Ian and Sick, Jonathan and Silvestri, Nicole and Slater, Colin T. and Smith, J. Allyn and Smith, R. Chris and Sobhani, Shahram and Soldahl, Christine and {Storrie-Lombardi}, Lisa and Stover, Edward and Strauss, Michael A. and Street, Rachel A. and Stubbs, Christopher W. and Sullivan, Ian S. and Sweeney, Donald and Swinbank, John D. and Szalay, Alexander and Takacs, Peter and Tether, Stephen A. and Thaler, Jon J. and Thayer, John Gregg and Thomas, Sandrine and Thornton, Adam J. and Thukral, Vaikunth and Tice, Jeffrey and Trilling, David E. and Turri, Max and Berg, Richard Van and Berk, Daniel Vanden and Vetter, Kurt and Virieux, Francoise and Vucina, Tomislav and Wahl, William and Walkowicz, Lucianne and Walsh, Brian and Walter, Christopher W. and Wang, Daniel L. and Wang, Shin-Yawn and Warner, Michael and Wiecha, Oliver and Willman, Beth and Winters, Scott E. and Wittman, David and Wolff, Sidney C. and {Wood-Vasey}, W. Michael and Wu, Xiuqin and Xin, Bo and Yoachim, Peter and Zhan, Hu},
  year = {2019},
  month = mar,
  journal = {ApJ},
  volume = {873},
  number = {2},
  pages = {111},
  publisher = {IOP Publishing},
  issn = {0004-637X},
  doi = {10.3847/1538-4357/ab042c},
  urldate = {2022-06-18},
  langid = {english}
}

@article{jedrzejewskiCCDSurfacePhotometry1987,
  title = {{{CCD}} Surface Photometry of Elliptical Galaxies -- {{I}}. {{Observations}}, Reduction and Results},
  author = {Jedrzejewski, Robert I.},
  year = {1987},
  month = jun,
  journal = {MNRAS},
  volume = {226},
  number = {4},
  pages = {747--768},
  issn = {0035-8711},
  doi = {10.1093/mnras/226.4.747},
  urldate = {2024-11-28}
}

@article{jiangSLOANDIGITALSKY2014,
  title = {{{THE SLOAN DIGITAL SKY SURVEY STRIPE}} 82 {{IMAGING DATA}}: {{DEPTH-OPTIMIZED CO-ADDS OVER}} 300 Deg2 {{IN FIVE FILTERS}}},
  shorttitle = {{{THE SLOAN DIGITAL SKY SURVEY STRIPE}} 82 {{IMAGING DATA}}},
  author = {Jiang, Linhua and Fan, Xiaohui and Bian, Fuyan and McGreer, Ian D. and Strauss, Michael A. and Annis, James and Buck, Zo{\"e} and Green, Richard and Hodge, Jacqueline A. and Myers, Adam D. and Rafiee, Alireza and Richards, Gordon},
  year = {2014},
  month = jun,
  journal = {ApJS},
  volume = {213},
  number = {1},
  pages = {12},
  publisher = {The American Astronomical Society},
  issn = {0067-0049},
  doi = {10.1088/0067-0049/213/1/12},
  urldate = {2025-04-27},
  langid = {english}
}

@article{jiLifetimeMergerFeatures2014,
  title = {Lifetime of Merger Features of Equal-Mass Disk Mergers},
  author = {Ji, Inchan and Peirani, S{\'e}bastien and Yi, Sukyoung K.},
  year = {2014},
  month = jun,
  journal = {A\&A},
  volume = {566},
  pages = {A97},
  issn = {0004-6361, 1432-0746},
  doi = {10.1051/0004-6361/201423530},
  urldate = {2022-06-09}
}

@article{johnstonInterpretingDebrisSatellite2001,
  title = {Interpreting {{Debris}} from {{Satellite Disruption}} in {{External Galaxies}}},
  author = {Johnston, Kathryn V. and Sackett, Penny D. and Bullock, James S.},
  year = {2001},
  month = aug,
  journal = {ApJ},
  volume = {557},
  number = {1},
  pages = {137},
  publisher = {IOP Publishing},
  issn = {0004-637X},
  doi = {10.1086/321644},
  urldate = {2022-06-11},
  langid = {english}
}

@article{kado-fongTidalFeatures0052018,
  title = {Tidal {{Features}} at 0.05 {$<$} z {$<$} 0.45 in the {{Hyper Suprime-Cam Subaru Strategic Program}}: {{Properties}} and {{Formation Channels}}},
  author = {{Kado-Fong}, E and Greene, J E and Hendel, D and {Price-Whelan}, A M and Greco, J P and Goulding, A D and Huang, S and Johnston, K V and Komiyama, Y and Lee, C.-H. and Lust, N B and Strauss, M A and Tanaka, M},
  year = {2018},
  journal = {ApJ},
  volume = {866},
  number = {2},
  pages = {103--103},
  publisher = {American Astronomical Society},
  doi = {10.3847/1538-4357/aae0f0}
}

@article{karademirOuterStellarHalos2019,
  title = {The Outer Stellar Halos of Galaxies: How Radial Merger Mass Deposition, Shells, and Streams Depend on Infall-Orbit Configurations},
  shorttitle = {The Outer Stellar Halos of Galaxies},
  author = {Karademir, Geray S and Remus, Rhea-Silvia and Burkert, Andreas and Dolag, Klaus and Hoffmann, Tadziu L and Moster, Benjamin P and Steinwandel, Ulrich P and Zhang, Jielai},
  year = {2019},
  month = jul,
  journal = {MNRAS},
  volume = {487},
  number = {1},
  pages = {318--332},
  issn = {0035-8711, 1365-2966},
  doi = {10.1093/mnras/stz1251},
  urldate = {2022-06-09},
  langid = {english}
}

@article{katzGalaxiesGasCold1992,
  title = {Galaxies and {{Gas}} in a {{Cold Dark Matter Universe}}},
  author = {Katz, Neal and Hernquist, Lars and Weinberg, David H.},
  year = {1992},
  month = nov,
  journal = {ApJ},
  volume = {399},
  pages = {L109},
  issn = {0004-637X},
  doi = {10.1086/186619},
  urldate = {2023-08-02}
}

@article{kawinwanichakijHyperSuprimeCamSubaru2021,
  title = {Hyper {{Suprime-Cam Subaru Strategic Program}}: {{A Mass-dependent Slope}} of the {{Galaxy Size}}-{{Mass Relation}} at z {$<$} 1},
  shorttitle = {Hyper {{Suprime-Cam Subaru Strategic Program}}},
  author = {Kawinwanichakij, Lalitwadee and Silverman, John D. and Ding, Xuheng and George, Angelo and Damjanov, Ivana and Sawicki, Marcin and Tanaka, Masayuki and Taranu, Dan S. and Birrer, Simon and Huang, Song and Li, Junyao and Onodera, Masato and Shibuya, Takatoshi and Yasuda, Naoki},
  year = {2021},
  month = oct,
  journal = {ApJ},
  volume = {921},
  number = {1},
  pages = {38},
  publisher = {The American Astronomical Society},
  issn = {0004-637X},
  doi = {10.3847/1538-4357/ac1f21},
  urldate = {2024-09-03},
  langid = {english}
}

@article{kennicuttGlobalSchmidtLaw1998,
  title = {The {{Global Schmidt Law}} in {{Star-forming Galaxies}}},
  author = {Kennicutt, Jr., Robert C.},
  year = {1998},
  month = may,
  journal = {ApJ},
  volume = {498},
  pages = {541--552},
  issn = {0004-637X},
  doi = {10.1086/305588},
  urldate = {2023-08-02}
}

@article{khalidCharacterizingTidalFeatures2024,
  title = {Characterizing Tidal Features around Galaxies in Cosmological Simulations},
  author = {Khalid, A and Brough, S and Martin, G and Kimmig, L C and Lagos, C D P and Remus, R -S and {Martinez-Lombilla}, C},
  year = {2024},
  month = jun,
  journal = {MNRAS},
  volume = {530},
  number = {4},
  pages = {4422--4445},
  issn = {0035-8711},
  doi = {10.1093/mnras/stae1064},
  urldate = {2024-06-20}
}

@article{kimmEscapeFractionIonizing2014,
  title = {Escape {{Fraction}} of {{Ionizing Photons}} during {{Reionization}}: {{Effects}} Due to {{Supernova Feedback}} and {{Runaway OB Stars}}},
  shorttitle = {Escape {{Fraction}} of {{Ionizing Photons}} during {{Reionization}}},
  author = {Kimm, Taysun and Cen, Renyue},
  year = {2014},
  month = jun,
  journal = {ApJ},
  volume = {788},
  pages = {121},
  issn = {0004-637X},
  doi = {10.1088/0004-637X/788/2/121},
  urldate = {2023-08-02}
}

@article{kimmImpactLymanAlpha2018,
  title = {Impact of {{Lyman}} Alpha Pressure on Metal-Poor Dwarf Galaxies},
  author = {Kimm, Taysun and Haehnelt, Martin and Blaizot, J{\'e}r{\'e}my and Katz, Harley and {Michel-Dansac}, L{\'e}o and Garel, Thibault and Rosdahl, Joakim and Teyssier, Romain},
  year = {2018},
  month = apr,
  journal = {MNRAS},
  volume = {475},
  pages = {4617--4635},
  issn = {0035-8711},
  doi = {10.1093/mnras/sty126},
  urldate = {2023-08-02}
}

@article{kimmSimulatingStarFormation2015,
  title = {Towards Simulating Star Formation in Turbulent High-z Galaxies with Mechanical Supernova Feedback},
  author = {Kimm, Taysun and Cen, Renyue and Devriendt, Julien and Dubois, Yohan and Slyz, Adrianne},
  year = {2015},
  month = aug,
  journal = {MNRAS},
  volume = {451},
  pages = {2900--2921},
  issn = {0035-8711},
  doi = {10.1093/mnras/stv1211},
  urldate = {2023-08-02}
}

@article{komatsuSevenyearWilkinsonMicrowave2011,
  title = {Seven-Year {{Wilkinson Microwave Anisotropy Probe}} ({{WMAP}}) {{Observations}}: {{Cosmological Interpretation}}},
  author = {Komatsu, E and Smith, K.{\textasciitilde}M. and Dunkley, J and Bennett, C.{\textasciitilde}L. and Gold, B and Hinshaw, G and Jarosik, N and Larson, D and Nolta, M.{\textasciitilde}R. and Page, L and Spergel, D.{\textasciitilde}N. and Halpern, M and Hill, R.{\textasciitilde}S. and Kogut, A and Limon, M and Meyer, S.{\textasciitilde}S. and Odegard, N and Tucker, G.{\textasciitilde}S. and Weiland, J.{\textasciitilde}L. and Wollack, E and Wright, E.{\textasciitilde}L.},
  year = {2011},
  month = feb,
  journal = {ApJS},
  volume = {192},
  number = {2},
  pages = {18--18},
  doi = {10.1088/0067-0049/192/2/18}
}

@article{kroupaVariationInitialMass2001,
  title = {On the Variation of the Initial Mass Function},
  author = {Kroupa, Pavel},
  year = {2001},
  month = apr,
  journal = {MNRAS},
  volume = {322},
  number = {2},
  pages = {231--246},
  issn = {0035-8711},
  doi = {10.1046/j.1365-8711.2001.04022.x},
  urldate = {2024-11-28}
}

@article{lablancheATLAS3DProjectXII2012,
  title = {The {{ATLAS3D}} Project --- {{XII}}. {{Recovery}} of the Mass-to-Light Ratio of Simulated Early-Type Barred Galaxies with Axisymmetric Dynamical Models},
  author = {Lablanche, Pierre-Yves and Cappellari, Michele and Emsellem, Eric and Bournaud, Fr{\'e}d{\'e}ric and {Michel-Dansac}, Leo and Alatalo, Katherine and Blitz, Leo and Bois, Maxime and Bureau, Martin and Davies, Roger L. and Davis, Timothy A. and {de Zeeuw}, P. T. and Duc, Pierre-Alain and Khochfar, Sadegh and Krajnovi{\'c}, Davor and Kuntschner, Harald and Morganti, Raffaella and McDermid, Richard M. and Naab, Thorsten and Oosterloo, Tom and Sarzi, Marc and Scott, Nicholas and Serra, Paolo and Weijmans, Anne-Marie and Young, Lisa M.},
  year = {2012},
  month = aug,
  journal = {MNRAS},
  volume = {424},
  number = {2},
  pages = {1495--1521},
  issn = {0035-8711},
  doi = {10.1111/j.1365-2966.2012.21343.x},
  urldate = {2025-07-08}
}

@article{lacknerBuildingGalaxiesAccretion2012,
  title = {Building Galaxies by Accretion and in Situ Star Formation},
  author = {Lackner, C. N. and Cen, R. and Ostriker, J. P. and Joung, M. R.},
  year = {2012},
  month = sep,
  journal = {MNRAS},
  volume = {425},
  number = {1},
  pages = {641--656},
  issn = {0035-8711},
  doi = {10.1111/j.1365-2966.2012.21525.x},
  urldate = {2024-11-13}
}

@article{lagosConnectionMassEnvironment2018,
  title = {The Connection between Mass, Environment, and Slow Rotation in Simulated Galaxies},
  author = {Lagos, Claudia del P and Schaye, Joop and Bah{\'e}, Yannick and {Van de Sande}, Jesse and Kay, Scott T and Barnes, David and Davis, Timothy A and Dalla Vecchia, Claudio},
  year = {2018},
  month = jun,
  journal = {MNRAS},
  volume = {476},
  number = {4},
  pages = {4327--4345},
  doi = {10.1093/mnras/sty489}
}

@article{lagosDiverseStarFormation2025,
  title = {The Diverse Star Formation Histories of Early Massive, Quenched Galaxies in Modern Galaxy Formation Simulations},
  author = {Lagos, Claudia del P and Valentino, Francesco and Wright, Ruby J and {de~Graaff}, Anna and Glazebrook, Karl and De~Lucia, Gabriella and Robotham, Aaron S G and Nanayakkara, Themiya and {Chandro-Gomez}, Angel and Bravo, Mat{\'i}as and Baugh, Carlton M and Harborne, Katherine E and Hirschmann, Michaela and Fontanot, Fabio and Xie, Lizhi and Chittenden, Harry},
  year = {2025},
  month = jan,
  journal = {MNRAS},
  volume = {536},
  number = {3},
  pages = {2324--2354},
  issn = {0035-8711},
  doi = {10.1093/mnras/stae2626},
  urldate = {2025-01-13}
}

@article{lagosQuantifyingImpactMergers2018,
  title = {Quantifying the Impact of Mergers on the Angular Momentum of Simulated Galaxies},
  author = {Lagos, Claudia del P. and Stevens, Adam R. H. and Bower, Richard G. and Davis, Timothy A. and Contreras, Sergio and Padilla, Nelson D. and Obreschkow, Danail and Croton, Darren and Trayford, James W. and Welker, Charlotte and Theuns, Tom},
  year = {2018},
  month = feb,
  journal = {MNRAS},
  volume = {473},
  number = {4},
  pages = {4956--4974},
  issn = {0035-8711},
  doi = {10.1093/mnras/stx2667},
  urldate = {2025-04-28}
}

@misc{larry_bradley_2023_8248020,
  title = {Astropy/Photutils: 1.9.0},
  author = {Bradley, Larry and Sip{\H o}cz, Brigitta and Robitaille, Thomas and Tollerud, Erik and Vin{\'i}cius, Z{\'e} and Deil, Christoph and Barbary, Kyle and Wilson, Tom J and Busko, Ivo and Donath, Axel and G{\"u}nther, Hans Moritz and Cara, Mihai and Lim, P. L. and Me{\ss}linger, Sebastian and Conseil, Simon and Bostroem, Azalee and Droettboom, Michael and Bray, E. M. and Bratholm, Lars Andersen and Jamieson, William and Ginsburg, Adam and Barentsen, Geert and Craig, Matt and Morris, Brett M. and Perrin, Marshall and Rathi, Shivangee and Pascual, Sergio and Perren, Gabriel and Georgiev, Iskren Y. and Kerzendorf, Wolfgang},
  year = {2023},
  month = aug,
  doi = {10.5281/zenodo.8248020},
  howpublished = {Zenodo}
}

@article{larsonEvolutionDiskGalaxies1980,
  title = {The Evolution of Disk Galaxies and the Origin of {{S0}} Galaxies},
  author = {Larson, R. B. and Tinsley, B. M. and Caldwell, C. N.},
  year = {1980},
  month = may,
  journal = {ApJ},
  volume = {237},
  pages = {692--707},
  publisher = {IOP},
  issn = {0004-637X},
  doi = {10.1086/157917},
  urldate = {2025-01-20}
}

@article{lotzGalaxyMergerMorphologies2008,
  title = {Galaxy Merger Morphologies and Time-Scales from Simulations of Equal-Mass Gas-Rich Disc Mergers},
  author = {Lotz, Jennifer M. and Jonsson, Patrik and Cox, T. J. and Primack, Joel R.},
  year = {2008},
  month = dec,
  journal = {MNRAS},
  volume = {391},
  number = {3},
  pages = {1137--1162},
  issn = {0035-8711},
  doi = {10.1111/j.1365-2966.2008.14004.x},
  urldate = {2025-06-12}
}

@article{lotzMAJORMINORGALAXY2011,
  title = {{{THE MAJOR AND MINOR GALAXY MERGER RATES AT}} {\emph{z}} {$<$} 1.5},
  author = {Lotz, Jennifer M. and Jonsson, Patrik and Cox, T. J. and Croton, Darren and Primack, Joel R. and Somerville, Rachel S. and Stewart, Kyle},
  year = {2011},
  month = dec,
  journal = {ApJ},
  volume = {742},
  number = {2},
  pages = {103},
  issn = {0004-637X, 1538-4357},
  doi = {10.1088/0004-637X/742/2/103},
  urldate = {2022-08-10},
  langid = {english}
}

@article{ludlowSpuriousHeatingStellar2021,
  title = {Spurious Heating of Stellar Motions in Simulated Galactic Discs by Dark Matter Halo Particles},
  author = {Ludlow, Aaron D and Fall, S Michael and Schaye, Joop and Obreschkow, Danail},
  year = {2021},
  month = dec,
  journal = {MNRAS},
  volume = {508},
  number = {4},
  pages = {5114--5137},
  issn = {0035-8711},
  doi = {10.1093/mnras/stab2770},
  urldate = {2025-01-10}
}

@article{ludlowSpuriousHeatingStellar2023,
  title = {Spurious Heating of Stellar Motions by Dark Matter Particles in Cosmological Simulations of Galaxy Formation},
  author = {Ludlow, Aaron D and Fall, S Michael and Wilkinson, Matthew J and Schaye, Joop and Obreschkow, Danail},
  year = {2023},
  month = nov,
  journal = {MNRAS},
  volume = {525},
  number = {4},
  pages = {5614--5630},
  issn = {0035-8711},
  doi = {10.1093/mnras/stad2615},
  urldate = {2025-01-10}
}

@article{malinCatalogEllipticalGalaxies1983,
  title = {A Catalog of Elliptical Galaxies with Shells.},
  author = {Malin, D. F. and Carter, D.},
  year = {1983},
  month = nov,
  journal = {ApJ},
  volume = {274},
  pages = {534--540},
  issn = {0004-637X},
  doi = {10.1086/161467},
  urldate = {2022-04-11}
}

@article{mancillasProbingMergerHistory2019,
  title = {Probing the Merger History of Red Early-Type Galaxies with Their Faint Stellar Substructures},
  author = {Mancillas, Brisa and Duc, Pierre-Alain and Combes, Fran{\c c}oise and Bournaud, Fr{\'e}d{\'e}ric and Emsellem, Eric and Martig, Marie and {Michel-Dansac}, Leo},
  year = {2019},
  month = dec,
  journal = {A\&A},
  volume = {632},
  pages = {A122},
  publisher = {EDP Sciences},
  issn = {0004-6361, 1432-0746},
  doi = {10.1051/0004-6361/201936320},
  urldate = {2022-06-09},
  copyright = {{\copyright} B. Mancillas et al. 2019},
  langid = {english}
}

@article{marianColorGradientsReflect2018,
  title = {Color Gradients Reflect an Inside-out Growth in Early-Type Galaxies of the Cluster {{MACS J1206}}.2-0847},
  author = {Marian, V. and Ziegler, B. and Kuchner, U. and Verdugo, M.},
  year = {2018},
  month = sep,
  journal = {A\&A},
  volume = {617},
  pages = {A34},
  publisher = {EDP Sciences},
  issn = {0004-6361, 1432-0746},
  doi = {10.1051/0004-6361/201832750},
  urldate = {2024-09-02},
  copyright = {{\copyright} ESO 2018},
  langid = {english}
}

@article{marinacciFirstResultsIllustrisTNG2018,
  title = {First Results from the {{IllustrisTNG}} Simulations: Radio Haloes and Magnetic Fields},
  shorttitle = {First Results from the {{IllustrisTNG}} Simulations},
  author = {Marinacci, Federico and Vogelsberger, Mark and Pakmor, R{\"u}diger and Torrey, Paul and Springel, Volker and Hernquist, Lars and Nelson, Dylan and Weinberger, Rainer and Pillepich, Annalisa and Naiman, Jill and Genel, Shy},
  year = {2018},
  month = nov,
  journal = {MNRAS},
  volume = {480},
  pages = {5113--5139},
  issn = {0035-8711},
  doi = {10.1093/mnras/sty2206},
  urldate = {2023-05-02}
}

@article{martinez-lombillaGalaxyMassAssembly2023,
  title = {Galaxy {{And Mass Assembly}} ({{GAMA}}): Extended Intragroup Light in a Group at z~=~0.2 from Deep {{Hyper Suprime-Cam}} Images},
  shorttitle = {Galaxy {{And Mass Assembly}} ({{GAMA}})},
  author = {{Mart{\'i}nez-Lombilla}, Cristina and Brough, Sarah and Montes, Mireia and {Baena-Gall{\'e}}, Roberto and Akhlaghi, Mohammad and {Infante-Sainz}, Ra{\'u}l and Driver, Simon P and Holwerda, Benne W and Pimbblet, Kevin A and Robotham, Aaron S G},
  year = {2023},
  month = jan,
  journal = {MNRAS},
  volume = {518},
  number = {1},
  pages = {1195--1213},
  issn = {0035-8711},
  doi = {10.1093/mnras/stac3119},
  urldate = {2025-01-29}
}

@article{martinLimitedRoleGalaxy2017,
  title = {The Limited Role of Galaxy Mergers in Driving Stellar Mass Growth over Cosmic Time},
  author = {Martin, G. and Kaviraj, S. and Devriendt, J. E. G. and Dubois, Y. and Laigle, C. and Pichon, C.},
  year = {2017},
  month = nov,
  journal = {MNRAS},
  volume = {472},
  number = {1},
  pages = {L50-L54},
  issn = {1745-3925},
  doi = {10.1093/mnrasl/slx136},
  urldate = {2024-12-18}
}

@article{martinPreparingLowSurface2022,
  title = {Preparing for Low Surface Brightness Science with the {{Vera C}}. {{Rubin Observatory}}: {{Characterization}} of Tidal Features from Mock Images},
  shorttitle = {Preparing for Low Surface Brightness Science with the {{Vera C}}. {{Rubin Observatory}}},
  author = {Martin, G. and Bazkiaei, A. E. and Spavone, M. and Iodice, E. and Mihos, J. C. and Montes, M. and Benavides, J. A. and Brough, S. and Carlin, J. L. and Collins, C. A. and Duc, P. A. and G{\'o}mez, F. A. and Galaz, G. and {Hern{\'a}ndez-Toledo}, H. M. and Jackson, R. A. and Kaviraj, S. and Knapen, J. H. and {Mart{\'i}nez-Lombilla}, C. and McGee, S. and O'Ryan, D. and Prole, D. J. and Rich, R. M. and Rom{\'a}n, J. and Shah, E. A. and Starkenburg, T. K. and Watkins, A. E. and Zaritsky, D. and Pichon, C. and Armus, L. and Bianconi, M. and Buitrago, F. and Bus{\'a}, I. and Davis, F. and Demarco, R. and Desmons, A. and Garc{\'i}a, P. and Graham, A. W. and Holwerda, B. and Hon, D. S. -H. and Khalid, A. and Klehammer, J. and Klutse, D. Y. and Lazar, I. and Nair, P. and {Noakes-Kettel}, E. A. and Rutkowski, M. and Saha, K. and Sahu, N. and Sola, E. and {V{\'a}zquez-Mata}, J. A. and {Vera-Casanova}, A. and Yoon, I.},
  year = {2022},
  month = jun,
  journal = {MNRAS},
  volume = {513},
  pages = {1459--1487},
  issn = {0035-8711},
  doi = {10.1093/mnras/stac1003},
  urldate = {2022-06-03}
}

@article{martinRoleMergersDriving2018,
  title = {The Role of Mergers in Driving Morphological Transformation over Cosmic Time},
  author = {Martin, G. and Kaviraj, S. and Devriendt, J. E. G. and Dubois, Y. and Pichon, C.},
  year = {2018},
  month = oct,
  journal = {MNRAS},
  volume = {480},
  number = {2},
  pages = {2266--2283},
  issn = {0035-8711},
  doi = {10.1093/mnras/sty1936},
  urldate = {2022-06-30}
}

@article{martinRoleMergersInteractions2021,
  title = {The Role of Mergers and Interactions in Driving the Evolution of Dwarf Galaxies over Cosmic Time},
  author = {Martin, G. and Jackson, R. A. and Kaviraj, S. and Choi, H. and Devriendt, J. E. G. and Dubois, Y. and Kimm, T. and Kraljic, K. and Peirani, S. and Pichon, C. and Volonteri, M. and Yi, S. K.},
  year = {2021},
  month = jan,
  journal = {MNRAS},
  volume = {500},
  pages = {4937--4957},
  issn = {0035-8711},
  doi = {10.1093/mnras/staa3443},
  urldate = {2024-02-01}
}

@article{mckeeTheoryStarFormation2007,
  title = {Theory of {{Star Formation}}},
  author = {McKee, Christopher F. and Ostriker, Eve C.},
  year = {2007},
  month = sep,
  journal = {ARA\&A},
  volume = {45},
  number = {Volume 45, 2007},
  pages = {565--687},
  publisher = {Annual Reviews},
  issn = {0066-4146, 1545-4282},
  doi = {10.1146/annurev.astro.45.051806.110602},
  urldate = {2024-11-17},
  langid = {english}
}

@inproceedings{mckinney-proc-scipy-2010,
  title = {Data {{Structures}} for {{Statistical Computing}} in {{Python}}},
  booktitle = {Proceedings of the 9th {{Python}} in {{Science Conference}}},
  author = {McKinney, Wes},
  editor = {{van der Walt}, St{\'e}fan and Millman, Jarrod},
  year = {2010},
  pages = {56--61},
  doi = {10.25080/Majora-92bf1922-00a}
}

@article{merrittDRAGONFLYNEARBYGALAXIES2016,
  title = {{{THE DRAGONFLY NEARBY GALAXIES SURVEY}}. {{II}}. {{ULTRA-DIFFUSE GALAXIES NEAR THE ELLIPTICAL GALAXY NGC}} 5485},
  author = {Merritt, Allison and {van Dokkum}, Pieter and Danieli, Shany and Abraham, Roberto and Zhang, Jielai and Karachentsev, I. D. and Makarova, L. N.},
  year = {2016},
  month = dec,
  journal = {The Astrophysical Journal},
  volume = {833},
  number = {2},
  pages = {168--168},
  publisher = {American Astronomical Society},
  doi = {10.3847/1538-4357/833/2/168}
}

@article{merrittMissingOutskirtsProblem2020,
  title = {A Missing Outskirts Problem? {{Comparisons}} between Stellar Haloes in the {{Dragonfly Nearby Galaxies Survey}} and the {{TNG100}} Simulation},
  author = {Merritt, Allison and Pillepich, Annalisa and {van Dokkum}, Pieter and Nelson, Dylan and Hernquist, Lars and Marinacci, Federico and Vogelsberger, Mark and Merritt, Allison and Pillepich, Annalisa and {van Dokkum}, Pieter and Nelson, Dylan and Hernquist, Lars and Marinacci, Federico and Vogelsberger, Mark},
  year = {2020},
  journal = {MNRAS},
  volume = {495},
  number = {4},
  pages = {4570--4604},
  publisher = {Oxford University Press},
  doi = {10.1093/MNRAS/STAA1164}
}

@article{millerColorGradientsHalfmass2023,
  title = {Color {{Gradients}} and {{Half-mass Radii}} of {{Galaxies Out}} to z = 2 in the {{CANDELS}}/{{3D-HST Fields}}: {{Further Evidence}} for {{Important Differences}} in the {{Evolution}} of {{Mass-weighted}} and {{Light-weighted Sizes}}},
  shorttitle = {Color {{Gradients}} and {{Half-mass Radii}} of {{Galaxies Out}} to z = 2 in the {{CANDELS}}/{{3D-HST Fields}}},
  author = {Miller, Tim B. and van Dokkum, Pieter and Mowla, Lamiya},
  year = {2023},
  month = mar,
  journal = {ApJ},
  volume = {945},
  number = {2},
  pages = {155},
  publisher = {The American Astronomical Society},
  issn = {0004-637X},
  doi = {10.3847/1538-4357/acbc74},
  urldate = {2024-08-01},
  langid = {english}
}

@article{miskolcziTidalStreamsGalaxies2011,
  title = {Tidal Streams around Galaxies in the {{SDSS DR7}} Archive: {{I}}. {{First}} Results},
  shorttitle = {Tidal Streams around Galaxies in the {{SDSS DR7}} Archive},
  author = {Miskolczi, A. and Bomans, D. J. and Dettmar, R.-J.},
  year = {2011},
  month = dec,
  journal = {A\&A},
  volume = {536},
  pages = {A66},
  issn = {0004-6361, 1432-0746},
  doi = {10.1051/0004-6361/201116716},
  urldate = {2022-06-09}
}

@article{miyazakiHyperSuprimeCamSystem2018,
  title = {Hyper {{Suprime-Cam}}: {{System}} Design and Verification of Image Quality},
  shorttitle = {Hyper {{Suprime-Cam}}},
  author = {Miyazaki, Satoshi and Komiyama, Yutaka and Kawanomoto, Satoshi and Doi, Yoshiyuki and Furusawa, Hisanori and Hamana, Takashi and Hayashi, Yusuke and Ikeda, Hiroyuki and Kamata, Yukiko and Karoji, Hiroshi and Koike, Michitaro and Kurakami, Tomio and Miyama, Shoken and Morokuma, Tomoki and Nakata, Fumiaki and Namikawa, Kazuhito and Nakaya, Hidehiko and Nariai, Kyoji and Obuchi, Yoshiyuki and Oishi, Yukie and Okada, Norio and Okura, Yuki and Tait, Philip and Takata, Tadafumi and Tanaka, Yoko and Tanaka, Masayuki and Terai, Tsuyoshi and Tomono, Daigo and Uraguchi, Fumihiro and Usuda, Tomonori and Utsumi, Yousuke and Yamada, Yoshihiko and Yamanoi, Hitomi and Aihara, Hiroaki and Fujimori, Hiroki and Mineo, Sogo and Miyatake, Hironao and Oguri, Masamune and Uchida, Tomohisa and Tanaka, Manobu M and Yasuda, Naoki and Takada, Masahiro and Murayama, Hitoshi and Nishizawa, Atsushi J and Sugiyama, Naoshi and Chiba, Masashi and Futamase, Toshifumi and Wang, Shiang-Yu and Chen, Hsin-Yo and Ho, Paul T P and Liaw, Eric J Y and Chiu, Chi-Fang and Ho, Cheng-Lin and Lai, Tsang-Chih and Lee, Yao-Cheng and Jeng, Dun-Zen and Iwamura, Satoru and Armstrong, Robert and Bickerton, Steve and Bosch, James and Gunn, James E and Lupton, Robert H and Loomis, Craig and Price, Paul and Smith, Steward and Strauss, Michael A and Turner, Edwin L and Suzuki, Hisanori and Miyazaki, Yasuhito and Muramatsu, Masaharu and Yamamoto, Koei and Endo, Makoto and Ezaki, Yutaka and Ito, Noboru and Kawaguchi, Noboru and Sofuku, Satoshi and Taniike, Tomoaki and Akutsu, Kotaro and Dojo, Naoto and Kasumi, Kazuyuki and Matsuda, Toru and Imoto, Kohei and Miwa, Yoshinori and Suzuki, Masayuki and Takeshi, Kunio and Yokota, Hideo},
  year = {2018},
  month = jan,
  journal = {PASJ},
  volume = {70},
  number = {SP1},
  pages = {S1},
  issn = {0004-6264},
  doi = {10.1093/pasj/psx063},
  urldate = {2025-01-10}
}

@article{monnetModellingStellarIntensity1992,
  title = {Modelling the Stellar Intensity and Radial Velocity Fields in Triaxial Galaxies by Sums of {{Gaussian}} Functions.},
  author = {Monnet, G. and Bacon, R. and Emsellem, E.},
  year = {1992},
  month = jan,
  journal = {A\&A},
  volume = {253},
  pages = {366--373},
  issn = {0004-6361},
  urldate = {2025-04-07}
}

@article{montesBuildupIntraclusterLight2021,
  title = {The {{Buildup}} of the {{Intracluster Light}} of {{A85}} as {{Seen}} by {{Subaru}}'s {{Hyper Suprime-Cam}}},
  author = {Montes, Mireia and Brough, Sarah and Owers, Matt S. and Santucci, Giulia},
  year = {2021},
  month = mar,
  journal = {ApJ},
  volume = {910},
  number = {1},
  pages = {45},
  publisher = {American Astronomical Society},
  issn = {0004-637X},
  doi = {10.3847/1538-4357/abddb6},
  urldate = {2022-06-18},
  langid = {english}
}

@article{mooreMorphologicalTransformationGalaxy1998,
  title = {Morphological {{Transformation}} from {{Galaxy Harassment}}},
  author = {Moore, Ben and Lake, George and Katz, Neal},
  year = {1998},
  month = mar,
  journal = {ApJ},
  volume = {495},
  number = {1},
  pages = {139},
  publisher = {IOP Publishing},
  issn = {0004-637X},
  doi = {10.1086/305264},
  urldate = {2025-01-20},
  langid = {english}
}

@misc{munMAGPISurveyRadial2024,
  title = {The {{MAGPI Survey}}: Radial Trends in Star Formation across Different Cosmological Simulations in Comparison with Observations at \$z {\textbackslash}sim\$ 0.3},
  shorttitle = {The {{MAGPI Survey}}},
  author = {Mun, Marcie and Wisnioski, Emily and Harborne, Katherine E. and Lagos, Claudia D. P. and Valenzuela, Lucas M. and Remus, Rhea-Silvia and Mendel, J. Trevor and Battisti, Andrew J. and Ellison, Sara L. and Foster, Caroline and Bravo, Matias and Brough, Sarah and Croom, Scott M. and Gao, Tianmu and Grasha, Kathryn and Gupta, Anshu and Mai, Yifan and Mailvaganam, Anilkumar and Muller, Eric G. M. and Sharma, Gauri and Sweet, Sarah M. and Taylor, Edward N. and Zafar, Tayyaba},
  year = {2024},
  month = nov,
  number = {arXiv:2411.17882},
  eprint = {2411.17882},
  primaryclass = {astro-ph},
  publisher = {arXiv},
  doi = {10.48550/arXiv.2411.17882},
  urldate = {2025-01-24},
  archiveprefix = {arXiv}
}

@article{naabMINORMERGERSSIZE2009,
  title = {{{MINOR MERGERS AND THE SIZE EVOLUTION OF ELLIPTICAL GALAXIES}}},
  author = {Naab, Thorsten and Johansson, Peter H. and Ostriker, Jeremiah P.},
  year = {2009},
  month = jun,
  journal = {ApJ},
  volume = {699},
  number = {2},
  pages = {L178},
  publisher = {The American Astronomical Society},
  issn = {0004-637X},
  doi = {10.1088/0004-637X/699/2/L178},
  urldate = {2024-09-09},
  langid = {english}
}

@article{naimanFirstResultsIllustrisTNG2018,
  title = {First Results from the {{IllustrisTNG}} Simulations: A Tale of Two Elements - Chemical Evolution of Magnesium and Europium},
  shorttitle = {First Results from the {{IllustrisTNG}} Simulations},
  author = {Naiman, Jill P. and Pillepich, Annalisa and Springel, Volker and {Ramirez-Ruiz}, Enrico and Torrey, Paul and Vogelsberger, Mark and Pakmor, R{\"u}diger and Nelson, Dylan and Marinacci, Federico and Hernquist, Lars and Weinberger, Rainer and Genel, Shy},
  year = {2018},
  month = jun,
  journal = {MNRAS},
  volume = {477},
  pages = {1206--1224},
  issn = {0035-8711},
  doi = {10.1093/mnras/sty618},
  urldate = {2023-05-02}
}

@article{nelsonFirstResultsIllustrisTNG2018,
  title = {First Results from the {{IllustrisTNG}} Simulations: The Galaxy Colour Bimodality},
  shorttitle = {First Results from the {{IllustrisTNG}} Simulations},
  author = {Nelson, Dylan and Pillepich, Annalisa and Springel, Volker and Weinberger, Rainer and Hernquist, Lars and Pakmor, R{\"u}diger and Genel, Shy and Torrey, Paul and Vogelsberger, Mark and Kauffmann, Guinevere and Marinacci, Federico and Naiman, Jill},
  year = {2018},
  month = mar,
  journal = {MNRAS},
  volume = {475},
  pages = {624--647},
  issn = {0035-8711},
  doi = {10.1093/mnras/stx3040},
  urldate = {2023-05-02}
}

@article{nelsonIllustrisTNGSimulationsPublic2019,
  title = {The {{IllustrisTNG}} Simulations: Public Data Release},
  shorttitle = {The {{IllustrisTNG}} Simulations},
  author = {Nelson, Dylan and Springel, Volker and Pillepich, Annalisa and {Rodriguez-Gomez}, Vicente and Torrey, Paul and Genel, Shy and Vogelsberger, Mark and Pakmor, Ruediger and Marinacci, Federico and Weinberger, Rainer and Kelley, Luke and Lovell, Mark and Diemer, Benedikt and Hernquist, Lars},
  year = {2019},
  month = may,
  journal = {Computational Astrophysics and Cosmology},
  volume = {6},
  pages = {2},
  doi = {10.1186/s40668-019-0028-x},
  urldate = {2022-06-02}
}

@misc{niemitaloAnswerHowDoes2015,
  title = {Answer to "{{How Does Gaussian Blur Affect Image Variance}}"},
  author = {Niemitalo, Olli},
  year = {2015},
  month = nov,
  journal = {Signal Processing Stack Exchange},
  urldate = {2025-09-18},
  howpublished = {https://dsp.stackexchange.com/a/26891}
}

@article{ohImpactGalaxyMergers2019,
  title = {Impact of Galaxy Mergers on the Colours of Cluster Galaxies},
  author = {Oh, Sree and Kim, Keunho and Lee, Joon Hyeop and Kim, Minjin and Sheen, Yun-Kyeong and Rhee, Jinsu and Ree, Chang H and Jeong, Hyunjin and Ho, Luis C and Kyeong, Jaemann and Sung, Eon-Chang and Park, Byeong-Gon and Yi, Sukyoung K},
  year = {2019},
  month = sep,
  journal = {MNRAS},
  volume = {488},
  number = {3},
  pages = {4169--4180},
  issn = {0035-8711},
  doi = {10.1093/mnras/stz1920},
  urldate = {2024-09-02}
}

@inproceedings{olivierOpticalDesignLSST2008a,
  title = {Optical Design of the {{LSST}} Camera},
  booktitle = {{{SPIE Astronomical Telescopes}} + {{Instrumentation}}},
  author = {Olivier, Scot S. and Seppala, Lynn and Gilmore, Kirk},
  editor = {{Atad-Ettedgui}, Eli and Lemke, Dietrich},
  year = {2008},
  month = jul,
  pages = {70182G},
  address = {Marseille, France},
  doi = {10.1117/12.790264},
  urldate = {2022-06-14}
}

@article{oserTWOPHASESGALAXY2010,
  title = {{{THE TWO PHASES OF GALAXY FORMATION}}},
  author = {Oser, Ludwig and Ostriker, Jeremiah P. and Naab, Thorsten and Johansson, Peter H. and Burkert, Andreas},
  year = {2010},
  month = dec,
  journal = {ApJ},
  volume = {725},
  number = {2},
  pages = {2312},
  publisher = {The American Astronomical Society},
  issn = {0004-637X},
  doi = {10.1088/0004-637X/725/2/2312},
  urldate = {2024-11-13},
  langid = {english}
}

@article{ostrikerGalaxyFormationIntergalactic1981,
  title = {Galaxy Formation in an Intergalactic Medium Dominated by Explosions},
  author = {Ostriker, J. P. and Cowie, L. L.},
  year = {1981},
  month = feb,
  journal = {ApJ},
  volume = {243},
  pages = {L127-L131},
  publisher = {IOP},
  issn = {0004-637X},
  doi = {10.1086/183458},
  urldate = {2025-01-10}
}

@article{pillepichFirstResultsIllustristng2018,
  title = {First Results from the Illustristng Simulations: {{The}} Stellar Mass Content of Groups and Clusters of Galaxies},
  shorttitle = {First Results from the {{IllustrisTNG}} Simulations},
  author = {Pillepich, Annalisa and Nelson, Dylan and Hernquist, Lars and Springe, Volker and {R{\"u}diger Pakmor} and Torrey, Paul and Weinberger, Rainer and Gene, Shy and Naiman, Jill P. and Marinacci, Federico and Vogelsberger, Mark},
  year = {2018},
  month = mar,
  journal = {MNRAS},
  volume = {475},
  number = {1},
  pages = {648--675},
  publisher = {Oxford University Press},
  issn = {0035-8711, 1365-2966},
  doi = {10.1093/mnras/stx3112},
  langid = {english}
}

@article{pillepichSimulatingGalaxyFormation2018,
  title = {Simulating Galaxy Formation with the {{IllustrisTNG}} Model},
  author = {Pillepich, Annalisa and Springel, Volker and Nelson, Dylan and Genel, Shy and Naiman, Jill and Pakmor, R{\"u}diger and Hernquist, Lars and Torrey, Paul and Vogelsberger, Mark and Weinberger, Rainer and Marinacci, Federico},
  year = {2018},
  month = jan,
  journal = {MNRAS},
  volume = {473},
  number = {3},
  pages = {4077--4106},
  issn = {0035-8711},
  doi = {10.1093/mnras/stx2656},
  urldate = {2022-06-13}
}

@article{popFormationIncidenceShell2018,
  title = {Formation and Incidence of Shell Galaxies in the {{Illustris}} Simulation},
  author = {Pop, Ana-Roxana and Pillepich, Annalisa and Amorisco, Nicola C and Hernquist, Lars},
  year = {2018},
  month = oct,
  journal = {MNRAS},
  volume = {480},
  number = {2},
  pages = {1715--1739},
  issn = {0035-8711, 1365-2966},
  doi = {10.1093/mnras/sty1932},
  urldate = {2022-06-09},
  langid = {english}
}

@article{popGalaxiesShellsIllustris2017,
  title = {Galaxies with {{Shells}} in the {{Illustris Simulation}}: {{Metallicity Signatures}}},
  shorttitle = {Galaxies with {{Shells}} in the {{Illustris Simulation}}},
  author = {Pop, Ana-Roxana and Pillepich, Annalisa and Amorisco, Nicola C. and Hernquist, Lars},
  year = {2017},
  month = aug,
  journal = {Galaxies},
  volume = {5},
  number = {3},
  eprint = {1708.01615},
  primaryclass = {astro-ph},
  pages = {34},
  issn = {2075-4434},
  doi = {10.3390/galaxies5030034},
  urldate = {2023-05-11},
  archiveprefix = {arXiv},
  langid = {english}
}

@article{pressFormationGalaxiesClusters1974,
  title = {Formation of {{Galaxies}} and {{Clusters}} of {{Galaxies}} by {{Self-Similar Gravitational Condensation}}},
  author = {Press, William H and Schechter, Paul},
  year = {1974},
  month = feb,
  journal = {ApJ},
  volume = {187},
  pages = {425--438},
  doi = {10.1086/152650}
}

@article{ragagninWebPortalHydrodynamical2017,
  title = {A Web Portal for Hydrodynamical, Cosmological Simulations},
  author = {Ragagnin, A. and Dolag, K. and Biffi, V. and Cadolle Bel, M. and Hammer, N. J. and Krukau, A. and Petkova, M. and Steinborn, D.},
  year = {2017},
  month = jul,
  journal = {Astronomy and Computing, Volume 20, p. 52-67.},
  volume = {20},
  pages = {52},
  issn = {2213-1337},
  doi = {10.1016/j.ascom.2017.05.001},
  urldate = {2022-04-12},
  langid = {english}
}

@article{remusAccretedNotAccreted2022,
  title = {Accreted or {{Not Accreted}}? {{The Fraction}} of {{Accreted Mass}} in {{Galaxies}} from the {{Magneticum Simulations}} and {{Observations}}},
  shorttitle = {Accreted or {{Not Accreted}}?},
  author = {Remus, Rhea-Silvia and Forbes, Duncan A.},
  year = {2022},
  month = aug,
  journal = {ApJ},
  volume = {935},
  number = {1},
  pages = {37},
  publisher = {The American Astronomical Society},
  issn = {0004-637X},
  doi = {10.3847/1538-4357/ac7b30},
  urldate = {2024-03-22},
  langid = {english}
}

@article{robertsonGalaxyFormationEvolution2019,
  title = {Galaxy Formation and Evolution Science in the Era of the {{Large Synoptic Survey Telescope}}},
  author = {Robertson, Brant E. and Banerji, Manda and Brough, Sarah and Davies, Roger L. and Ferguson, Henry C. and Hausen, Ryan and Kaviraj, Sugata and Newman, Jeffrey A. and Schmidt, Samuel J. and Tyson, J. Anthony and Wechsler, Risa H.},
  year = {2019},
  month = jun,
  journal = {Nature Reviews Physics},
  volume = {1},
  pages = {450--462},
  doi = {10.1038/s42254-019-0067-x},
  urldate = {2023-05-11}
}

@article{robothamGalaxyMassAssembly2014,
  title = {Galaxy {{And Mass Assembly}} ({{GAMA}}): Galaxy Close Pairs, Mergers and the Future Fate of Stellar Mass},
  shorttitle = {Galaxy {{And Mass Assembly}} ({{GAMA}})},
  author = {Robotham, A. S. G. and Driver, S. P. and Davies, L. J. M. and Hopkins, A. M. and Baldry, I. K. and Agius, N. K. and Bauer, A. E. and {Bland-Hawthorn}, J. and Brough, S. and Brown, M. J. I. and Cluver, M. and De Propris, R. and Drinkwater, M. J. and Holwerda, B. W. and Kelvin, L. S. and {Lara-Lopez}, M. A. and Liske, J. and {L{\'o}pez-S{\'a}nchez}, {\'A}. R. and Loveday, J. and Mahajan, S. and {McNaught-Roberts}, T. and Moffett, A. and Norberg, P. and Obreschkow, D. and Owers, M. S. and Penny, S. J. and Pimbblet, K. and Prescott, M. and Taylor, E. N. and {van Kampen}, E. and Wilkins, S. M.},
  year = {2014},
  month = nov,
  journal = {Monthly Notices of the Royal Astronomical Society},
  volume = {444},
  number = {4},
  pages = {3986--4008},
  issn = {0035-8711},
  doi = {10.1093/mnras/stu1604},
  urldate = {2022-06-10}
}

@article{rodriguez-gomezOpticalMorphologiesGalaxies2019,
  title = {The Optical Morphologies of Galaxies in the {{IllustrisTNG}} Simulation: A Comparison to {{Pan-STARRS}} Observations},
  shorttitle = {The Optical Morphologies of Galaxies in the {{IllustrisTNG}} Simulation},
  author = {{Rodriguez-Gomez}, Vicente and Snyder, Gregory F and Lotz, Jennifer M and Nelson, Dylan and Pillepich, Annalisa and Springel, Volker and Genel, Shy and Weinberger, Rainer and Tacchella, Sandro and Pakmor, R{\"u}diger and Torrey, Paul and Marinacci, Federico and Vogelsberger, Mark and Hernquist, Lars and Thilker, David A},
  year = {2019},
  month = mar,
  journal = {MNRAS},
  volume = {483},
  number = {3},
  pages = {4140--4159},
  issn = {0035-8711},
  doi = {10.1093/mnras/sty3345},
  urldate = {2024-11-25}
}

@article{rodriguez-gomezStellarMassAssembly2016,
  title = {The Stellar Mass Assembly of Galaxies in the {{Illustris}} Simulation: Growth by Mergers and the Spatial Distribution of Accreted Stars},
  shorttitle = {The Stellar Mass Assembly of Galaxies in the {{Illustris}} Simulation},
  author = {{Rodriguez-Gomez}, Vicente and Pillepich, Annalisa and Sales, Laura V. and Genel, Shy and Vogelsberger, Mark and Zhu, Qirong and Wellons, Sarah and Nelson, Dylan and Torrey, Paul and Springel, Volker and Ma, Chung-Pei and Hernquist, Lars},
  year = {2016},
  month = may,
  journal = {MNRAS},
  volume = {458},
  number = {3},
  pages = {2371--2390},
  issn = {0035-8711, 1365-2966},
  doi = {10.1093/mnras/stw456},
  urldate = {2022-06-09},
  langid = {english}
}

@article{romanGalacticCirriDeep2020,
  title = {Galactic Cirri in Deep Optical Imaging},
  author = {Rom{\'a}n, Javier and Trujillo, Ignacio and Montes, Mireia},
  year = {2020},
  month = dec,
  journal = {A\&A},
  volume = {644},
  pages = {A42},
  issn = {0004-6361, 1432-0746},
  doi = {10.1051/0004-6361/201936111},
  urldate = {2022-07-13},
  langid = {english}
}

@article{rutherfordSAMIGalaxySurvey2024,
  title = {The {{SAMI Galaxy Survey}}: Using Tidal Streams and Shells to Trace the Dynamical Evolution of Massive Galaxies},
  shorttitle = {The {{SAMI Galaxy Survey}}},
  author = {Rutherford, Tomas H. and {van de Sande}, Jesse and Croom, Scott M. and Valenzuela, Lucas M. and Remus, Rhea-Silvia and D'Eugenio, Francesco and Vaughan, Sam P. and Zovaro, Henry R. M. and Casura, Sarah and Barsanti, Stefania and {Bland-Hawthorn}, Joss and Brough, Sarah and Bryant, Julia J. and Goodwin, Michael and Lorente, Nuria and Oh, Sree and Ristea, Andrei},
  year = {2024},
  month = apr,
  journal = {MNRAS},
  volume = {529},
  pages = {810--830},
  issn = {0035-8711},
  doi = {10.1093/mnras/stae398},
  urldate = {2024-03-12}
}

@article{rydenGALAXYFORMATIONGRAVITATIONAL1987,
  title = {{{GALAXY FORMATION BY GRAVITATIONAL COLLAPSE}}},
  author = {Ryden, Barbara S and Gunn, James E},
  year = {1987},
  journal = {ApJ},
  volume = {318},
  pages = {15--31}
}

@article{santucciSAMIGalaxySurvey2020,
  title = {The {{SAMI Galaxy Survey}}: {{Stellar Population Gradients}} of {{Central Galaxies}}},
  shorttitle = {The {{SAMI Galaxy Survey}}},
  author = {Santucci, Giulia and Brough, Sarah and Scott, Nicholas and Montes, Mireia and Owers, Matt S. and van Sande, Jesse and {Bland-Hawthorn}, Joss and Bryant, Julia J. and Croom, Scott M. and Ferreras, Ignacio and Lawrence, Jon S. and {L{\'o}pez-S{\'a}nchez}, {\'A}ngel R. and Richards, Samuel N.},
  year = {2020},
  month = jun,
  journal = {ApJ},
  volume = {896},
  number = {1},
  pages = {75},
  publisher = {The American Astronomical Society},
  issn = {0004-637X},
  doi = {10.3847/1538-4357/ab92a9},
  urldate = {2024-11-26},
  langid = {english}
}

@article{schayeEAGLEProjectSimulating2015,
  title = {The {{EAGLE}} Project: {{Simulating}} the Evolution and Assembly of Galaxies and Their Environments},
  author = {Schaye, Joop and Crain, Robert A. and Bower, Richard G. and Furlong, Michelle and Schaller, Matthieu and Theuns, Tom and Dalla Vecchia, Claudio and Frenk, Carlos S. and Mccarthy, I. G. and Helly, John C. and Jenkins, Adrian and {Rosas-Guevara}, Y. M. and White, Simon D.M. and Baes, Maarten and Booth, C. M. and Camps, Peter and Navarro, Julio F. and Qu, Yan and Rahmati, Alireza and Sawala, Till and Thomas, Peter A. and Trayford, James},
  year = {2015},
  month = jan,
  journal = {MNRAS},
  volume = {446},
  number = {1},
  pages = {521--554},
  publisher = {Oxford University Press},
  doi = {10.1093/mnras/stu2058}
}

@article{schayeRelationSchmidtKennicuttSchmidt2008,
  title = {On the Relation between the {{Schmidt}} and {{Kennicutt-Schmidt}} Star Formation Laws and Its Implications for Numerical Simulations},
  author = {Schaye, Joop and Dalla Vecchia, Claudio},
  year = {2008},
  month = jan,
  journal = {MNRAS},
  volume = {383},
  pages = {1210--1222},
  issn = {0035-8711},
  doi = {10.1111/j.1365-2966.2007.12639.x},
  urldate = {2023-08-02}
}

@article{schayeStarFormationThresholds2004,
  title = {Star {{Formation Thresholds}} and {{Galaxy Edges}}: {{Why}} and {{Where}}},
  shorttitle = {Star {{Formation Thresholds}} and {{Galaxy Edges}}},
  author = {Schaye, Joop},
  year = {2004},
  month = jul,
  journal = {ApJ},
  volume = {609},
  pages = {667--682},
  issn = {0004-637X},
  doi = {10.1086/421232},
  urldate = {2023-08-02}
}

@article{schlegelMapsDustInfrared1998,
  title = {Maps of {{Dust Infrared Emission}} for {{Use}} in {{Estimation}} of {{Reddening}} and {{Cosmic Microwave Background Radiation Foregrounds}}},
  author = {Schlegel, David J. and Finkbeiner, Douglas P. and Davis, Marc},
  year = {1998},
  month = jun,
  journal = {ApJ},
  volume = {500},
  number = {2},
  pages = {525},
  publisher = {IOP Publishing},
  issn = {0004-637X},
  doi = {10.1086/305772},
  urldate = {2025-01-14},
  langid = {english}
}

@article{schmidtRateStarFormation1959,
  title = {The {{Rate}} of {{Star Formation}}.},
  author = {Schmidt, Maarten},
  year = {1959},
  month = mar,
  journal = {ApJ},
  volume = {129},
  pages = {243},
  issn = {0004-637X},
  doi = {10.1086/146614},
  urldate = {2023-08-02}
}

@article{scottSAURONProjectXIV2009,
  title = {The {{SAURON Project}} - {{XIV}}. {{No}} Escape from {{Vesc}}: {{A}} Global and Local Parameter in Early-Type Galaxy Evolution},
  author = {Scott, Nicholas and Cappellari, Michele and Davies, Roger L. and Bacon, R. and De Zeeuw, P. T. and Emsellem, Eric and {Falc{\'o}n-Barroso}, J{\'e}sus and Krajnovi{\'c}, Davor and Kuntschner, Harald and McDermid, Richard M. and Peletier, Reynier F. and Pipino, Antonio and Sarzi, Marc and Van Den Bosch, Remco C.E. and Van De Ven, Glenn and Van Scherpenzeel, Eveline},
  year = {2009},
  month = oct,
  journal = {MNRAS},
  volume = {398},
  number = {4},
  pages = {1835--1857},
  doi = {10.1111/j.1365-2966.2009.15275.x}
}

@article{sijackiUnifiedModelAGN2007,
  title = {A Unified Model for {{AGN}} Feedback in Cosmological Simulations of Structure Formation},
  author = {Sijacki, Debora and Springel, Volker and Di Matteo, Tiziana and Hernquist, Lars},
  year = {2007},
  month = sep,
  journal = {MNRAS},
  volume = {380},
  pages = {877--900},
  issn = {0035-8711},
  doi = {10.1111/j.1365-2966.2007.12153.x},
  urldate = {2023-08-02}
}

@article{solaCharacterizationLowSurface2022,
  title = {Characterization of Low Surface Brightness Structures in Annotated Deep Images},
  author = {Sola, Elisabeth and Duc, Pierre-Alain and Richards, Felix and Paiement, Adeline and Urbano, Mathias and Klehammer, Julie and B{\'i}lek, Michal and Cuillandre, Jean-Charles and Gwyn, Stephen and McConnachie, Alan},
  year = {2022},
  month = jun,
  journal = {A\&A},
  volume = {662},
  pages = {A124},
  publisher = {EDP Sciences},
  issn = {0004-6361, 1432-0746},
  doi = {10.1051/0004-6361/202142675},
  urldate = {2024-07-21},
  copyright = {{\copyright} E. Sola et al. 2022},
  langid = {english}
}

@article{solaLowSurfaceBrightness2025a,
  title = {Low Surface Brightness Structures from Annotated Deep {{CFHT}} Images: Effects of the Host Galaxy's Properties and Environment},
  shorttitle = {Low Surface Brightness Structures from Annotated Deep {{CFHT}} Images},
  author = {Sola, Elisabeth and Duc, Pierre-Alain and Urbano, Mathias and Richards, Felix and Paiement, Adeline and B{\'i}lek, Michal and Y{\i}ld{\i}z, Mustafa K and Boselli, Alessandro and C{\^o}t{\'e}, Patrick and Cuillandre, Jean-Charles and Ferrarese, Laura and Gwyn, Stephen and Marchal, Olivier and McConnachie, Alan W and Baumann, Matthieu and Boch, Thomas and Durret, Florence and Fossati, Matteo and Habas, Rebecca and Marleau, Francine R and M{\"u}ller, Oliver and Poulain, M{\'e}lina and Belokurov, Vasily},
  year = {2025},
  month = aug,
  journal = {MNRAS},
  volume = {541},
  number = {4},
  pages = {3015--3042},
  issn = {0035-8711},
  doi = {10.1093/mnras/staf1139},
  urldate = {2025-09-30}
}

@article{sparreZoomingMajorMergers2016,
  title = {Zooming in on Major Mergers: Dense, Starbursting Gas in Cosmological Simulations},
  shorttitle = {Zooming in on Major Mergers},
  author = {Sparre, Martin and Springel, Volker},
  year = {2016},
  month = nov,
  journal = {MNRAS},
  volume = {462},
  number = {3},
  pages = {2418--2430},
  issn = {0035-8711},
  doi = {10.1093/mnras/stw1793},
  urldate = {2024-11-26}
}

@book{spitzerPhysicsFullyIonized1962,
  title = {Physics of {{Fully Ionized Gases}}},
  author = {Spitzer, Lyman},
  year = {1962},
  publisher = {Interscience Publishers},
  googlebooks = {CilRAAAAMAAJ},
  isbn = {978-0-470-81723-0},
  langid = {english}
}

@article{springelCosmologicalSmoothedParticle2003,
  title = {Cosmological Smoothed Particle Hydrodynamics Simulations: A Hybrid Multiphase Model for Star Formation},
  shorttitle = {Cosmological Smoothed Particle Hydrodynamics Simulations},
  author = {Springel, Volker and Hernquist, Lars},
  year = {2003},
  month = feb,
  journal = {MNRAS},
  volume = {339},
  number = {2},
  pages = {289--311},
  issn = {0035-8711},
  doi = {10.1046/j.1365-8711.2003.06206.x},
  urldate = {2024-03-22}
}

@article{springelFirstResultsIllustrisTNG2018,
  title = {First Results from the {{IllustrisTNG}} Simulations: Matter and Galaxy Clustering},
  shorttitle = {First Results from the {{IllustrisTNG}} Simulations},
  author = {Springel, Volker and Pakmor, R{\"u}diger and Pillepich, Annalisa and Weinberger, Rainer and Nelson, Dylan and Hernquist, Lars and Vogelsberger, Mark and Genel, Shy and Torrey, Paul and Marinacci, Federico and Naiman, Jill},
  year = {2018},
  month = mar,
  journal = {MNRAS},
  volume = {475},
  pages = {676--698},
  issn = {0035-8711},
  doi = {10.1093/mnras/stx3304},
  urldate = {2023-05-02}
}

@article{springelModellingFeedbackStars2005,
  title = {Modelling Feedback from Stars and Black Holes in Galaxy Mergers},
  author = {Springel, Volker and Di Matteo, Tiziana and Hernquist, Lars},
  year = {2005},
  month = aug,
  journal = {MNRAS},
  volume = {361},
  pages = {776--794},
  issn = {0035-8711},
  doi = {10.1111/j.1365-2966.2005.09238.x},
  urldate = {2023-08-02}
}

@article{stoughtonSloanDigitalSky2002,
  title = {Sloan {{Digital Sky Survey}}: {{Early Data Release}}},
  shorttitle = {Sloan {{Digital Sky Survey}}},
  author = {Stoughton, Chris and Lupton, Robert H. and Bernardi, Mariangela and Blanton, Michael R. and Burles, Scott and Castander, Francisco J. and Connolly, A. J. and Eisenstein, Daniel J. and Frieman, Joshua A. and Hennessy, G. S. and Hindsley, Robert B. and Ivezi{\'c}, {\v Z}eljko and Kent, Stephen and Kunszt, Peter Z. and Lee, Brian C. and Meiksin, Avery and Munn, Jeffrey A. and Newberg, Heidi Jo and Nichol, R. C. and Nicinski, Tom and Pier, Jeffrey R. and Richards, Gordon T. and Richmond, Michael W. and Schlegel, David J. and Smith, J. Allyn and Strauss, Michael A. and SubbaRao, Mark and Szalay, Alexander S. and Thakar, Aniruddha R. and Tucker, Douglas L. and Berk, Daniel E. Vanden and Yanny, Brian and Adelman, Jennifer K. and John E. Anderson, Jr and Anderson, Scott F. and Annis, James and Bahcall, Neta A. and Bakken, J. A. and Bartelmann, Matthias and Bastian, Steven and Bauer, Amanda and Berman, Eileen and B{\"o}hringer, Hans and Boroski, William N. and Bracker, Steve and Briegel, Charlie and Briggs, John W. and Brinkmann, J. and Brunner, Robert and Carey, Larry and Carr, Michael A. and Chen, Bing and Christian, Damian and Colestock, Patrick L. and Crocker, J. H. and Csabai, Istv{\'a}n and Czarapata, Paul C. and Dalcanton, Julianne and Davidsen, Arthur F. and Davis, John Eric and Dehnen, Walter and Dodelson, Scott and Doi, Mamoru and Dombeck, Tom and Donahue, Megan and Ellman, Nancy and Elms, Brian R. and Evans, Michael L. and Eyer, Laurent and Fan, Xiaohui and Federwitz, Glenn R. and Friedman, Scott and Fukugita, Masataka and Gal, Roy and Gillespie, Bruce and Glazebrook, Karl and Gray, Jim and Grebel, Eva K. and Greenawalt, Bruce and Greene, Gretchen and Gunn, James E. and de Haas, Ernst and Haiman, Zolt{\'a}n and Haldeman, Merle and Hall, Patrick B. and Hamabe, Masaru and Hansen, Brad and Harris, Frederick H. and Harris, Hugh and Harvanek, Michael and Hawley, Suzanne L. and Hayes, J. J. E. and Heckman, Timothy M. and Helmi, Amina and Henden, Arne and Hogan, Craig J. and Hogg, David W. and Holmgren, Donald J. and Holtzman, Jon and Huang, Chih-Hao and Hull, Charles and Ichikawa, Shin-Ichi and Ichikawa, Takashi and Johnston, David E. and Kauffmann, Guinevere and Kim, Rita S. J. and Kimball, Tim and Kinney, E. and Klaene, Mark and Kleinman, S. J. and Klypin, Anatoly and Knapp, G. R. and Korienek, John and Krolik, Julian and Kron, Richard G. and Krzesi{\'n}ski, Jurek and Lamb, D. Q. and Leger, R. French and Limmongkol, Siriluk and Lindenmeyer, Carl and Long, Daniel C. and Loomis, Craig and Loveday, Jon and MacKinnon, Bryan and Mannery, Edward J. and Mantsch, P. M. and Margon, Bruce and McGehee, Peregrine and McKay, Timothy A. and McLean, Brian and Menou, Kristen and Merelli, Aronne and Mo, H. J. and Monet, David G. and Nakamura, Osamu and Narayanan, Vijay K. and Nash, Thomas and Eric H. Neilsen, Jr and Newman, Peter R. and Nitta, Atsuko and Odenkirchen, Michael and Okada, Norio and Okamura, Sadanori and Ostriker, Jeremiah P. and Owen, Russell and Pauls, A. George and Peoples, John and Peterson, R. S. and Petravick, Donald and Pope, Adrian and Pordes, Ruth and Postman, Marc and Prosapio, Angela and Quinn, Thomas R. and Rechenmacher, Ron and Rivetta, Claudio H. and Rix, Hans-Walter and Rockosi, Constance M. and Rosner, Robert and Ruthmansdorfer, Kurt and Sandford, Dale and Schneider, Donald P. and Scranton, Ryan and Sekiguchi, Maki and Sergey, Gary and Sheth, Ravi and Shimasaku, Kazuhiro and Smee, Stephen and Snedden, Stephanie A. and Stebbins, Albert and Stubbs, Christopher and Szapudi, Istv{\'a}n and Szkody, Paula and Szokoly, Gyula P. and Tabachnik, Serge and Tsvetanov, Zlatan and Uomoto, Alan and Vogeley, Michael S. and Voges, Wolfgang and Waddell, Patrick and Walterbos, Ren{\'e} and Wang, Shu-i and Watanabe, Masaru and Weinberg, David H. and White, Richard L. and White, Simon D. M. and Wilhite, Brian and Wolfe, David and Yasuda, Naoki and York, Donald G. and Zehavi, Idit and Zheng, Wei},
  year = {2002},
  month = jan,
  journal = {AJ},
  volume = {123},
  number = {1},
  pages = {485},
  publisher = {IOP Publishing},
  issn = {1538-3881},
  doi = {10.1086/324741},
  urldate = {2024-11-10},
  langid = {english}
}

@article{sutherlandCoolingFunctionsLowDensity1993,
  title = {Cooling {{Functions}} for {{Low-Density Astrophysical Plasmas}}},
  author = {Sutherland, Ralph S. and Dopita, M. A.},
  year = {1993},
  month = sep,
  journal = {ApJS},
  volume = {88},
  pages = {253},
  issn = {0067-0049},
  doi = {10.1086/191823},
  urldate = {2023-08-02}
}

@article{tahmasebzadehDeprojectionExternalBarred2021,
  title = {Deprojection of External Barred Galaxies from Photometry},
  author = {Tahmasebzadeh, Behzad and Zhu, Ling and Shen, Juntai and Gerhard, Ortwin and Qin, Yujing},
  year = {2021},
  month = dec,
  journal = {MNRAS},
  volume = {508},
  number = {4},
  pages = {6209--6222},
  issn = {0035-8711},
  doi = {10.1093/mnras/stab3002},
  urldate = {2025-07-08}
}

@article{taylorGalaxyMassAssembly2015,
  title = {Galaxy {{And Mass Assembly}} ({{GAMA}}): Deconstructing Bimodality -- {{I}}. {{Red}} Ones and Blue Ones},
  shorttitle = {Galaxy {{And Mass Assembly}} ({{GAMA}})},
  author = {Taylor, Edward N. and Hopkins, Andrew M. and Baldry, Ivan K. and {Bland-Hawthorn}, Joss and Brown, Michael J. I. and Colless, Matthew and Driver, Simon and Norberg, Peder and Robotham, Aaron S. G. and Alpaslan, Mehmet and Brough, Sarah and Cluver, Michelle E. and Gunawardhana, Madusha and Kelvin, Lee S. and Liske, Jochen and Conselice, Christopher J. and Croom, Scott and Foster, Caroline and Jarrett, Thomas H. and {Lara-Lopez}, Maritza and Loveday, Jon},
  year = {2015},
  month = jan,
  journal = {MNRAS},
  volume = {446},
  number = {2},
  pages = {2144--2185},
  issn = {0035-8711},
  doi = {10.1093/mnras/stu1900},
  urldate = {2024-09-13}
}

@article{tekluConnectingAngularMomentum2015,
  title = {Connecting {{Angular Momentum}} and {{Galactic Dynamics}}: {{The Complex Interplay}} between {{Spin}}, {{Mass}}, and {{Morphology}}},
  shorttitle = {Connecting {{Angular Momentum}} and {{Galactic Dynamics}}},
  author = {Teklu, Adelheid F. and Remus, Rhea-Silvia and Dolag, Klaus and Beck, Alexander M. and Burkert, Andreas and Schmidt, Andreas S. and Schulze, Felix and Steinborn, Lisa K.},
  year = {2015},
  month = oct,
  journal = {ApJ},
  volume = {812},
  pages = {29},
  issn = {0004-637X},
  doi = {10.1088/0004-637X/812/1/29},
  urldate = {2023-08-02}
}

@article{terrazasRelationshipBlackHole2020,
  title = {The Relationship between Black Hole Mass and Galaxy Properties: Examining the Black Hole Feedback Model in {{IllustrisTNG}}},
  shorttitle = {The Relationship between Black Hole Mass and Galaxy Properties},
  author = {Terrazas, Bryan A and Bell, Eric F and Pillepich, Annalisa and Nelson, Dylan and Somerville, Rachel S and Genel, Shy and Weinberger, Rainer and Habouzit, M{\'e}lanie and Li, Yuan and Hernquist, Lars and Vogelsberger, Mark},
  year = {2020},
  month = apr,
  journal = {MNRAS},
  volume = {493},
  number = {2},
  pages = {1888--1906},
  issn = {0035-8711},
  doi = {10.1093/mnras/staa374},
  urldate = {2025-01-13}
}

@article{teyssierMassDistributionGalaxy2011,
  title = {Mass Distribution in Galaxy Clusters: The Role of {{Active Galactic Nuclei}} Feedback},
  shorttitle = {Mass Distribution in Galaxy Clusters},
  author = {Teyssier, Romain and Moore, Ben and Martizzi, Davide and Dubois, Yohan and Mayer, Lucio},
  year = {2011},
  month = jun,
  journal = {MNRAS},
  volume = {414},
  pages = {195--208},
  issn = {0035-8711},
  doi = {10.1111/j.1365-2966.2011.18399.x},
  urldate = {2023-08-02}
}

@misc{thepandasdevelopmentteamPandasdevPandasPandas2023,
  title = {Pandas-Dev/Pandas: {{Pandas}}},
  shorttitle = {Pandas-Dev/Pandas},
  author = {{The pandas development team}},
  year = {2023},
  month = sep,
  doi = {10.5281/zenodo.8364959},
  urldate = {2023-09-29},
  howpublished = {Zenodo}
}

@article{toomreGalacticBridgesTails1972,
  title = {Galactic {{Bridges}} and {{Tails}}},
  author = {Toomre, Alar and Toomre, Juri},
  year = {1972},
  month = dec,
  journal = {ApJ},
  volume = {178},
  pages = {623--666},
  issn = {0004-637X},
  doi = {10.1086/151823},
  urldate = {2022-05-19}
}

@article{tornatoreChemicalEnrichmentGalaxy2007,
  title = {Chemical Enrichment of Galaxy Clusters from Hydrodynamical Simulations},
  author = {Tornatore, L. and Borgani, S. and Dolag, K. and Matteucci, F.},
  year = {2007},
  month = dec,
  journal = {MNRAS},
  volume = {382},
  pages = {1050--1072},
  issn = {0035-8711},
  doi = {10.1111/j.1365-2966.2007.12070.x},
  urldate = {2023-08-02}
}

@article{torreyMETALLICITYEVOLUTIONINTERACTING2012,
  title = {{{THE METALLICITY EVOLUTION OF INTERACTING GALAXIES}}},
  author = {Torrey, Paul and Cox, T. J. and Kewley, Lisa and Hernquist, Lars},
  year = {2012},
  month = jan,
  journal = {ApJ},
  volume = {746},
  number = {1},
  pages = {108},
  publisher = {The American Astronomical Society},
  issn = {0004-637X},
  doi = {10.1088/0004-637X/746/1/108},
  urldate = {2024-11-24},
  langid = {english}
}

@article{tortoraColourStellarPopulation2010,
  title = {Colour and Stellar Population Gradients in Galaxies: Correlation with Mass},
  shorttitle = {Colour and Stellar Population Gradients in Galaxies},
  author = {Tortora, C. and Napolitano, N. R. and Cardone, V. F. and Capaccioli, M. and Jetzer, {\relax Ph}. and Molinaro, R.},
  year = {2010},
  month = sep,
  journal = {MNRAS},
  volume = {407},
  number = {1},
  pages = {144--162},
  issn = {0035-8711},
  doi = {10.1111/j.1365-2966.2010.16938.x},
  urldate = {2024-11-26}
}

@article{tortoraStellarPopulationGradients2011,
  title = {Stellar Population Gradients from Cosmological Simulations: Dependence on Mass and Environment in Local Galaxies},
  shorttitle = {Stellar Population Gradients from Cosmological Simulations},
  author = {Tortora, C. and Romeo, A. D. and Napolitano, N. R. and {Antonuccio-Delogu}, V. and Meza, A. and {Sommer-Larsen}, J. and Capaccioli, M.},
  year = {2011},
  month = feb,
  journal = {MNRAS},
  volume = {411},
  number = {1},
  pages = {627--634},
  issn = {0035-8711},
  doi = {10.1111/j.1365-2966.2010.17708.x},
  urldate = {2024-11-26}
}

@article{trayfordColoursLuminosities012015,
  title = {Colours and Luminosities of z = 0.1 Galaxies in the {{EAGLE}} Simulation},
  author = {Trayford, James W. and Theuns, Tom and Bower, Richard G. and Schaye, Joop and Furlong, Michelle and Schaller, Matthieu and Frenk, Carlos S. and Crain, Robert A. and Dalla Vecchia, Claudio and McCarthy, Ian G.},
  year = {2015},
  month = sep,
  journal = {MNRAS},
  volume = {452},
  pages = {2879--2896},
  publisher = {OUP},
  issn = {0035-8711},
  doi = {10.1093/mnras/stv1461},
  urldate = {2024-08-20}
}

@article{trebitschOBELISKSimulationGalaxies2021,
  title = {The {{OBELISK}} Simulation: {{Galaxies}} Contribute More than {{AGN}} to {{H I}} Reionization of Protoclusters},
  shorttitle = {The {{OBELISK}} Simulation},
  author = {Trebitsch, Maxime and Dubois, Yohan and Volonteri, Marta and Pfister, Hugo and Cadiou, Corentin and Katz, Harley and Rosdahl, Joakim and Kimm, Taysun and Pichon, Christophe and Beckmann, Ricarda S. and Devriendt, Julien and Slyz, Adrianne},
  year = {2021},
  month = sep,
  journal = {A\&A},
  volume = {653},
  pages = {A154},
  issn = {0004-6361},
  doi = {10.1051/0004-6361/202037698},
  urldate = {2023-08-02}
}

@article{trujilloStrongSizeEvolution2007,
  title = {Strong Size Evolution of the Most Massive Galaxies since z {$\sim$} 2},
  author = {Trujillo, Ignacio and Conselice, C. J. and Bundy, Kevin and Cooper, M. C. and Eisenhardt, P. and Ellis, Richard S.},
  year = {2007},
  month = nov,
  journal = {MNRAS},
  volume = {382},
  number = {1},
  pages = {109--120},
  issn = {0035-8711},
  doi = {10.1111/j.1365-2966.2007.12388.x},
  urldate = {2024-09-09}
}

@article{valenzuelaStreamComeTrue2024,
  title = {A Stream Come True: {{Connecting}} Tidal Tails, Shells, Streams, and Planes with Galaxy Kinematics and Formation History},
  shorttitle = {A Stream Come True},
  author = {Valenzuela, Lucas M. and Remus, Rhea-Silvia},
  year = {2024},
  month = jun,
  journal = {A\&A},
  volume = {686},
  pages = {A182},
  publisher = {EDP Sciences},
  issn = {0004-6361, 1432-0746},
  doi = {10.1051/0004-6361/202244758},
  urldate = {2024-09-09},
  copyright = {{\copyright} The Authors 2024},
  langid = {english}
}

@article{vandenboschUniversalMassAccretion2002,
  title = {The Universal Mass Accretion History of Cold Dark Matter Haloes},
  author = {{van den Bosch}, Frank C.},
  year = {2002},
  month = mar,
  journal = {MNRAS},
  volume = {331},
  pages = {98--110},
  issn = {0035-8711},
  doi = {10.1046/j.1365-8711.2002.05171.x},
  urldate = {2023-05-02}
}

@article{vanderwel3DHST+CANDELSEvolutionGalaxy2014,
  title = {{{3D-HST}}+{{CANDELS}}: {{The}} Evolution of the Galaxy Size-Mass Distribution since z = 3},
  author = {{van der Wel}, A. and Franx, M. and Van Dokkum, P. G. and Skelton, R. E. and Momcheva, I. G. and Whitaker, K. E. and Brammer, G. B. and Bell, E. F. and Rix, H. W. and Wuyts, S. and Ferguson, H. C. and Holden, B. P. and Barro, G. and Koekemoer, A. M. and Chang, Yu Yen and McGrath, E. J. and H{\"a}ussler, B. and Dekel, A. and Behroozi, P. and Fumagalli, M. and Leja, J. and Lundgren, B. F. and Maseda, M. V. and Nelson, E. J. and Wake, D. A. and Patel, S. G. and Labb{\'e}, I. and Faber, S. M. and Grogin, N. A. and Kocevski, D. D.},
  year = {2014},
  month = jun,
  journal = {ApJ},
  volume = {788},
  number = {1},
  publisher = {Institute of Physics Publishing},
  doi = {10.1088/0004-637X/788/1/28}
}

@article{virtanenSciPy10Fundamental2020,
  title = {{{SciPy}} 1.0: Fundamental Algorithms for Scientific Computing in {{Python}}},
  shorttitle = {{{SciPy}} 1.0},
  author = {Virtanen, Pauli and Gommers, Ralf and Oliphant, Travis E. and Haberland, Matt and Reddy, Tyler and Cournapeau, David and Burovski, Evgeni and Peterson, Pearu and Weckesser, Warren and Bright, Jonathan and {van der Walt}, St{\'e}fan J. and Brett, Matthew and Wilson, Joshua and Millman, K. Jarrod and Mayorov, Nikolay and Nelson, Andrew R. J. and Jones, Eric and Kern, Robert and Larson, Eric and Carey, C. J. and Polat, {\.I}lhan and Feng, Yu and Moore, Eric W. and VanderPlas, Jake and Laxalde, Denis and Perktold, Josef and Cimrman, Robert and Henriksen, Ian and Quintero, E. A. and Harris, Charles R. and Archibald, Anne M. and Ribeiro, Ant{\^o}nio H. and Pedregosa, Fabian and {van Mulbregt}, Paul},
  year = {2020},
  month = mar,
  journal = {Nature Methods},
  volume = {17},
  number = {3},
  pages = {261--272},
  publisher = {Nature Publishing Group},
  issn = {1548-7105},
  doi = {10.1038/s41592-019-0686-2},
  urldate = {2023-09-29},
  copyright = {2020 The Author(s)},
  langid = {english}
}

@article{vogelsbergerModelCosmologicalSimulations2013,
  title = {A Model for Cosmological Simulations of Galaxy Formation Physics},
  author = {Vogelsberger, Mark and Genel, Shy and Sijacki, Debora and Torrey, Paul and Springel, Volker and Hernquist, Lars},
  year = {2013},
  month = dec,
  journal = {MNRAS},
  volume = {436},
  pages = {3031--3067},
  issn = {0035-8711},
  doi = {10.1093/mnras/stt1789},
  urldate = {2023-08-02}
}

@article{watkinsStrategiesOptimalSky2024,
  title = {Strategies for Optimal Sky Subtraction in the Low Surface Brightness Regime},
  author = {Watkins, Aaron E and Kaviraj, Sugata and Collins, Chris C and Knapen, Johan H and Kelvin, Lee S and Duc, Pierre-Alain and Rom{\'a}n, Javier and Mihos, J Christopher},
  year = {2024},
  month = mar,
  journal = {MNRAS},
  volume = {528},
  number = {3},
  pages = {4289--4306},
  issn = {0035-8711},
  doi = {10.1093/mnras/stae236},
  urldate = {2024-03-24}
}

@article{weigelStellarMassFunctions2016,
  title = {Stellar Mass Functions: Methods, Systematics and Results for the Local {{Universe}}},
  shorttitle = {Stellar Mass Functions},
  author = {Weigel, Anna K. and Schawinski, Kevin and Bruderer, Claudio},
  year = {2016},
  month = jun,
  journal = {MNRAS},
  volume = {459},
  number = {2},
  pages = {2150--2187},
  issn = {0035-8711},
  doi = {10.1093/mnras/stw756},
  urldate = {2025-01-15}
}

@article{weinbergerSimulatingGalaxyFormation2017,
  title = {Simulating Galaxy Formation with Black Hole Driven Thermal and Kinetic Feedback},
  author = {Weinberger, Rainer and Springel, Volker and Hernquist, Lars and Pillepich, Annalisa and Marinacci, Federico and Pakmor, R{\"u}diger and Nelson, Dylan and Genel, Shy and Vogelsberger, Mark and Naiman, Jill and Torrey, Paul},
  year = {2017},
  month = mar,
  journal = {MNRAS},
  volume = {465},
  number = {3},
  pages = {3291--3308},
  issn = {0035-8711},
  doi = {10.1093/mnras/stw2944},
  urldate = {2022-06-13}
}

@article{weinbergerSupermassiveBlackHoles2018,
  title = {Supermassive Black Holes and Their Feedback Effects in the {{IllustrisTNG}} Simulation},
  author = {Weinberger, Rainer and Springel, Volker and Pakmor, R{\"u}diger and Nelson, Dylan and Genel, Shy and Pillepich, Annalisa and Vogelsberger, Mark and Marinacci, Federico and Naiman, Jill and Torrey, Paul and Hernquist, Lars},
  year = {2018},
  month = sep,
  journal = {MNRAS},
  volume = {479},
  number = {3},
  pages = {4056--4072},
  issn = {0035-8711},
  doi = {10.1093/mnras/sty1733},
  urldate = {2024-11-26}
}

@article{wiersmaEffectPhotoionizationCooling2009,
  title = {The Effect of Photoionization on the Cooling Rates of Enriched, Astrophysical Plasmas},
  author = {Wiersma, Robert P. C. and Schaye, Joop and Smith, Britton D.},
  year = {2009},
  month = feb,
  journal = {MNRAS},
  volume = {393},
  pages = {99--107},
  issn = {0035-8711},
  doi = {10.1111/j.1365-2966.2008.14191.x},
  urldate = {2023-08-02}
}

@article{yoonFrequencyTidalFeatures2020,
  title = {Frequency of {{Tidal Features Correlates}} with {{Age}} and {{Internal Structure}} of {{Early-type Galaxies}}},
  author = {Yoon, Yongmin and Lim, Gu},
  year = {2020},
  month = dec,
  journal = {ApJ},
  volume = {905},
  pages = {154},
  issn = {0004-637X},
  doi = {10.3847/1538-4357/abc621},
  urldate = {2024-02-12}
}

@article{yoonImpactGalaxyMergers2023,
  title = {Impact of {{Galaxy Mergers}} on {{Stellar Population Profiles}} of {{Early-type Galaxies}}},
  author = {Yoon, Yongmin and Ko, Jongwan and Kim, Jae-Woo},
  year = {2023},
  month = mar,
  journal = {ApJ},
  volume = {946},
  number = {1},
  pages = {41},
  publisher = {The American Astronomical Society},
  issn = {0004-637X},
  doi = {10.3847/1538-4357/acbcc5},
  urldate = {2024-11-24},
  langid = {english}
}

@article{yorkSloanDigitalSky2000,
  title = {The {{Sloan Digital Sky Survey}}: {{Technical Summary}}},
  shorttitle = {The {{Sloan Digital Sky Survey}}},
  author = {York, Donald G. and Adelman, J. and Anderson, Jr., John E. and Anderson, Scott F. and Annis, James and Bahcall, Neta A. and Bakken, J. A. and Barkhouser, Robert and Bastian, Steven and Berman, Eileen and Boroski, William N. and Bracker, Steve and Briegel, Charlie and Briggs, John W. and Brinkmann, J. and Brunner, Robert and Burles, Scott and Carey, Larry and Carr, Michael A. and Castander, Francisco J. and Chen, Bing and Colestock, Patrick L. and Connolly, A. J. and Crocker, J. H. and Csabai, Istv{\'a}n and Czarapata, Paul C. and Davis, John Eric and Doi, Mamoru and Dombeck, Tom and Eisenstein, Daniel and Ellman, Nancy and Elms, Brian R. and Evans, Michael L. and Fan, Xiaohui and Federwitz, Glenn R. and Fiscelli, Larry and Friedman, Scott and Frieman, Joshua A. and Fukugita, Masataka and Gillespie, Bruce and Gunn, James E. and Gurbani, Vijay K. and {de Haas}, Ernst and Haldeman, Merle and Harris, Frederick H. and Hayes, J. and Heckman, Timothy M. and Hennessy, G. S. and Hindsley, Robert B. and Holm, Scott and Holmgren, Donald J. and Huang, Chi-hao and Hull, Charles and Husby, Don and Ichikawa, Shin-Ichi and Ichikawa, Takashi and Ivezi{\'c}, {\v Z}eljko and Kent, Stephen and Kim, Rita S. J. and Kinney, E. and Klaene, Mark and Kleinman, A. N. and Kleinman, S. and Knapp, G. R. and Korienek, John and Kron, Richard G. and Kunszt, Peter Z. and Lamb, D. Q. and Lee, B. and Leger, R. French and Limmongkol, Siriluk and Lindenmeyer, Carl and Long, Daniel C. and Loomis, Craig and Loveday, Jon and Lucinio, Rich and Lupton, Robert H. and MacKinnon, Bryan and Mannery, Edward J. and Mantsch, P. M. and Margon, Bruce and McGehee, Peregrine and McKay, Timothy A. and Meiksin, Avery and Merelli, Aronne and Monet, David G. and Munn, Jeffrey A. and Narayanan, Vijay K. and Nash, Thomas and Neilsen, Eric and Neswold, Rich and Newberg, Heidi Jo and Nichol, R. C. and Nicinski, Tom and Nonino, Mario and Okada, Norio and Okamura, Sadanori and Ostriker, Jeremiah P. and Owen, Russell and Pauls, A. George and Peoples, John and Peterson, R. L. and Petravick, Donald and Pier, Jeffrey R. and Pope, Adrian and Pordes, Ruth and Prosapio, Angela and Rechenmacher, Ron and Quinn, Thomas R. and Richards, Gordon T. and Richmond, Michael W. and Rivetta, Claudio H. and Rockosi, Constance M. and Ruthmansdorfer, Kurt and Sandford, Dale and Schlegel, David J. and Schneider, Donald P. and Sekiguchi, Maki and Sergey, Gary and Shimasaku, Kazuhiro and Siegmund, Walter A. and Smee, Stephen and Smith, J. Allyn and Snedden, S. and Stone, R. and Stoughton, Chris and Strauss, Michael A. and Stubbs, Christopher and SubbaRao, Mark and Szalay, Alexander S. and Szapudi, Istvan and Szokoly, Gyula P. and Thakar, Anirudda R. and Tremonti, Christy and Tucker, Douglas L. and Uomoto, Alan and Vanden Berk, Dan and Vogeley, Michael S. and Waddell, Patrick and Wang, Shu-i and Watanabe, Masaru and Weinberg, David H. and Yanny, Brian and Yasuda, Naoki},
  year = {2000},
  month = sep,
  journal = {AJ},
  volume = {120},
  number = {3},
  pages = {1579--1587},
  issn = {00046256},
  doi = {10.1086/301513},
  urldate = {2022-06-09}
}

@article{zwickyMultipleGalaxies1956,
  title = {Multiple {{Galaxies}}},
  author = {Zwicky, F.},
  year = {1956},
  month = jan,
  journal = {Ergebnisse der exakten Naturwissenschaften},
  volume = {29},
  pages = {344--385},
  urldate = {2025-05-13}
}



\appendix

\section{Masking}
\label{app:masking}

\begin{figure}
    \centering
    \includegraphics[width=\linewidth]{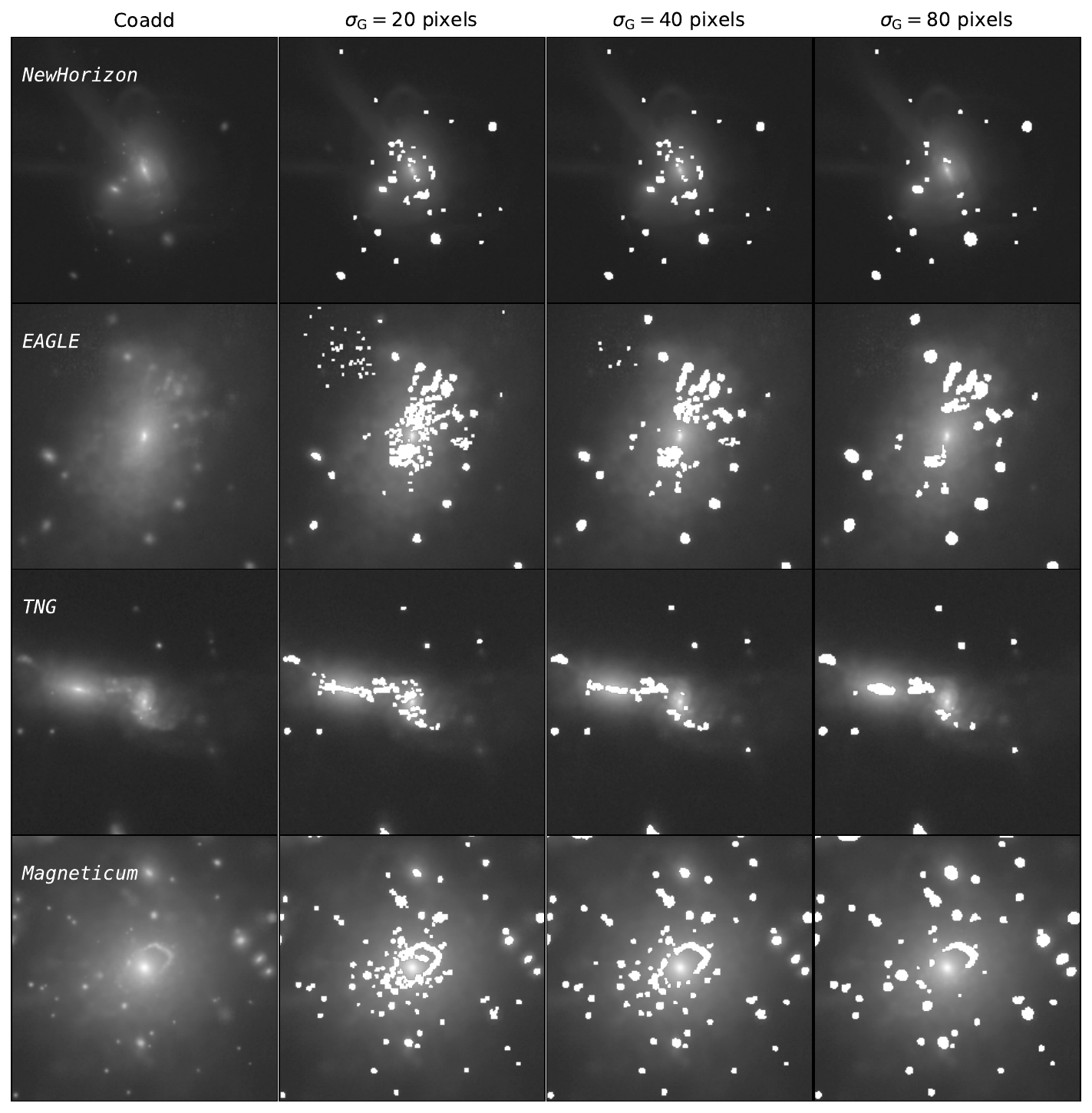}
    \caption{Examples of different parameter settings for hot mask process. From top to bottom, shows galaxies from NewHorizon, EAGLE and Magneticum. From left to right we have the resulting masked image for $\sigma_\mathrm{G}=20$, 40 and 80 pixels. The mask selected was $\sigma_\mathrm{G}=40$ pixels. Each image covers a region $\sim240$ pkpc.}
    \label{fig:hot_mask_demo}
\end{figure}

\begin{figure}
    \centering
    \includegraphics[width=\linewidth]{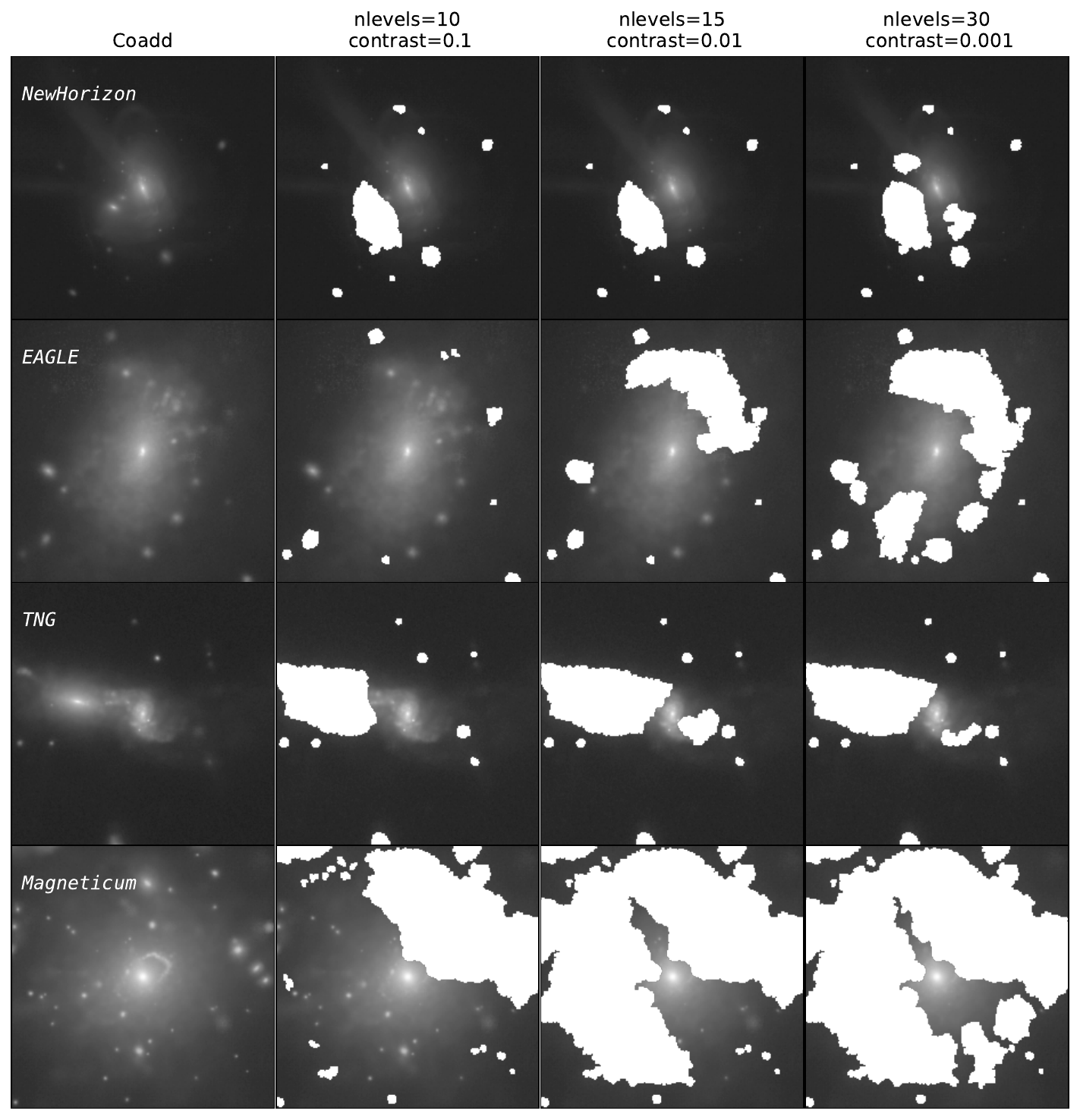}
    \caption{Examples of different parameter choices for cold masking process. From top to bottom, shows galaxies from NewHorizon, EAGLE and Magneticum. From left to right we have the resulting masked image for different combinations of \texttt{nlevels} and \texttt{contrast}. The mask selected was \texttt{nlevels}=30 and \texttt{contrast}=0.001. Each image covers a region of $\sim240$ pkpc.}
    \label{fig:cold_mask_demo}
\end{figure}

This section briefly outlines some of the testing we performed to arrive at a masking parameters suitable for an automatic process given in Section \ref{subsubsec:masks}. We were aiming to find a set of parameters that resulted in a conservative mask ensuring that light from neighbouring objects was limited to a minimum, as tidal features are faint and their contribution to light profiles is likely to be missed if other galaxies are unmasked. Fig.~\ref{fig:hot_mask_demo}, illustrates our exploration of hot masking parameters, for a range of $\sigma_\mathrm{G}$, the standard deviation used for the Gaussian-smoothed image used in the creation of the unsharped image. For detection, we set \texttt{npixels}$=\sigma_\mathrm{G}$ as we do not expect any sources to be detectable beneath the smoothing scale of the image. With an increasing smoothing scale, we have fewer detections, which is to be expected as fewer sources are above \texttt{npixels}, the size threshold. For $\sigma_\mathrm{G}=20$ pixels, most of the galaxy is covered in detections that are not seen in the coadded image. This indicates that spatial fluctuations on a scale of 20 pixels correspond mostly to the internal structure of the galaxy. While the masking corresponding to $\sigma_\mathrm{G}=40$ and 80 pixels covers similar regions, we opt for 40 pixels as this is more conservative and captures the dense compact sources near the nucleus of the galaxy (unlike $\sigma_\mathrm{G}=80$ pixels) without overmasking (see the third row). As with any automated procedure, the choice is not perfect for all sources and there are cases where the $\sigma_\mathrm{G}=40$ pixel mask covers pixels within the galaxy that are associated with diffuse light/tidal features. However, this is preferred to the alternative where less of the feature is masked at the risk of light from compact source impacting the measured colour.

For the cold mask, using a similar approach to the pixel scales for hot masking, we selected \texttt{npixels}$=400$, as that provided the most appropriate scale for the detection of extended sources. Individually increasing $\texttt{nlevels}$ or decreasing \texttt{contrast} tends to result in more deblended sources. While in our testing we explored a range of masking parameters, we summarise our combined findings in  Fig.~\ref{fig:cold_mask_demo}. Fig.~\ref{fig:cold_mask_demo}, from left to right shows the combination of choices that lead to less deblending to those that lead to more deblending. In general, we again find no ideal solution for the entire dataset. We opt for \texttt{nlevels}$=30$ and \texttt{contrast}$=0.001$ (equivalent to the 4th column of Fig.~\ref{fig:cold_mask_demo}). This leads to the largest amount of deblending and results in the detection of many overlapping sources when compared to settings that lead to less deblending. While this can occasionally lead to the eventual masking of tidal features, e.g. the tidal tail in the third row, this is again preferable to leaving large numbers of sources unidentified and therefore unmasked as they might contribute more significantly to our measurement than the tidal features, which are usually much fainter.

\section{The impact of projection, inclination and galaxy morphology on MGE size measurements}
\label{app:size_measurement}
\begin{figure*}
    \centering
    \includegraphics[width=\linewidth]{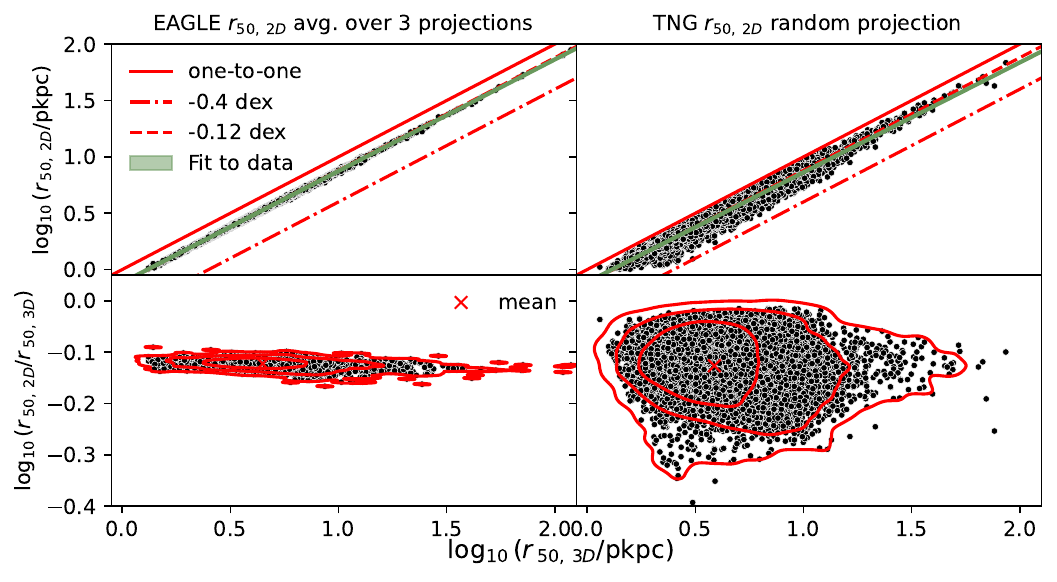}
    \caption{The top panels show the 2D projected size measurements, where the size of the galaxy is the radius of a circular aperture containing half of the stellar mass, from EAGLE and TNG as a function of the 3D size, where the size is the radius of the spherical aperture containing half of the stellar mass in the galaxy. The available results from EAGLE were averaged over three axes, whereas the results from TNG were taken at a random projection with respect to the galaxy (observed along the $z$ axis of the box). The solid line shows the one-to-one relation, the dashed-dotted line shows the $-0.4$ dex line, and the dashed line shows the $-0.12$ dex line. The green shaded region shows the best linear fit and $1\sigma$ uncertainty to the relationship between $\log_{10}(r_\mathrm{50,\:2D}/\mathrm{pkpc})$ and $\log_{10}(r_\mathrm{50,\:3D}/\mathrm{pkpc})$.  The bottom panels show the difference in dex between 2D projected size and 3D size ($\log_{10}(r_\mathrm{50,\:2D}/r_\mathrm{50,\:3D})$) as a function of the 3D size. The red cross marks the mean value of $\log_{10}(r_\mathrm{50,\:2D}/\mathrm{pkpc})$ and $\log_{10}(r_\mathrm{50,\:3D}/\mathrm{pkpc})$, the contour lines show the $1\sigma$, $2\sigma$ and $3\sigma$ contours for the Gaussian kernel density estimates of the distribution of points. The fit to the EAGLE and TNG results both illustrate that the projected size is, on average, $\sim0.12$ dex smaller than the 3D size of a galaxy. However, the random projection measurements used for TNG indicate that an additional scatter of $\sigma\sim0.08$ is introduced on this average size difference when a random projection is selected.}
    \label{fig:size_3d_vs_size2d}
\end{figure*}

Given that the MGE modelling works best in early-type galaxies, we test how the measurement of galaxy size varies across our sample as a function of both colour and the ellipticity of our galaxies. In order to quantify the impact of galaxy colour and ellipticity on the MGE measured sizes, we compare the measured sizes to the intrinsic $r_\mathrm{50,\:3D}$, the radius of the sphere containing half of the stellar mass within the \textsc{SubFind} identified galaxy.

Unlike our mock images, the 3D spherical aperture sizes ($r_\mathrm{50,\: 3D}$) are not affected by the random orientation of the galaxy to the observer. We therefore first measure the impact of this on any difference we see. We estimate the impact of random inclination on the measured size for TNG using the \citet{genelSizeEvolutionStarforming2018} measurements of 2D projected spherical aperture measurements of stellar radius, $r_\mathrm{50,\:2D}$, to the 3D measurements, $r_\mathrm{50,\:3D}$, for snapshot 99 of TNG. The randomly projected (galaxy observed along the $z$ axis of the simulation, i.e. the x-y projection) half-stellar mass radius is only available for TNG, and snapshot 99 is the closest to the one used in our work (snapshot 95). We select all the galaxies with $r_\mathrm{50,\:3D}$ greater than 1 pkpc and $M_\star \geq 10^{9.5} \mathrm{M}_\odot$, to match the sample used here. We also compare the projected radii to the 3D radii for all EAGLE galaxies with $r_\mathrm{50,\: 3D}\geq1$ pkpc and $M_\star \geq 10^{9.5} \mathrm{M}_\odot$ for comparison. However, in a slight difference to TNG, the EAGLE projected measurements have been averaged over the x-y, y-z and x-z projections of the galaxies.

Fig.~\ref{fig:size_3d_vs_size2d} shows the results of our comparison between $r_\mathrm{50,\:2D}$ and $r_\mathrm{50,\:3D}$. The top panel shows the relationship between $\log_{10}(r_\mathrm{50,\: 3D}/\mathrm{pkpc})$ and $\log_{10}(r_\mathrm{50,\: 2D}/\mathrm{pkpc})$ for EAGLE and TNG, along with red lines showing the one-to-one relation (solid), 0.12 dex less than the one-to-one relation (dashed) and 0.4 dex less than the one-to-one relation dot dashed. The top panel for TNG highlights that the projected radii are either equivalent to or smaller than the 3D radii. A least squares linear fit to the TNG data, gives the equation $\log_{10}(r_\mathrm{50,\:2D})=(0.974\pm0.002)\log_{10}(r_\mathrm{50,\:3D})-(0.112\pm0.001)$ with Pearson correlation $r=0.97$ and a p-value, $p=0.00$ (to floating point precision), indicates a strong linear correlation between the logged radii that is consistent with a $-0.12$ dex difference on average. The fit to the EAGLE data, which has a projected radius averaged over three random projections of the galaxy (with line-of-sights along the \textit{x}, \textit{y} and \textit{z} axes of the simulation), gives the equation $\log_{10}(r_\mathrm{50,\:2D})=(0.9899\pm0.0004)\log_{10}(r_\mathrm{50,\:3D})-(0.1157\pm0.0003)$ and shows an even tighter relation along the $-0.12$ dex line, with $r=0.99$ and $p=0.00$ (to floating point precision). Further supporting the idea that, on average, the projected radii are $-0.12$ dex smaller than the unprojected sizes, mostly independent of simulation physics.

The bottom panel of Fig.~\ref{fig:size_3d_vs_size2d} shows the difference in log-radii and the 1, 2 and 3 $\sigma$ Gaussian kernel density contours. The EAGLE results have minimal scatter around $-0.12$ dex, whereas the TNG results, where a single random projection (collapsed along the \textit{z} axis of the simulation) has been selected, range from $-0.014$ dex to $-0.39$ dex of scatter. Given that the EAGLE results have been averaged over 3 random projections, assuming Gaussian uncertainties in the projected radii, we would expect them to be a factor of $\sqrt{3}$ smaller than the values for TNG. We check this using 10 bins in 3D galaxy radius between 1 and 30 pkpc and find that the mean standard deviation for TNG over this range is $1.2\mathrm{\:pkpc}$, whereas, for EAGLE this is $0.7\mathrm{\:pkpc}$, $\sqrt{3}$ times this standard deviation is $1.2\mathrm{\:pkpc}$, consistent with TNG. Comparison between the spread of the EAGLE and TNG results highlights that selecting a random projection results in an additional $\sigma\sim0.08$ dex scatter on the average $-0.12$ dex lower projected size than the 3D size.

\begin{figure*}
    \centering
    \includegraphics[width=\linewidth]{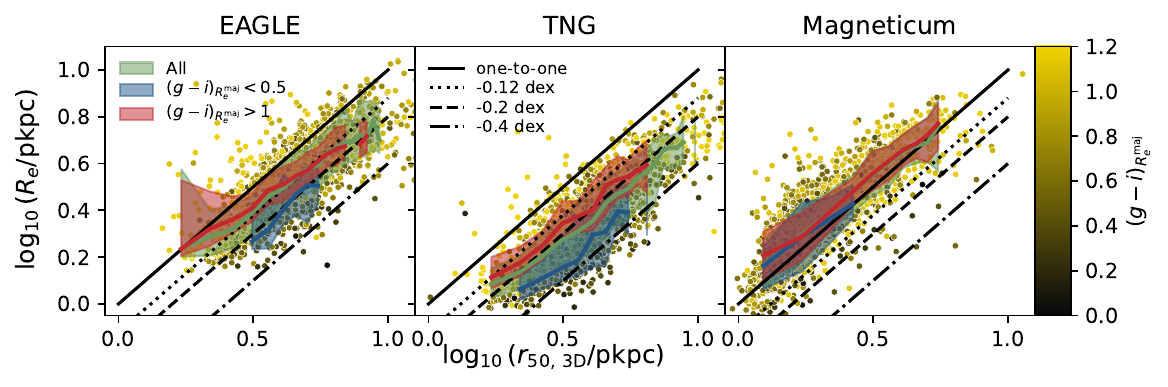}
    \caption{The MGE-measured galaxy radius as a function of the simulation-measured $r_\mathrm{50,\:3D}$. The data points are coloured by $(g-i)_{R_e^\mathrm{maj}}$. The solid black line shows the one-to-one relation, the dotted line shows the $-0.12$ dex relation, the dashed line shows the $-0.2$ dex relation, and the dot-dashed line shows the $-0.4$ dex relation. The solid green line and shaded region show the median and $16^\mathrm{th}$ to $84^\mathrm{th}$ percentiles for all galaxies, whereas the blue and red solid lines and shaded regions do the same for galaxies with $(g-i)_{R_e^\mathrm{maj}}<0.5$ and $(g-i)_{R_e^\mathrm{maj}}>1$, respectively. We see that for EAGLE and TNG, there is a tendency for the blue cloud galaxies to have smaller MGE-measured sizes for a given $r_\mathrm{50,\:3D}$, whereas, in Magneticum, all galaxies have MGE sizes that are similar, if not slightly larger than the 3D simulation-measured size.}
    \label{fig:rmge_vs_r50}
\end{figure*}

Having quantified the impact of projection on the measured sizes of galaxies in Fig.~\ref{fig:size_3d_vs_size2d}, we now compare our MGE-measured sizes to the 3D simulation-measured, $r_\mathrm{50,\:3D}$. In order to make a fair comparison, we transform our MGE measurements from the semi-major axis, $R_e^\mathrm{maj}$, to the radius of a circle of equivalent area, $R_e$, with the following equation:

\begin{equation}
    R_e=R_e^\mathrm{maj}\sqrt{1-\overline{\varepsilon}}.
\end{equation}

Where $\overline{\varepsilon}$ is the average ellipticity of the galaxy as measured by the \texttt{find\_galaxy} function in the \textsc{MGEfit} package. In Fig.~\ref{fig:rmge_vs_r50}, we show the relation between $R_e$ and $r_\mathrm{50,\:3D}$ for each of our simulations. EAGLE and TNG both have negative offsets from the one-to-one relation, with EAGLE being most similar to the $-0.12$ dex expected from projection alone, whereas TNG shows the largest offset, below the $-0.2$ dex relation. For EAGLE and TNG our results show, that bluer galaxies ($(g-i)_{R_e^\mathrm{maj}}<0.5$) have systematically smaller MGE-measured sizes than red galaxies, for a given $r_\mathrm{50,\:3D}$, whereas, Magneticum shows no differences between populations of different colours and remains largely consistent with the one-to-one relation. Differences between the sizes and shapes of the populations of galaxies are not surprising, as the different galaxy formation models lead to different, morphological characteristics across the simulations.

\begin{figure*}
    \centering
    \includegraphics[width=\linewidth]{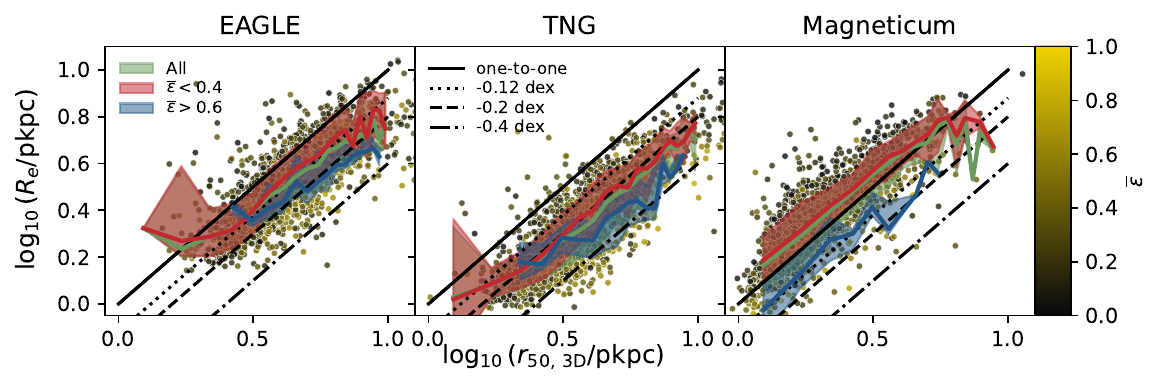}
    \caption{The MGE-measured galaxy radius as a function of the simulation-measured $r_\mathrm{50,\:3D}$. The data points are coloured by the average ellipticity $\overline{\varepsilon}$. The solid line shows the one-to-one relation, the dotted line shows the $-0.12$ dex relation, the dashed line shows the $-0.2$ dex relation, and the dot-dashed line shows the $-0.4$ dex relation. The solid green line and shaded regions show the median, $16^\mathrm{th}$ and $84^\mathrm{th}$ percentiles for all galaxies, the blue and red lines and shaded regions do the same for $\overline{\varepsilon}>0.6$ and $\overline{\varepsilon}<0.4$, respectively. Across all simulations, we clearly see for a given $r_\mathrm{50,\:3D}$, we have decreasing MGE-measured sizes with increasing $\overline{\varepsilon}$.}
    \label{fig:rmge_vs_r50_eps}
\end{figure*}

In Fig.~\ref{fig:rmge_vs_r50_eps}, we show how the shape of the galaxy influences the MGE measurement. We show a similar plot to Fig.~\ref{fig:rmge_vs_r50}, only this time we colour the points by average ellipticity, as measured by \textsc{MGEfit}, $\varepsilon$. It is evident that with increasing $\overline{\varepsilon}$, there is a decrease in the MGE-measured size for a given $r_\mathrm{50,\:3D}$. The trend with ellipticity is stronger than that seen with colour in Fig.~\ref{fig:rmge_vs_r50}. Given the clear trend with ellipticity, it seems likely that this is contributing to the trends with colour and MGE measured sizes that we see in EAGLE and TNG (Fig.~\ref{fig:rmge_vs_r50}). The reason we do not see similar differences in Magneticum is that the blue galaxies have lower ellipticities than in the other simulations. This is likely being driven by the majority of these galaxies being in the $10^{9.5}<M_{\star,\mathrm{\:30\:pkpc}}/\mathrm{M}_\odot<10^{10}$ bin. The shapes of these galaxies are likely harder to differentiate from circular, particularly given the lower particle resolution of Magneticum, resulting in, on average, more circular galaxy shapes. The differences we see are likely well accounted for by the average ellipticity of the galaxy, along with the distribution of light within the galaxy itself.

\begin{figure*}
    \centering
    \includegraphics[width=\linewidth]{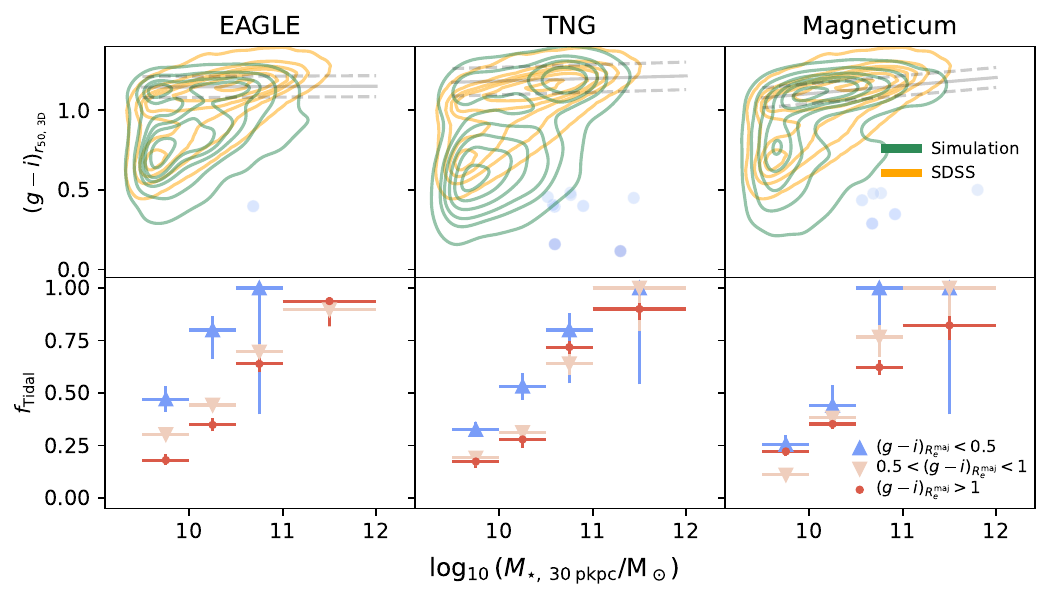}
    \caption{Identical to Fig.~\ref{fig:col_mag_total}, however, using $r_\mathrm{50,\:3D}$ measured sizes instead of $R_e^\mathrm{maj}$. The conclusions remain qualitatively unchanged.}
    \label{fig:col_mag_total_r50}
\end{figure*}

\begin{figure}
    \centering
    \includegraphics[width=\linewidth]{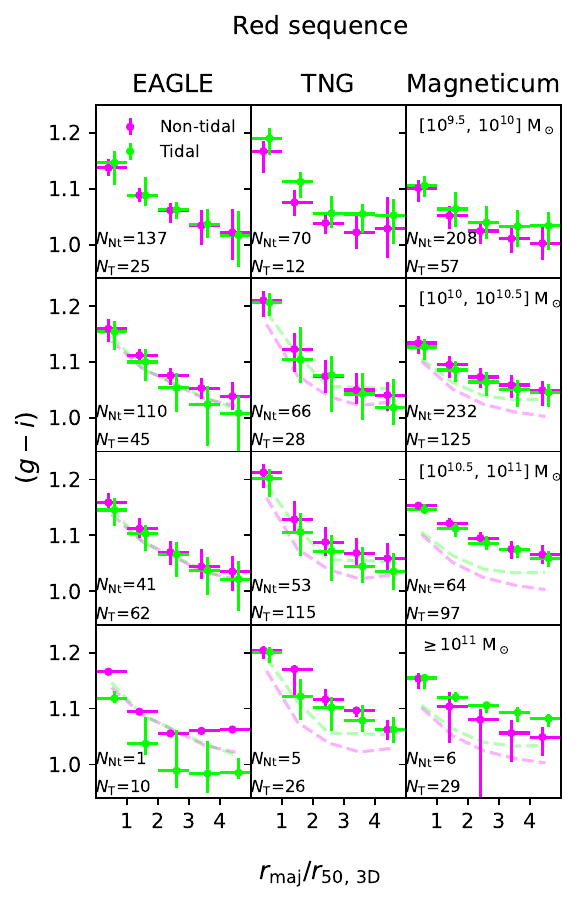}
    \caption{Identical to Fig.~\ref{fig:col_profile_rs}, however, using $r_\mathrm{50,\:3D}$ measured sizes instead of $R_e^\mathrm{maj}$. The conclusions remain qualitatively unchanged.}
    \label{fig:col_profile_rs_r50}
\end{figure}

\begin{figure}
    \centering
    \includegraphics[width=\linewidth]{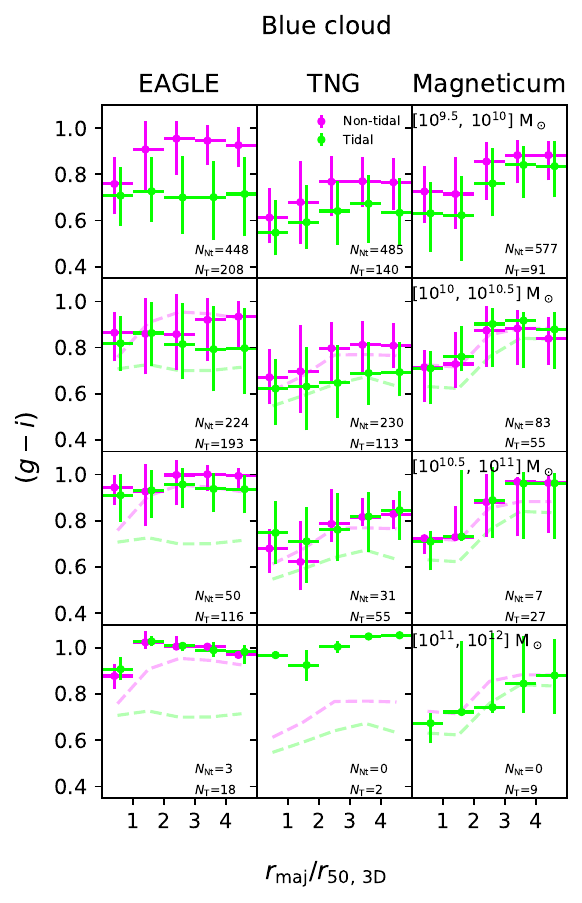}
    \caption{Identical to Fig.~\ref{fig:col_profile_bc}, however, using $r_\mathrm{50,\:3D}$ measured sizes instead of $R_e^\mathrm{maj}$. The conclusions remain qualitatively unchanged.}
    \label{fig:col_profile_bc_r50}
\end{figure}

For assurance that the differences in the MGE-measured sizes across galaxy colours do not play a significant role in our conclusions, we check how our conclusions change if we use $r_\mathrm{50,\:3D}$ sizes instead of the MGE-measured $R_e^\mathrm{maj}$. We find that the conclusions remain qualitatively unchanged (see Fig.~\ref{fig:col_mag_total_r50}, Fig.~\ref{fig:col_profile_rs_r50} and Fig.~\ref{fig:col_profile_bc_r50}).

\section{Colour threshold for red sequence and blue cloud selection}
\label{app:col_thresh}

\begin{figure*}
    \centering
    \includegraphics[width=\linewidth]{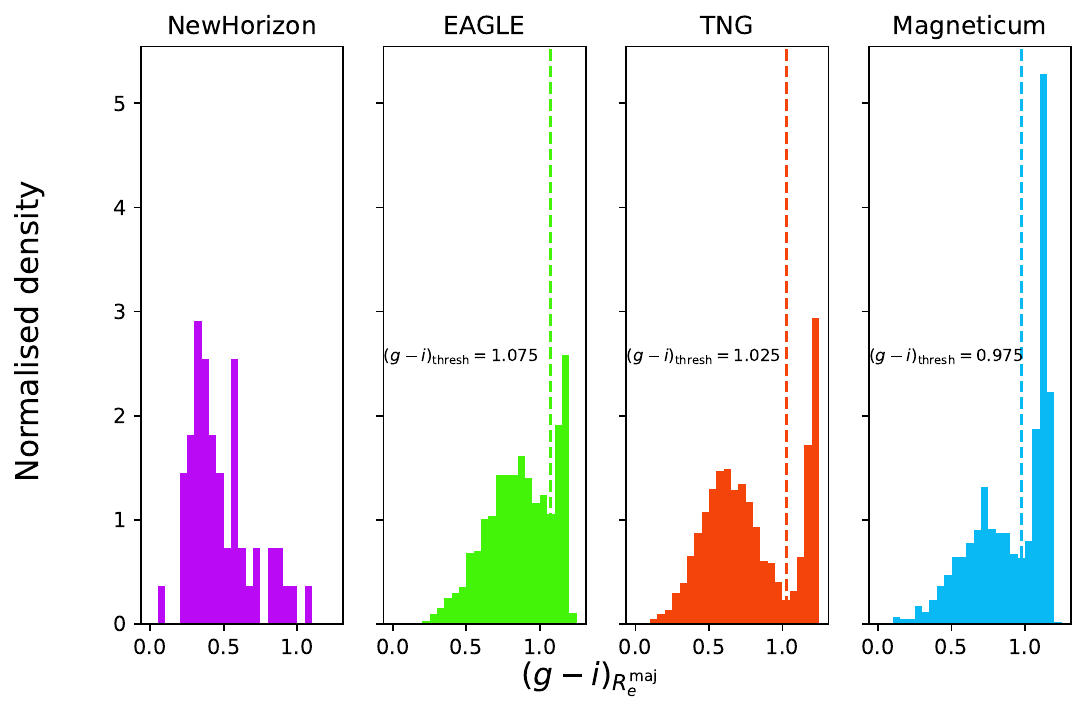}
    \caption{Colour distribution for galaxies with $M_{\star\mathrm{,\:30\:pkpc}}\geq10^{9.5}\:\mathrm{M}_{\scriptstyle \odot}$. The vertical line delineates the threshold value corresponding to the cut between red and blue galaxies, with all galaxies below the threshold being assigned to the blue cloud and all galaxies above the threshold being used in the fitting and selection of the red sequence.}
    \label{fig:col_dist}
\end{figure*}

In this section, we describe how our threshold for blue/red is determined in the selection of the red sequence and blue cloud galaxies in Section \ref{subsec:delta_col}. We show coarse histograms of the colour distributions for our samples in Fig.~\ref{fig:col_dist}. We locate the minimum value in this distribution between the peak for the blue cloud and the red sequence. This minimum value provides a coarse primary cut between red and blue galaxies, the red end of this cut is further refined into the selection of red sequence galaxies by the method outlined in Section \ref{subsec:measure_colour}.

We also note here that the NewHorizon simulation is dominated by blue cloud galaxies. While this blue cloud is shifted to even bluer values than we see in the other simulations, it is worth noting that NewHorizon only simulates a small region (16 cMpc)$^3$, and therefore, may have simulated an environment conducive to particularly blue galaxies.





\bsp	
\label{lastpage}
\end{document}